\documentclass[11pt]{article}
\pdfoutput = 1
\usepackage[utf8]{inputenc}

\usepackage[margin = 2.2cm]{geometry}
 \makeatletter \g@addto@macro\@floatboxreset\centering \makeatother
\usepackage{color}
\definecolor{darkgreen}{rgb}{0,0.5,0}
\definecolor{darkblue}{rgb}{0,0,0.6}
\definecolor{purple}{rgb}{0.4,.2,0.7}

\usepackage[hyperfootnotes = false, bookmarks = false, colorlinks = true, linkcolor = blue, citecolor = purple]{hyperref}

\usepackage{aas_macros,empheq,fancybox,graphicx,multirow}
\usepackage[bottom]{footmisc}
\usepackage{aas_macros}
\usepackage{esint}
\usepackage[all,cmtip]{xy}
\usepackage{amsfonts}
\usepackage{amsmath}
\usepackage{amssymb}
\usepackage{appendix}
\usepackage{cite}
\usepackage{enumitem}
\usepackage[mathcal]{eucal}
\usepackage{float}
\usepackage{graphicx}
\usepackage{mathabx}
\usepackage{mathrsfs}
\usepackage{pifont}
\usepackage{setspace}
\usepackage{slashed}
\usepackage{titlesec}
\usepackage{subcaption}
\usepackage{comment}
\usepackage{wrapfig}
\usepackage{tikz}
\usetikzlibrary{decorations.markings}
\usepackage{xcolor,colortbl}
\numberwithin{equation}{section}
\newcommand{\ap}{\alpha_+}
\newcommand{\am}{\alpha_-}

\usetikzlibrary{shapes.misc}

\tikzset{cross/.style={cross out, draw=black, minimum size=2*(#1-\pgflinewidth), inner sep=0pt, outer sep=0pt},
cross/.default={1pt}}

\begin{document}

\thispagestyle{empty}

\begin{center}
    ~
    \vskip10mm

     {\LARGE  {\textsc{Comments on the Quantum Field Theory of the Coulomb Gas Formalism }}}
    \vskip10mm
    
 Daniel Kapec$^{a}$ and  Raghu Mahajan$^{b}$\\
    \vskip1em
    
    {\it
        $^a$ School of Natural Sciences, Institute for Advanced Study, Princeton, NJ 08540, USA \\ \vskip1mm
        $^b$ Department of Physics, Stanford University, Stanford, CA 94305, USA\\ \vskip1mm
    }
    \vskip5mm
    \tt{kapec@ias.edu, raghumahajan@stanford.edu}
\end{center}
\vspace{8mm}

\begin{abstract}
\noindent
The holomorphic Coulomb gas formalism, as developed by Feigin-Fuchs, Dotsenko-Fateev and Felder, is a set of rules for computing minimal model observables using free field techniques. We attempt to derive  and clarify these rules using standard techniques of quantum field theory. We begin with a careful examination of the timelike linear dilaton. Although the background charge of the model breaks the scalar field's continuous shift symmetry, the exponential of the action remains invariant under a discrete shift because the background charge is imaginary. Gauging this symmetry makes the dilaton compact and introduces winding modes into the spectrum. One of these winding operators corresponds to the anti-holomorphic completion of the BRST current first introduced by Felder, and the full left/right cohomology of this BRST charge isolates the irreducible representations of the Virasoro algebra within the degenerate Fock space of the linear dilaton. The ``supertrace" in the BRST complex reproduces the minimal model partition function and exhibits delicate cancellations between states with both momentum and winding. The model at the radius $R=\sqrt{pp'}$ has two marginal operators corresponding to the Dotsenko-Fateev ``screening charges." Deforming by them, we obtain a model that might be called a ``BRST quotiented compact timelike Liouville theory." The Hamiltonian of the zero-mode quantum mechanics of this model is not Hermitian, but it is $PT$-symmetric and exactly solvable. Its eigenfunctions have support on an infinite number of plane waves, suggesting an infinite reduction in the number of independent states in the full quantum field theory.  Applying conformal  perturbation theory to the exponential interactions reproduces the Coulomb gas calculations of minimal model correlation functions. In contrast to spacelike Liouville, these ``resonance correlators" are finite because the zero mode is compact. We comment on subtleties regarding the reflection operator identification, as well as naive violations of truncation in correlators with multiple reflection operators inserted. This work is part of an attempt to understand the relationship between the JT model of two dimensional gravity and the worldsheet description of the $(2,p)$ minimal string as suggested by Seiberg and Stanford.

\end{abstract}
\pagebreak


\setcounter{tocdepth}{2}
{\hypersetup{linkcolor=black}
\small
\tableofcontents
}

\newpage
\section{Introduction}

The minimal models of two dimensional conformal field theory are among the most studied and best understood of all quantum field theories\cite{Polyakov:1984yq,Friedan:1983xq,Friedan:1986kd}. They serve as the prototypical examples of rational conformal field theories\cite{Moore:1988uz,Moore:1988qv}, and their solution via the conformal bootstrap exemplifies the power of the abstract, non-Lagrangian approach to conformal field theory and critical phenomena.

Although these models are defined abstractly by their spectrum of local operators  and OPE coefficients, they are perhaps most interesting when viewed in relation to other two-dimensional systems. Each minimal model controls the long-distance dynamics of a rich universality class of interesting and physically realizable lattice models \cite{Andrews:1984af, Huse:1984mn}. Within the realm of continuum field theory, these universality classes are known to include the Landau-Ginzburg models\cite{Zamolodchikov:1986db}, and probably include some version of highly curved, small-AdS quantum gravity in three dimensions \cite{Castro:2011zq,Gaberdiel:2010pz,Gaberdiel:2012uj}. There are interesting renormalization group flows between the different models\cite{Zamolodchikov:1987ti}, and even their massive deformations exhibit fascinating properties \cite{Zamolodchikov:1989hfa}. 

In each of the previous examples, the minimal model describes the (ultraviolet or infrared) asymptotics of another scale-dependent system. The minimal models also have a number of exact descriptions in terms of other seemingly unrelated conformal field theories, each of which involves a gauging procedure or the imposition of constraints. The Goddard-Kent-Olive coset construction of the minimal models\cite{Goddard:1984vk, Goddard:1986ee,Gepner:1986ip,Bowcock:1987mw,Douglas:1987cv} played a key role in demonstrating unitarity of the $(m,m+1)$ series, and the work of Bershadsky and Ooguri relates the models to a Hamiltonian reduction of the $SL(2,\mathbb{R})$ WZW model. The focus of this paper will be on the oldest, and in some sense most puzzling, exact construction of the minimal models: the Coulomb gas formalism. Our main conclusion is that, at least for the $(2,p)$ models, the Coulomb Gas formalism fits comfortably within the standard framework of local quantum field theory, and can in fact be derived from standard QFT manipulations of a somewhat exotic ``compact time-like linear dilaton theory" with a peculiar operator spectrum and a BRST symmetry. The framework could equivalently be described as a ``BRST quotient of the compact time-like Liouville theory." The goal is to subject these models to a careful modern analysis and to clarify their relation to the minimal models.

It was known very early on that a large class of two dimensional lattice models renormalize onto a scalar field at criticality. Indeed, a two dimensional scalar field with a background coupling to curvature can realize any central charge, and the exponential vertex operators of such a model can realize almost any conformal dimension. This fact alone suggests that many fixed points could be described or embedded within such a model, and most known rational conformal field theories do have realizations in terms of multiple free fields. The heuristic suggests that, just as a highly curved manifold can always be embedded in a flat space of sufficiently high dimension, it might be possible to conformally embed any interacting 2d CFT in a set of free (flat) fields subject to some constraints (gauging, deformation by a marginal operator, BRST quotient, etc.). In this sense the Coulomb gas formalism plays a role similar to the Whitney embedding theorem for 2d CFT \cite{Douglas:2010ic}.

The standard free field realization of the minimal models involves a single scalar field, the so-called timelike linear dilaton.\footnote{The terminology ``timelike" derives from the string theory literature, where it is common to use a dilaton with a wrong-sign kinetic term and a real background charge rather than  a dilaton with a conventional kinetic term and an imaginary background charge. The two descriptions are related by the field redefinition $\phi(x)\to i\phi(x)$.}
The linear dilaton first appeared in the physics literature in a paper by Chodos and Thorn \cite{Chodos:1973gt} as part of an attempt to change the critical dimension of the bosonic string. A decade later, Thorn returned to the model to study the Kac determinant formula and to work out explicit expressions for the singular vectors\cite{Thorn:1984sn} (see also the works of Tsuchiya-Kanie\cite{Tsuchiya:1986},  Kato-Matsuda\cite{Kato:1985vq,Kato:1986rq,Kato:1987qda,Matsuda:1987tt}, and Gervais-Neveu\cite{Gervais:1984hw,Gervais:1984sn,Gervais:1984cy}).
In a separate attempt to understand the structure of the reducible Verma modules over the Virasoro algebra (and to prove the Kac determinant formula), Feigin and Fuchs were led to study a series of auxiliary modules (importantly inequivalent to any Verma module), the spaces of semi-infinite forms. Phrased in physical terms, this construction embedded states of the degenerate Virasoro representation in the Fock space of an anticommuting $bc$ ghost system \cite{Feigin:1981st,Feigin:1982tg}. Later work ``bosonized" the construction \cite{Feigin:1986}, leading to the modern presentation of the Coulomb gas formalism. 
In the language of this paper, Feigin and Fuchs investigated the Hilbert space of the timelike linear dilaton, but they did not study its marginal deformations  since they were concerned with kinematics rather than dynamics.

These mathematical developments first made contact with the minimal models through the work of Dotsenko and Fateev\cite{Dotsenko:1984nm,Dotsenko:1984ad,Dotsenko:1985hi,Dotsenko:1986ca}, who invented a seemingly ad-hoc prescription to compute observables in the minimal models using free field correlation functions. Working with the holomorphic sector of the timelike linear dilaton theory
\begin{equation*}
    S[\phi,g]=\frac{1}{4\pi}\int d^2x \sqrt{g} \left[g^{ab}\partial_a\phi \partial_b \phi +Q\phi R(g)\right] \; , \hspace{.5 in} Q=i\frac{p-p'}{\sqrt{pp'}}\equiv i(\alpha_++\alpha_-) \; ,
\end{equation*}
Dotsenko and Fateev were led to identify the minimal model primaries $M_{r,s}(z)$ with the exponential operators $e^{i\alpha_{r,s}\phi(z)}$ of the same conformal dimension (the indices $r,s$ label entries in the Kac table). Genus zero correlation functions of these exponential operators are given by free field expectation values with a modified selection rule $\sum \alpha_i=-2iQ$ on the momenta, and obviously do not reproduce the correlations of the interacting minimal models.
As a remedy, Dotsenko and Fateev identified a pair of dimension $h=1$ primary operators $e^{2i\alpha_{\pm}\phi(z)}$ (importantly, neither corresponds to an operator in the minimal model). Closed line integrals of these operators commute with the Virasoro algebra, but not with the anomalous $U(1)$ current $\partial \phi(z)$. They can therefore be used to ``screen" the background charge asymmetry, producing non-zero values for correlation functions on the sphere that would otherwise vanish. The prescription for calculating the holomorphic parts of correlators takes the schematic form
\begin{equation*}
    \langle M_{r_1,s_1}(z_1)\dots M_{r_n,s_n}(z_n) \rangle \sim  \langle e^{i\alpha_{r_1,s_1}\phi(z_1)}\dots e^{i\alpha_{r_n,s_n}\phi(z_n)}\prod_{i=1}^{n_+} \oint_{\mathcal{C}_i} e^{2i\alpha_+\phi(w_i)}dw_i \prod_{j=1}^{n_-} \oint_{\mathcal{C}_j} e^{2i\alpha_-\phi(w_j)}dw_j \rangle \; ,
\end{equation*}
where the number of screening charges is determined by the selection rule on the momenta. Without any a priori justification for the screening charge prescription, the choice of contours in this formula is not fixed. Instead, the correct combination of contours is determined by combining the holomorphic and anti-holomorphic correlators and requiring locality (absence of branch cuts) in the final answer. With the four-point functions in hand, Dotsenko and Fateev worked backwards to derive the set of OPE coefficients that define each model.

This convoluted procedure devised by Dotsenko and Fateev is commonly known as the Coulomb gas formalism. At the time of its invention, the framework appeared to be a clever trick for finding integral representations of conformal blocks, but the physical interpretation remained murky. Its relation to more central ideas in quantum field theory began with the work of Felder\cite{Felder:1988zp}, who identified a hidden BRST structure in the construction and used it to compute certain genus one observables. 

The most obvious question raised by the work of Dotsenko and Fateev regards the special status of the ``minimal model exponentials" in the linear dilaton theory. In particular, what is it that singles them out from the much larger space of exponential vertex operators in the model? Felder's answer to this question involved an ingenious combination of the work of Thorn, Feigin-Fuchs, and Dotsenko-Fateev. His basic innovation was to reinterpret the construction of Feigin and Fuchs as a BRST quotient of the linear dilaton Hilbert space. Combining Thorn's explicit expressions for the Fock space singular vectors  \cite{Thorn:1984sn,Tsuchiya:1986,Kato:1985vq,Kato:1986rq,Kato:1987qda,Matsuda:1987tt} with the structure of the Feigin-Fuchs bosonic resolution \cite{Feigin:1986}, Felder constructed a series of nilpotent BRST operators from multiple nested line integrals of the Dotsenko-Fateev screening operators.
The cohomology of these charges in the ``minimal model Fock spaces" $F_{r,s}$ yields the corresponding irreducible representation of the Virasoro algebra. Felder's holomorphic BRST complex is infinite in both directions, so his construction involves the extraneous vertex operators present in the linear dilaton but not the minimal model. Crucially, he demonstrated that all of the cohomology of the complex 
\begin{equation*}
\dots \xrightarrow{Q} F_{*} \xrightarrow{Q}  F_{r,s}\xrightarrow{Q} F_{*} \xrightarrow{Q} \dots
\end{equation*}
is concentrated in $F_{r,s}$: the extra operators in the model are all either BRST exact or not BRST closed. Since a trace in the cohomology is the same as the alternating trace in the full complex, Felder was able to easily reproduce the Rocha-Caridi form of the minimal model character \cite{Rocha-Caridi:1983}. A second major component of Felder's work developed the notion of ``screened vertex operators," in which one attaches line integrals of the Dotsenko-Fateev screening operators to the exponential operators. Inside of correlation functions, these integrals can be deformed onto the Dotsenko-Fateev contours, reproducing the Coulomb gas calculation. This construction seems far less natural from the point of view of local quantum field theory, and we will not need it in our formulation. In its place we substitute conformal perturbation theory. 

Felder's work sparked followups in many directions (see \cite{Bouwknegt:1990wa, Bouwknegt:1991mv} for useful reviews). In particular, the ability to project out unwanted states through the BRST quotient opened up the possibility to calculate observables on higher genus surfaces using the Coulomb gas formalism\cite{Felder:1989ve,Felder:1989vx,Felder:1992sv} (see also \cite{Frau:1989mu,Frau:1989yb,DiVecchia:1988cy,Bagger:1989bm,Shin:1992rw,Bagger:1988yc,Jayaraman:1988ex,Crnkovic:1988wc,Sciuto:1992vr,Foda:1988in,Foda:1989ui}).  Importantly, in some of these higher genus calculations it was noted that the scalar field should be compact, although attention to such global issues was scarce. As we will see, this is a crucial ingredient in the Coulomb gas quantum field theory: without it, Felder's BRST current (which is a winding operator in the full model) would not be in the spectrum. The Dotsenko-Fateev contour prescription becomes much more complicated on a surface with nontrivial cycles, and many of the works on higher genus had difficulty explicitly demonstrating modular invariance. This would be an afterthought if the rules could be derived from a description where locality was manifest.

The combined work of Dotsenko-Fateev and Felder is highly suggestive.  
In fact, already in their second paper Dotsenko and Fateev noted that their final answers could be reproduced by surface integrals of a pair of marginal operators. The precise physical interpretation of this fact was left open due to certain technical puzzles that we will describe, but this observation clearly suggests that the minimal model might be obtained as a marginal deformation of the timelike linear dilaton.

The aim of this paper is to combine this observation of Dotsenko-Fateev with Felder's BRST construction, paying close attention to global issues and clarifying some subtle points which have been glossed over in previous treatments.
The main takeaway of the analysis is that compactness of the linear dilaton is crucial, and the BRST structure of the model is indispensable.  Taken together, they conspire to fix several pathologies arising in the naive interpretation of the theory, including some apparent violations of the minimal model fusion rules that actually result from calculating non-BRST invariant quantities.

It is useful to compare the techniques of this paper with the standard treatment of spacelike Liouville theory, which historically also developed from Coulomb gas techniques. The screening charge prescription of Goulian-Li \cite{Goulian:1990qr} calculates a special class of resonance correlators in spacelike Liouville. For this set of correlators, which can be screened by an integer number of Liouville interactions, the zero-mode of the field feels no effective potential. Integrating over this undamped non-compact direction in field space produces divergences in the correlation functions which are identified as poles in the DOZZ formula \cite{Dorn:1994xn,Zamolodchikov:1995aa}, and the residues of these poles match the naive screening charge calculation. In spacelike Liouville theory, most of the correlation functions of interest cannot be screened in this way, and are instead obtained through a clever analytic continuation. In the case at hand it is precisely these resonance correlators that one wants to calculate and compare to the minimal model. We obtain finite answers, along with a normalizable $SL(2,\mathbb{C})$ invariant groundstate and normalizable operators, because our zero mode is compact.
We should also note that there has been scattered work on timelike Liouville theory\cite{Strominger:2003fn, Ribault:2015sxa, Giribet:2011zx, Harlow:2011ny, McElgin:2007ak, Bautista:2019jau,Zamolodchikov:2005fy,Delfino:2010xm,Gori:2017cyq,Santachiara:2013gna,He:2020rfk}. These treatments typically do not compactify the Liouville scalar, nor do they implement the BRST quotient. Taking these global issues into account might shed light on some of the puzzles encountered in these works, especially the failure of truncation noted in \cite{Zamolodchikov:2005fy}.

Finally, we should mention that our primary motivation for this work is to better understand the relationship between the JT model \cite{Jackiw:1984je,Teitelboim:1983ux,Almheiri:2014cka, Engelsoy:2016xyb, Maldacena:2016upp, Jensen:2016pah} of two-dimensional quantum gravity and the worldsheet description of the $(2,p)$ minimal string. The JT model is a profoundly simple theory of two-dimensional gravity in which one only path-integrates over metrics of constant negative curvature (plus some extrinsic wiggles in the asymptotic regions). The path integral of this model has recently been computed to all orders in the genus expansion \cite{Saad:2019lba}, where it was also linked to the type of double-scaled matrix integral known to describe minimal matter coupled to the Liouville field \cite{Brezin:1990rb,Douglas:1989ve,Gross:1989vs,DiFrancesco:1993cyw}. In particular, calculations in the JT description seem to imply a relationship with the $p\to \infty$ limit of the $(2,p)$ minimal model coupled to Liouville. The heuristic explanation for this correspondence
was suggested by Seiberg and Stanford \cite{Saad:2019lba}: as $p\to \infty$ the central charge of the minimal model $c_M\to-\infty$ so that the Liouville central charge tends to $c_L\to+\infty$. This is a semiclassical limit for the conformal factor in which only the saddles (which are constant curvature surfaces) contribute. Formal manipulations \cite{Saad:2019lba, Mertens:2020hbs} involving the Feigin-Fuchs description of the minimal model coupled to Liouville seem to suggest that a direct, rigorous limit might be taken, and this paper is a first step in that direction.

The outline of this paper is as follows. In section \ref{sec:Minimal_Models}, we briefly review well known facts about the minimal models and the structure of degenerate Verma modules. Section \ref{Sec:CoulombGas} presents the standard treatment of the Coulomb gas formalism, including the work of Feigin-Fuchs and Dotsenko-Fateev. Section \ref{sec:Felder} explores Felder's BRST construction in some instructive examples before presenting the general case. In section \ref{sec:LinDil}, we introduce and begin to analyze the compact timelike linear dilaton, paying special attention to the BRST structure of the model as well as its marginal deformations and currents. In particular, we demonstrate that a (slightly nontrivial) generalization of Felder's holomorphic BRST construction reduces the spectrum of the compact timelike linear dilaton to that of the minimal model.  In section \ref{sec:PerturbationTheory}, we perturb the timelike linear dilaton by its marginal deformations, and relate the resulting observables to the screening charge calculations of the Coulomb gas formalism. Along the way we clarify some subtle points regarding the truncation of the OPE and the ``reflection identification" of operators with the same scaling dimension but different $U(1)$ charge.

\section{Minimal models}\label{sec:Minimal_Models}
In this section we quickly review well known facts about the representation theory of the Virasoro algebra at central charge $c< 1$, as well as the operator content and fusion rules of the minimal models. Each minimal model is labeled by a distinct pair of coprime integers $(p,p')$, and our convention is $p>p'$. We will not restrict to the unitary series for which $p=p'+1$, but we will only consider the diagonal models.

\subsection{Structure of reducible Verma modules  }
 In this section we provide a review of the representation theory of the Virasoro algebra
\begin{equation}\label{eq:Virasoro}
    [L_n,L_m]=(n-m)L_{m+n}+ \frac{c}{12}(n^3-n)\delta_{n,-m} 
\end{equation}
when $c< 1$. Textbook treatments can be found in \cite{DiFrancesco:1997nk,Kac:1987gg}. Verma modules $M(h,c)$ over the Virasoro algebra are generated by the action of the $L_{n<0}$ on a highest weight vector $|h,c\rangle$. At each descendant level $n$, the matrix of overlaps is calculated using the adjoints $L_n^\dagger=L_{-n}$ and the commutation relations \eqref{eq:Virasoro}. The determinant of this matrix was worked out by Kac \cite{Kac:1978,Kac:1987gg} and takes the form
\begin{equation}\label{eq:KacDet}
    \text{det}_n(c,h)=K_n\prod_{\substack{r,s\in \mathbb{N}\\ rs\leq n}}(h-h_{r,s})^{p(n-rs)} \;, \hspace{.5 in} K_n=\prod_{\substack{r,s\in \mathbb{N}\\ rs\leq n}}((2r)^ss!)^{p(n-rs)-p(n-r(s+1))}\; .
\end{equation}
In this formula, $p(N)$ is  the number of partitions of $N$. The zeroes of this polynomial are given by
\begin{equation}
    h_{r,s}(c)=\frac{1}{24}(c-1)+\frac{1}{4}(r\alpha_+ + s\alpha_-)^2 \; , \hspace{.5 in} \alpha_{\pm}=\frac{\sqrt{1-c}\pm \sqrt{25-c}}{\sqrt{24}} \; .
\end{equation}
If our aim was to isolate the unitary models, we would require all norms to be positive semi-definite. However, the quantum field theory description of a statistical system at a second order phase transition need not be unitary, and the requirement that we impose instead is that the representations be of type $\text{III}_-$ in the classification of \cite{Feigin:1986}. In this case, there exists a rational relation
\begin{equation}
    p'\alpha_+ + p\alpha_-=0 \; , 
\end{equation}
and we take
\begin{equation}
    \alpha_+=\sqrt{\frac{p}{p'}} \; , \hspace{.5 in} \alpha_-=-\sqrt{\frac{p'}{p}} \; . 
\end{equation}
In terms of these coprime integers, the central charge takes the rational form
\begin{equation}
    c_{p,p'}=1-6\frac{(p-p')^2}{pp'} 
\end{equation}
and the degenerate weights are simply
\begin{equation}\label{eq:MMSpec}
    h_{r,s}=-\frac14(\alpha_++\alpha_-)^2+\frac14(r\alpha_++s\alpha_-)^2 \; . 
\end{equation}
 It is important to note that $h_{r,s}=h_{r+p',s+p}=h_{p'-r,p-s}$. Returning to \eqref{eq:KacDet}, this implies that the Verma module $M(h_{r,s},c_{p,p'})\equiv M(h_{r,s})$ has distinct singular vectors appearing at level $rs$ and level $(p'-r)(p-s)$. It is easy to calculate the conformal dimensions of these null descendants, and one finds that they too are highest weight states for type $\text{III}_-$ Verma modules: 
\begin{align}
    h_{r,s}+rs=h_{p'+r,p-s}=h_{p'-r,p+s}  \; , \hspace{.5 in} h_{r,s}+(p'-r)(p-s)=h_{r,2p-s}=h_{2p'-r,s} \; . 
\end{align}
In order to form the irreducible representation, one must quotient by the maximal sub-representation $M(h_{p'+r,p-s})+ M(h_{r,2p-s})$.  The irreducible state space of the model is therefore
\begin{equation}
     M_{r,s}=M(h_{r,s})/\{M(h_{p'+r,p-s})+ M(h_{r,2p-s}) \} \; . 
\end{equation}
Unfortunately, $ M(h_{p'+r,p-s})\cap M(h_{r,2p-s})\neq 0$, so it is incorrect to simply remove a copy of each state in the two modules $M(h_{p'+r,p-s})$ and $M(h_{r,2p-s})$. 
The Verma module $M(h_{p'+r,p-s})$ has two null descendants occurring at level $l=(p'+r)(p-s)$ and at level $l=(p'-r)(p+s)$, each of which also generates a type $\text{III}_-$ module:
\begin{align}
    h_{p'+r,p-s}+(p'+r)(p-s)&=h_{2p'+r,s}=h_{p'-r,3p-s} \; , \\
    h_{p'+r,p-s}+(p'-r)(p+s)&=h_{3p'-r,p-s}=h_{r,2p+s} \; .
\end{align}
Importantly, these are the same null states appearing in the Verma module $M(h_{r,2p-s})$ at the levels $l=r(2p-s)$ and $l=s(2p'-r)$:
\begin{align}
    h_{r,2p-s}+r(2p-s)=h_{2p'+r,s}=h_{p'-r,3p-s} \; , \hspace{.5 in}
    h_{r,2p-s}+(2p'-r)s=h_{3p'-r,p-s}=h_{r,2p+s} \; . 
\end{align}
Since both $M(h_{p'+r,p-s})$ and $M(h_{r,2p-s})$ contain the same two submodules, the sum is a quotient
\begin{equation}
    M(h_{p'+r,p-s})+ M(h_{r,2p-s}) =\{M(h_{p'+r,p-s})\oplus M(h_{r,2p-s})\}/ \{M(h_{2p'+r,s})+ M(h_{r,2p+s}) \} \; . 
\end{equation}
Therefore, if one removes a copy of each state in the two modules $M(h_{p'+r,p-s})$ and $M(h_{r,2p-s})$ from $M(h_{r,s})$, one must add back in the states contained in $M(h_{2p'+r,s})+ M(h_{r,2p+s})$.

\begin{figure}
\begin{tikzpicture}

\draw[->,gray, very thick] (.15,-.15) -- (.85,-.85);
\draw[->,gray, very thick] (-.15,-.15) -- (-.85,-.85);
\filldraw[black] (0,0) circle (2pt) node[anchor=south] { $(r,s)$};

\draw[->,gray, very thick] (1,-1.15) -- (1,-2.25);
\draw[->,gray, very thick] (-1,-1.15) -- (-1,-2.25);
\draw[->,gray, very thick] (.85,-1.15) -- (-.85,-2.25);
\draw[->,gray, very thick] (-.85,-1.15) -- (.85,-2.25);

\filldraw[black] (1,-1) circle (2pt) node[anchor=west] { $(p'+r,p-s)$};
\filldraw[black] (1,-2.4) circle (2pt) node[anchor=west] { $(2p'+r,s)$};

\filldraw[black] (-1,-1) circle (2pt) node[anchor=east] { $(r,2p-s)$};
\filldraw[black] (-1,-2.4) circle (2pt) node[anchor=east] { $(r,2p+s)$};

\draw[->,gray, very thick] (-1,-2.55) -- (-1,-3.65);
\draw[->,gray, very thick] (1,-2.55) -- (1,-3.65);

\draw[->,gray, very thick] (-.85,-2.55) -- (.85,-3.65);
\draw[->,gray, very thick] (.85,-2.55) -- (-.85,-3.65);

\filldraw[black] (-1,-3.8) circle (2pt) node[anchor=east] { };
\filldraw[black] (1,-3.8) circle (2pt) node[anchor=east] { };

\filldraw[gray] (0,-4.4) circle (1pt) node[anchor=south] {};
\filldraw[gray] (0,-4.6) circle (1pt) node[anchor=south] {};
\filldraw[gray] (0,-4.8) circle (1pt) node[anchor=south] {};

\filldraw[black] (-1,-5.2) circle (2pt) node[anchor=east] { $(r,kp+(-1)^ks+p[1-(-1)^k]/2)$};
\filldraw[black] (1,-5.2) circle (2pt) node[anchor=west] {$(r+kp',(-1)^ks +p[1-(-1)^k]/2)$};

\draw[->,gray, very thick] (1,-5.35) -- (1,-6.45);
\draw[->,gray, very thick] (-1,-5.35) -- (-1,-6.45);

\draw[->,gray, very thick] (.85,-5.35) -- (-.85,-6.45);
\draw[->,gray, very thick] (-.85,-5.35) -- (.85,-6.45);
\filldraw[black] (-1,-6.6) circle (2pt) node[anchor=east] { };
\filldraw[black] (1,-6.6) circle (2pt) node[anchor=west] {};

\filldraw[gray] (0,-7.2) circle (1pt) node[anchor=south] {};
\filldraw[gray] (0,-7.4) circle (1pt) node[anchor=south] {};
\filldraw[gray] (0,-7.6) circle (1pt) node[anchor=south] {};

\end{tikzpicture}
\caption{Structure of reducible Verma modules. An arrow from one point to the next indicates that the highest weight vector of the second module is a singular vector of the first module.}\label{fig:Verma}
\end{figure}
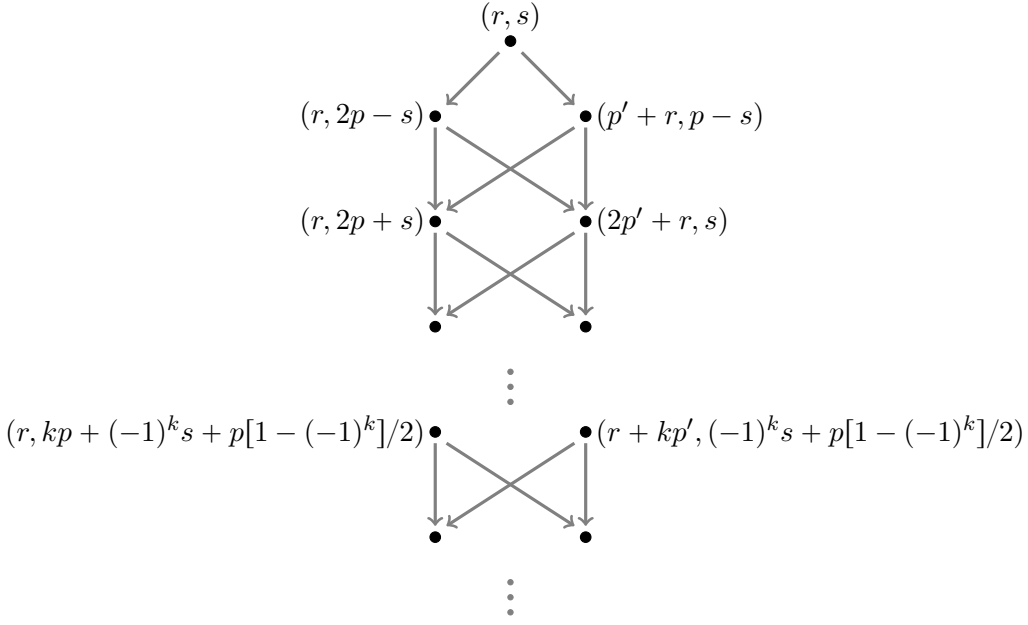

Repeating this analysis allows one to calculate the sum  $M(h_{2p'+r,s})+ M(h_{r,2p+s})$ in terms of the direct sum $M(h_{2p'+r,s})\oplus M(h_{r,2p+s})$ :
\begin{equation}
   M(h_{2p'+r,s})+ M(h_{r,2p+s}) =\{ M(h_{2p'+r,s})\oplus M(h_{r,2p+s}) \}/ \{M(h_{3p'+r,p-s})+ M(h_{r,4p-s}) \} \; .
\end{equation}
This pattern persists indefinitely, and is illustrated in figure \ref{fig:Verma}.

\subsection{Minimal model spectrum, OPE coefficients and torus partition function}
The diagonal minimal models contain scalar primary operators $M_{r,s}(x)$ of weight $h=\bar{h}=h_{r,s}$ with $r$ and $s$ in the range 
\begin{equation}
    1\leq r \leq p'-1 \; , \hspace{.5 in} 1\leq s \leq p-1 \; , 
\end{equation}
subject to the identification $M_{p'-r,p-s}=M_{r,s}$.

The fusion rules of the $(p,p')$ minimal model take the form 
\begin{equation}
    M_{r_1,s_1}\times M_{r_2,s_2}=\sum_{r_3=1+|r_1-r_2|}^{r_{\text{max}}} \sum_{s_3=1+|s_1-s_2|}^{s_{\text{max}}} M_{r_3,s_3} \; ,
\end{equation}
with $r_1+r_2+r_3$ and $s_1+s_2+s_3$ simultaneously odd and
\begin{align}
    r_{\text{max}}&=\min (r_1+r_2-1,2p'-1-r_1-r_2)\; , \\
    s_{\text{max}}&=\min (s_1+s_2-1,2p-1-s_1-s_2) \; .
\end{align}
The OPE coefficients, denoted in this section by $D_{r_1,s_1;r_2,s_2}^{r_3,s_3}$, were originally derived by Dotsenko and Fateev using the Coulomb gas formalism \cite{Dotsenko:1985hi}. For operators with canonically normalized two-point functions, they are given by
\begin{equation}
    D_{r_1,s_1;r_2,s_2}^{r_3,s_3}= \left(\frac{R(r_1,s_1)R(r_2,s_2)}{R(r_3,s_3)}\right)^{-\frac12}C_{r_1,s_1;r_2,s_2}^{r_3,s_3} \; , 
\end{equation}
where
\begin{align}
    C_{r_1,s_1;r_2,s_2}^{r_3,s_3}&=\mu_{l',l}\prod_{i=0}^{l'-2}\prod_{j=0}^{l-2}\frac{([r_2-1-i]-\rho[s_2-1-j])^2([r_1-1-i]-\rho[s_1-1-j])^2}{([r_3+1+i]-\rho[s_3+1+j])^2}\\
    &\times \prod_{i=0}^{l'-2}\frac{\Gamma(-\rho'[r_2-1-i]+s_2)\Gamma(-\rho'[r_1-1-i]+s_1)\Gamma(\rho'[r_3+1+i]-s_3)}{\Gamma(1+\rho'[r_2-1-i]-s_2)\Gamma(1+\rho'[r_1-1-i]-s_1)\Gamma(1-\rho'[r_3+1+i]+s_3)}\notag\\
    &\times \prod_{j=0}^{l-2} \frac{\Gamma(-\rho[s_2-1-j]+r_2)\Gamma(-\rho[s_1-1-j]+r_1)\Gamma(\rho[s_3+1+j]-r_3)}{\Gamma(1+\rho[s_2-1-j]-r_2)\Gamma(1+\rho[s_1-1-j]-r_1)\Gamma(1-\rho[s_3+1+j]+r_3)}\; ,
    \notag
\end{align}
\begin{equation}\label{eq:Reflection}
    \frac{1}{R(r,s)}=\prod_{i=1}^{r-1}\prod_{j=1}^{s-1}\frac{(1+i-\rho(1+j))^2}{(i-j\rho)^2}\prod_{i=1}^{r-1}\frac{\Gamma(i\rho')\Gamma(2-\rho'(1+i))}{\Gamma(1-i\rho')\Gamma(-1+\rho'(1+i))}
    \prod_{j=1}^{s-1}\frac{\Gamma(j\rho)\Gamma(2-\rho(1+j))}{\Gamma(1-j\rho)\Gamma(-1+\rho(1+j))} \;,
\end{equation}
and
\begin{equation}
    \mu_{r,s}(\rho)=\rho^{4(r-1)(s-1)}\prod_{i=1}^{r-1}\prod_{j=1}^{s-1}\frac{1}{(i-\rho j)^2}\prod_{i=1}^{r-1}\frac{\Gamma(i\rho')}{\Gamma(1-i\rho')}\prod_{j=1}^{s-1}\frac{\Gamma(j\rho)}{\Gamma(1-j\rho)} \; .
\end{equation}
In these formulas
\begin{equation}
    l'=\frac12(r_1+r_2-r_3+1) \; , \hspace{.25 in} l=\frac12(s_1+s_2-s_3+1) \; , \hspace{.25 in} \rho=\alpha_+^2 \; , \hspace{.25 in} \rho'=\alpha_-^2 \; .
\end{equation}
For the OPE coefficients in the non-diagonal models, see \cite{Fuchs:1988xm,Fuchs:1989kz}.

\subsubsection*{Representations of the torus partition function}
The character associated to each irreducible representation $M_{r,s}$ follows directly from the structure of inclusions in the degenerate Verma module:
 \begin{equation}
     \chi_{r,s}(q)=\frac{q^{(1-c)/24}}{\eta(q)}\left[q^{ h_{r,s}}+\sum_{k=1}^{\infty}(-1)^k\left(q^{ h_{r+kp',(-1)^ks+\frac12 (1-(-1)^k)p}}+ q^{h_{r,kp+(-1)^ks+\frac12 (1-(-1)^k)p}}\right)\right] \; .
 \end{equation}
Here $\eta(q)=q^{1/24}\prod_{n=1}^\infty(1-q^n)$ is the Dedekind eta function and $q=e^{2\pi i \tau}$ is the elliptic nome. There is also a simpler expression, originally due to Rocha-Caridi\cite{Rocha-Caridi:1983}, which will arise more naturally in the Coulomb gas construction:
\begin{equation}
    \chi_{r,s}(q)=K_{r,s}^{(p,p')}(q)- K_{r,-s}^{(p,p')}(q) \; , \hspace{1 in}
      K_{r,s}^{(p,p')}(q)=\frac{1}{\eta(q)}\sum_{n\in \mathbb{Z}}q^{(2pp'n+pr-p's)^2/4pp'} \; .
\end{equation}
This will be the form of the trace obtained through Felder's BRST construction.
Summing over all of the primaries in the Kac table, the partition function of the diagonal minimal model can be written in terms of the partition function $Z(R)$ of the $c=1$ compact boson \cite{Cappelli:1986hf,diFrancesco:1987qf,DiFrancesco:1987gwq}
\begin{equation}\label{eq:Partition}
    2Z_{(p,p')}=Z(\sqrt{pp'})-Z(\sqrt{p'/p}) \; .
\end{equation}
This can of course be rewritten several ways using $T$-duality.

\section{Review of the Coulomb gas formalism}\label{Sec:CoulombGas}
The conformal data of the minimal models reviewed in section \ref{sec:Minimal_Models} completely characterizes them as abstract conformal field theories. Historically, this data (the OPE coefficients in particular, but also the structure of the degenerate Verma modules) was actually derived using a set of free field techniques known as the Coulomb gas formalism \cite{Dotsenko:1984nm,Dotsenko:1984ad,Dotsenko:1985hi,Dotsenko:1986ca}.
This relationship with the free field progressed in several stages. The relationship between degenerate Verma modules over the Virasoro algebra and the timelike linear dilaton Fock space was proven in all generality by Feigin and Fuchs, who built on earlier work by Kac. We briefly review their results in section \ref{sec:FeiginFuchs}, focusing on simple examples that will play a role in the rest of the paper. These mathematical results were followed by the work of Dotsenko and Fateev, who first understood how to calculate genus zero minimal model correlation functions using the free field representation. This well known technique is reviewed in section \ref{sec:DotsenkoFateev}. The extension of the Coulomb gas formalism to higher genus surfaces required understanding the role of the numerous operators present in the linear dilaton description but absent in the minimal model. This problem was essentially solved by Felder, who unearthed a BRST structure in the model that can be used to project out the unphysical states that would otherwise contribute on higher genus surfaces. We describe this construction in section \ref{sec:Felder}.

\subsection{$U(1)$ current algebra and the Feigin-Fuchs free field resolution}\label{sec:FeiginFuchs}
The Feigin-Fuchs free field resolution of the degenerate Virasoro representations has its roots in the oscillator description of the old dual resonance models \cite{Mandelstam:1974fq,Thorn:1984sn}. Since the work of Feigin and Fuchs is completely kinematical and deals only with the representation theory of the Virasoro algebra, it can be completely described with a free Lagrangian field theory. The relation of this model to the interacting minimal models will be the focus of sections \ref{sec:LinDil}-\ref{sec:PerturbationTheory}.

The basic Lagrangian model for the Feigin-Fuchs construction is the linear dilaton with action
\begin{equation}\label{eq:LinDil1}
    S[\phi,g]=\frac{1}{4\pi}\int \sqrt{g}d^2x \left[g^{ab}\partial_a\phi \partial_b \phi +Q\phi R(g)\right] 
\end{equation}
and an imaginary background charge $Q=i(\alpha_++\alpha_-)$. Although the system is conformally invariant, it exhibits a number of peculiar, seemingly pathological features. 
The model will be thoroughly analyzed in section \ref{sec:LinDil} as a proper quantum field theory. In this section, as in the work of Feigin and Fuchs, we will view it simply as a vehicle for studying degenerate representations of the Virasoro algebra. In particular, we will work holomorphically and will not need to consider global issues like the compactness of $\phi(x)$.

\subsubsection*{Free scalar without background charge }
It will be useful to collect formulas for the theory with $Q=0$ in order to highlight the effect of the background charge.
 In our conventions,\footnote{Our conventions are those of \cite{Polchinski:1998rq} with $\alpha '=1$.} the two-point function of the scalar field is given by
\begin{equation}
   \langle \phi(x) \phi (y)\rangle = -\frac{1}{2}\log(x-y)^2 \; . 
\end{equation}
On any two dimensional surface, the model without a background charge has an exactly conserved current whose holomorphic component is a primary operator with the mode expansion
\begin{equation}
  J(z)=  i\partial \phi(z) \equiv\frac{1}{\sqrt{2}}\sum_{m=-\infty}^{\infty}\frac{j_m}{z^{m+1}} \; . 
\end{equation}
The modes of this current are simply the Fock space creation and annihilation operators
\begin{equation}
    j_m=\sqrt{2}\oint \frac{dz}{2\pi}z^m\partial \phi(z) \; , 
\end{equation}
and they satisfy the  commutation relations of the Heisenberg algebra
\begin{equation}
    \left[j_m,j_n\right]=m\delta_{m,-n} \; . 
\end{equation}
The holomorphic energy momentum tensor and its moments take the Sugawara form\footnote{These operators are defined by normal ordering, which we suppress.}
\begin{equation}
    T(z)=-\partial \phi(z) \partial \phi(z) =J(z)J(z) \; , \hspace{1 in} L_m=\frac{1}{2}\sum_{n=-\infty}^{\infty}j_{m-n}j_n \; , 
\end{equation}
from which it follows that
\begin{equation}
    \left[L_m,j_n\right]=-nj_{m+n} \; . 
\end{equation}
The spectrum of the model contains holomorphic vertex operators $e^{i\alpha \phi(z)}$ with weight $h=\frac{\alpha^2}{4}$, and
the state-operator correspondence assigns a state $|\alpha\rangle$ on the cylinder to each such primary operator inserted at the origin of the plane. These states satisfy
\begin{equation}\label{eq:HighestWeight}
    L_0|\alpha\rangle=\frac{\alpha^2}{4}|\alpha \rangle \; , \hspace{.5 in} j_0|\alpha\rangle=\frac{\alpha}{\sqrt{2}}|\alpha\rangle \; , \hspace{.5 in} L_{n>0}|\alpha\rangle=j_{n>0}|\alpha\rangle=0 \; . 
\end{equation}
For each such state we can consider the Fock space of current algebra descendants generated by vectors of the form
\begin{equation}
  j_{-n_1}\dots j_{-n_k}|\alpha\rangle \; . 
\end{equation}
The adjoint with respect to the standard CFT$_2$ inner product is
\begin{equation}
    L_{n}^\dagger=L_{-n}\;, \hspace{1 in} j_n^\dagger=j_{-n} \; . 
\end{equation}
Of particular importance is the state $j_{-1}|0\rangle$ corresponding to the conserved current. Since $[L_1,j_{-1}]=\nolinebreak j_0$ and $[L_2,j_{-1}]=j_1$, this state is annihilated by all $L_{n>0}$ and is therefore a Virasoro singular vector.\footnote{In what follows, we use the term ``singular vector" to denote a Virasoro singular vector since we are ultimately only interested in Virasoro representations. The terminology includes a current algebra descendant which is Virasoro primary.} The current algebra Fock space is an irreducible representation of the Heisenberg algebra but contains multiple representations of the Virasoro algebra. Note that although it is orthogonal to all Virasoro descendants of the identity, the level one current algebra descendant is of course not a zero norm state:
\begin{equation}
    \langle 0 | j_1 j_{-1}|0\rangle =1 \; . 
\end{equation}
As we will see, this discussion is related to the reducibility of the Verma module $M(h,c)$ with $c=1$ and $h=m^2/4$ with $m\in \mathbb{Z}$. For instance, for $m=1$ it is easy to verify that
\begin{equation}
    [L_{-2} -L_{-1}^2]|\alpha=1\rangle 
\end{equation}
is annihilated by $L_{1}$ and $L_{2}$, and therefore by all $L_{n>0}$. More generally, the exponential $e^{in\phi(z)}$ has an infinite set of singular vectors with $L_0=\frac{1}{4}(n+2k)^2$ within its highest weight module.
We will also need the commutators (for $n\geq -1$)
\begin{equation}\label{eq:VertexComm}
    \left[L_n, e^{i\alpha \phi(z)}\right]= \left(z^{n+1}\frac{d}{dz}+\frac{\alpha^2}{4}(n+1)z^n\right)e^{i\alpha \phi(z)} \; , \hspace{.5 in} \left[j_n,e^{i\alpha \phi(z)}\right]= \frac{\alpha}{\sqrt{2}} z^n e^{i\alpha \phi(z)} \; . 
\end{equation}

\subsubsection*{Non-zero background charge}
The nontrivial coupling of the model \eqref{eq:LinDil1} to the background curvature affects both the spectrum and correlation functions of the model. The effect on correlation functions is the focus of the work by Dotsenko and Fateev and will be described in the next section. The  effect which is important for the Feigin-Fuchs analysis is the change in the spectrum. The background coupling to curvature adds an improvement term to the energy momentum tensor, which becomes
\begin{equation}
    T(z)=-\partial \phi(z) \partial \phi(z) + Q \partial^2\phi(z) \; . 
\end{equation}
It is useful to view this as a twist of the original theory's energy momentum tensor by the  $U(1)$ current
\begin{equation}
    T(z)\to \hat{T}(z)=T(z)-iQ\partial J(z)\; . 
\end{equation}
This twist mixes the Virasoro and current algebra generators
\begin{align}\label{eq:Lhat}
    L_m \to \hat{L}_m&=L_m +\frac{iQ(m+1)}{\sqrt{2}}j_m \\
    &=\frac{1}{2}\sum_{n=-\infty}^{\infty}j_{m-n}j_n+\frac{iQ(m+1)}{\sqrt{2}}j_m
    \; , 
\end{align}
and the new generators $\hat{L}_m$ satisfy the Virasoro algebra with central charge $c=1+6Q^2$. The new conformal dimensions of the exponential operators follow simply from \eqref{eq:Lhat} and \eqref{eq:HighestWeight}:
\begin{equation}
    \left[\hat{L}_0,e^{i\alpha \phi(z)}\right]= \left(\frac{\alpha^2}{4} + iQ\frac{\alpha}{2}\right)e^{i\alpha \phi(z)} \; . 
\end{equation}
Perhaps most importantly, the twist also alters the Virasoro Kac-Moody commutation relation
\begin{equation}
    \left[\hat{L}_m,j_n\right]=-nj_{m+n}+ \frac{iQ}{\sqrt{2}}m(m+1)\delta_{m,-n}  \; . 
\end{equation}
In particular, the relation $[\hat{L}_1,j_{-1}]=j_0+i\sqrt{2}Q$ implies that $j_{-1}|0\rangle$ is no longer a singular vector:
\begin{equation}
  \hat{L}_1j_{-1}|0\rangle= i\sqrt{2}Q|0\rangle \neq 0 \; .
\end{equation}
This is equivalent to the statement that the current $J(z)=i\partial \phi(z)$ is no longer a primary operator in the twisted model due to the anomalous OPE with the energy-momentum tensor 
\begin{equation}
    T(z)J(w) \sim \frac{iQ}{(z-w)^3} + \frac{J}{(z-w)^2}+ \frac{\partial J}{z-w} + \dots
\end{equation}
 Taking the adjoint of the equation $[\hat{L}_1,j_{-1}]=j_0+i\sqrt{2}Q$ and requiring $\hat{L}^\dagger_n=\hat{L}_{-n}$ along with $j_n^\dagger=j_{-n}$ for $n\neq0$ demands that $j_0^\dagger=j_0+\sqrt{2}iQ$. Evaluating $\langle 0| j_0 \prod^n_{i=1} e^{i\alpha_i \phi(z_i)} |0\rangle$ in two ways then gives the genus zero selection rule
\begin{equation}
    \sum_{i=1}^n \alpha_i =-2iQ=2\alpha_++2\alpha_- \; 
\end{equation}
and requires $|\alpha\rangle^\dagger=\langle2\alpha_++2\alpha_--\alpha|$ for charged states. Although the primary motivation for Feigin and Fuchs was to determine the structure of the reducible Verma modules over the Virasoro algebra, they were led to study this model (more precisely, its fermionic version, or the space of semi-infinite forms) as an intermediate step in proving the Kac determinant formula. Although the structure of the Fock space is not the same as that of the Verma module, it is easier to study in some respects because of the extra structure (current algebra) that it carries. The Kac determinant can be viewed as the determinant of the map from a Verma module to its contragradient Verma module\footnote{If we denote the Verma module grading as $M=\oplus M_j$, then the contragradient Verma module satisfies $M^c_j=\text{Hom}(M_j,\mathbb{C}$.) } $M(h,c)\to M^c(h,c)$ given by the matrix of overlaps of descendants of the highest weight vector. Because the Fock space $F_{Q,\alpha}$ with $c=1+6Q^2$ has a highest weight state with $h=\frac{\alpha}{2}(\frac{\alpha}{2}+iQ)$ and furnishes a representation of the Virasoro algebra, there are also unique homomorphisms from $M(h,c)$ to $F_{Q,\alpha}$ and from $F_{Q,\alpha}$ to $M^c(h,c)$. We will denote these maps $\Gamma$ and $L$. The situation is depicted in the following diagram  \cite{iohara2010representation}:
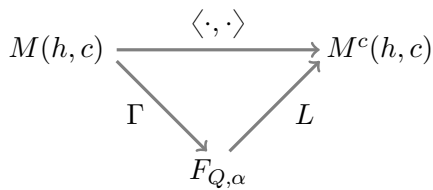
\begin{figure}[h]
\begin{tikzpicture}
\draw[->,gray, very thick] (.15,0) -- (2.85,0);
\draw[->,gray, very thick] (.15,-.15) -- (1.35,-1.35);
\draw[->,gray, very thick] (1.65,-1.35) -- (2.85,-.15);

\node[] at (-.65,0) {$M(h,c)$};
\node[] at (3.65,0) {$M^c(h,c)$};
\node[] at (1.5,-1.65) {$F_{Q,\alpha}$};

\node[] at (1.5,.35) {$\langle \cdot , \cdot \rangle$};
\node[] at (.4,-.85) {$\Gamma$};
\node[] at (2.65,-.85) {$L$};

\end{tikzpicture}
\caption{Route to the Kac determinant.}
\end{figure}

There is a natural grading according to level in each space and the maps restrict to the fixed-level subspaces. Importantly, the maps $\Gamma,L$ need not be isomorphisms. Since we are only interested in the case of the type $\text{III}_-$ Verma modules, there are two interesting possibilities. If $\Gamma$ is an isomorphism, then the Fock space is also reducible with two singular vectors at levels $rs$ and $(p'-r)(p-s)$. However, it could also be the case that $\Gamma$ is not surjective, in which case the Fock space is still reducible (the image of $\Gamma$ is a submodule), but the image of the second singular vector in $F_{Q,\alpha}$ might be zero. In this case the  role of the second singular vector is played by a current algebra descendant which is not a Virasoro descendant. It is this second scenario that will be relevant in what follows. The composition of $\Gamma$ and $L$ is a homomorphism from $M \to M^c$ that sends highest weight to highest weight. Such a map is unique (up to a scalar) by the universal property of Verma modules, so computing the determinants of $\Gamma$ and $L$ produces the Kac determinant and indirectly determines the singular vector structure of the original Verma module.

\begin{figure}[h]
\begin{tikzpicture}

\draw[<-,gray, very thick] (.15,-.15) -- (.85,-.85);
\draw[->,gray, very thick] (-.15,-.15) -- (-.85,-.85);
\filldraw[black] (0,0) circle (2pt) node[anchor=south] { $v_0$};

\draw[->,gray, very thick] (1,-1.15) -- (1,-2.25);
\draw[->,gray, very thick] (-1,-2.25) -- (-1,-1.15) ;
\draw[<-,gray, very thick] (-.85,-2.25) -- (.85,-1.15);
\draw[<-,gray, very thick] (-.85,-1.15) -- (.85,-2.25);

\filldraw[blue] (1,-1) circle (2pt) node[anchor=west] { $w_0$};
\filldraw[red] (1,-2.4) circle (2pt) node[anchor=west] { $v_1$};

\filldraw[black] (-1,-1) circle (2pt) node[anchor=east] { $u_1$};
\filldraw[red] (-1,-2.4) circle (2pt) node[anchor=east] { $v_{-1}$};

\draw[->,gray, very thick] (-1,-2.55) -- (-1,-3.65);
\draw[->,gray, very thick] (1,-3.65) -- (1,-2.55);

\draw[<-,gray, very thick] (-.85,-2.55) -- (.85,-3.65);
\draw[->,gray, very thick] (.85,-2.55) -- (-.85,-3.65);

\filldraw[black] (-1,-3.8) circle (2pt) node[anchor=east] { $u_2$};
\filldraw[blue] (1,-3.8) circle (2pt) node[anchor=west] { $w_1$};

\draw[->,gray, very thick] (1,-3.95) -- (1,-5.05);
\draw[<-,gray, very thick] (-1,-3.95) -- (-1,-5.05);

\filldraw[red] (-1,-5.2) circle (2pt) node[anchor=east] { $v_{-2}$};
\filldraw[red] (1,-5.2) circle (2pt) node[anchor=west] {$v_2$};

\draw[<-,gray, very thick] (1,-5.35) -- (1,-6.45);
\draw[->,gray, very thick] (-1,-5.35) -- (-1,-6.45);

\draw[->,gray, very thick] (.85,-3.95) -- (-.85,-5.05);
\draw[->,gray, very thick] (.85,-5.05) -- (-.85,-3.95);

\draw[->,gray, very thick] (.85,-5.35) -- (-.85,-6.45);
\draw[<-,gray, very thick] (-.85,-5.35) -- (.85,-6.45);
\filldraw[black] (-1,-6.6) circle (2pt) node[anchor=east] { };
\filldraw[black] (1,-6.6) circle (2pt) node[anchor=west] {};

\filldraw[gray] (0,-7.2) circle (1pt) node[anchor=south] {};
\filldraw[gray] (0,-7.4) circle (1pt) node[anchor=south] {};
\filldraw[gray] (0,-7.6) circle (1pt) node[anchor=south] {};

\filldraw[black] (-1,-6.6) circle (2pt) node[anchor=east] { $u_{3}$};
\filldraw[blue] (1,-6.6) circle (2pt) node[anchor=west] {$w_2$};

\end{tikzpicture}
\caption{Feigin-Fuchs characterization of the Fock space structure.}\label{fig:FockStructure}
\end{figure}
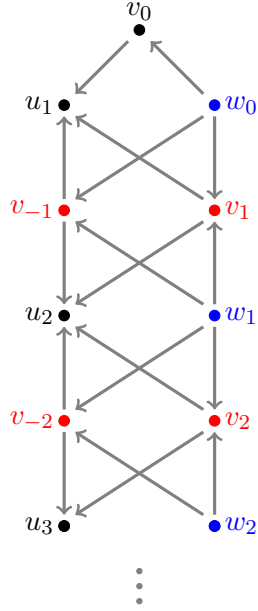

\begin{figure}[h]
\begin{tikzpicture}

\draw[<-,gray, very thick] (.15,-.15) -- (.85,-.85);
\draw[->,gray, very thick] (-.15,-.15) -- (-.85,-.85);
\filldraw[black] (0,0) circle (2pt) node[anchor=south] { $v_0$};

\draw[->,gray, very thick] (1,-1.15) -- (1,-2.25);
\draw[->,gray, very thick] (-1,-2.25) -- (-1,-1.15) ;
\draw[<-,gray, very thick] (-.85,-2.25) -- (.85,-1.15);
\draw[<-,gray, very thick] (-.85,-1.15) -- (.85,-2.25);

\filldraw[blue] (1,-1) circle (2pt) node[anchor=west] { $w_0$};
\filldraw[red] (1,-2.4) circle (2pt) node[anchor=west] { $v_1$};

\filldraw[black] (-1,-1) circle (2pt) node[anchor=east] { $u_1$};
\filldraw[red] (-1,-2.4) circle (2pt) node[anchor=east] { $v_{-1}$};

\draw[->,gray, very thick] (-1,-2.55) -- (-1,-3.65);
\draw[->,gray, very thick] (1,-3.65) -- (1,-2.55);

\draw[<-,gray, very thick] (-.85,-2.55) -- (.85,-3.65);
\draw[->,gray, very thick] (.85,-2.55) -- (-.85,-3.65);

\filldraw[black] (-1,-3.8) circle (2pt) node[anchor=east] { $u_2$};
\filldraw[blue] (1,-3.8) circle (2pt) node[anchor=west] { $w_1$};

\draw[->,gray, very thick] (1,-3.95) -- (1,-5.05);
\draw[<-,gray, very thick] (-1,-3.95) -- (-1,-5.05);

\filldraw[red] (-1,-5.2) circle (2pt) node[anchor=east] { $v_{-2}$};
\filldraw[red] (1,-5.2) circle (2pt) node[anchor=west] {$v_2$};

\draw[<-,gray, very thick] (1,-5.35) -- (1,-6.45);
\draw[->,gray, very thick] (-1,-5.35) -- (-1,-6.45);

\draw[->,gray, very thick] (.85,-3.95) -- (-.85,-5.05);
\draw[->,gray, very thick] (.85,-5.05) -- (-.85,-3.95);

\draw[->,gray, very thick] (.85,-5.35) -- (-.85,-6.45);
\draw[<-,gray, very thick] (-.85,-5.35) -- (.85,-6.45);
\filldraw[black] (-1,-6.6) circle (2pt) node[anchor=east] { };
\filldraw[black] (1,-6.6) circle (2pt) node[anchor=west] {};

\filldraw[gray] (0,-7.2) circle (1pt) node[anchor=south] {};
\filldraw[gray] (0,-7.4) circle (1pt) node[anchor=south] {};
\filldraw[gray] (0,-7.6) circle (1pt) node[anchor=south] {};

\filldraw[black] (-1,-6.6) circle (2pt) node[anchor=east] { $u_{3}$};
\filldraw[blue] (1,-6.6) circle (2pt) node[anchor=west] {$w_2$};

\draw[<-,gray, very thick] (5.15,-.15) -- (5.85,-.85);
\filldraw[black] (5,0) circle (2pt) node[anchor=south] { $v_0$};

\draw[->,gray, very thick] (6,-1.15) -- (6,-2.25);
\draw[<-,gray, very thick] (4.15,-2.25) -- (5.85,-1.15);

\filldraw[blue] (6,-1) circle (2pt) node[anchor=west] { $w_0$};
\filldraw[black] (6,-2.4) circle (2pt) node[anchor=west] { $v_1$};

\draw (4,-1) node[cross=3pt,black] {};
\filldraw[black] (4,-2.4) circle (2pt) node[anchor=east] { $v_{-1}$};

\draw[->,gray, very thick] (6,-3.65) -- (6,-2.55);

\draw[<-,gray, very thick] (4.15,-2.55) -- (5.85,-3.65);

\draw (4,-3.8) node[cross=3pt,black] {};
\filldraw[blue] (6,-3.8) circle (2pt) node[anchor=west] { $w_1$};

\draw[->,gray, very thick] (6,-3.95) -- (6,-5.05);

\filldraw[black] (4,-5.2) circle (2pt) node[anchor=east] { $v_{-2}$};
\filldraw[black] (6,-5.2) circle (2pt) node[anchor=west] {$v_2$};

\draw[<-,gray, very thick] (6,-5.35) -- (6,-6.45);

\draw[->,gray, very thick] (5.85,-3.95) -- (4.15,-5.05);

\draw[<-,gray, very thick] (4.15,-5.35) -- (5.85,-6.45);

\filldraw[black] (6,-6.6) circle (2pt) node[anchor=west] {};

\filldraw[gray] (5,-7.2) circle (1pt) node[anchor=south] {};
\filldraw[gray] (5,-7.4) circle (1pt) node[anchor=south] {};
\filldraw[gray] (5,-7.6) circle (1pt) node[anchor=south] {};

\draw (4,-6.6) node[cross=3pt,black] {};
\filldraw[blue] (6,-6.6) circle (2pt) node[anchor=west] {$w_2$};

\draw (10,0) node[cross=3pt,black] {};

\filldraw[black] (11,-1) circle (2pt) node[anchor=west] { $w_0$};
\draw (11,-2.4) node[cross=3pt,black] {};

\draw (9,-1) node[cross=3pt,black] {};
\draw (9,-2.4) node[cross=3pt,black] {};

\draw (9,-3.8) node[cross=3pt,black] {};
\filldraw[black] (11,-3.8) circle (2pt) node[anchor=west] { $w_1$};

\draw (9,-5.2) node[cross=3pt,black] {};
\draw (11,-5.2) node[cross=3pt,black] {};

\filldraw[black] (11,-6.6) circle (2pt) node[anchor=west] {};

\filldraw[gray] (10,-7.2) circle (1pt) node[anchor=south] {};
\filldraw[gray] (10,-7.4) circle (1pt) node[anchor=south] {};
\filldraw[gray] (10,-7.6) circle (1pt) node[anchor=south] {};

\draw (9,-6.6) node[cross=3pt,black] {};
\filldraw[black] (11,-6.6) circle (2pt) node[anchor=west] {$w_2$};

\end{tikzpicture}
\caption{Successive quotients of the Fock space.}\label{fig:FockQuotient}
\end{figure}
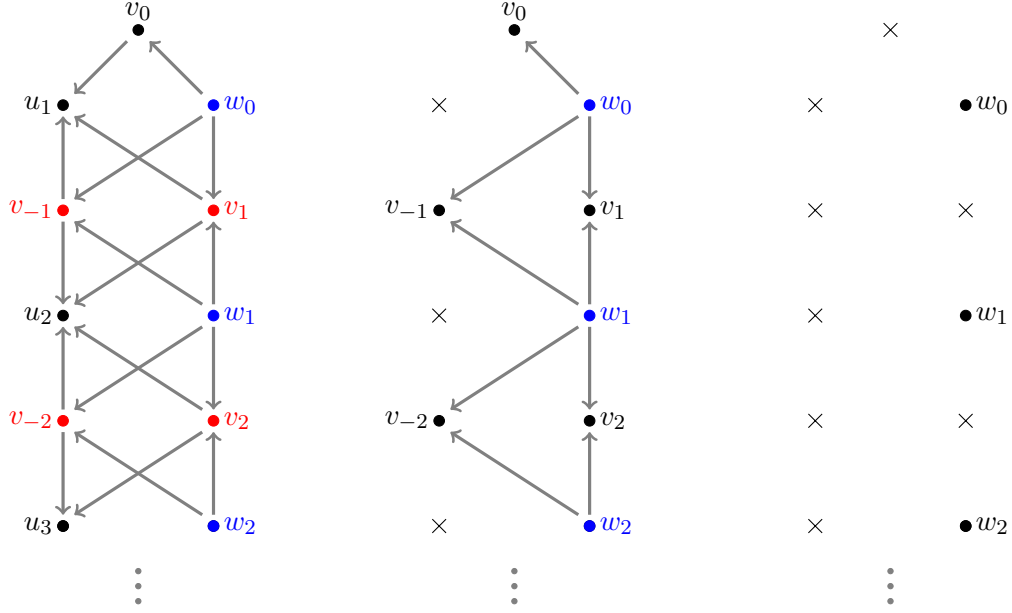

The determination of the singular vector structure of the Fock module is technical. The computations are most easily carried out in the fermionic oscillator description, but a bosonized representation is implicit in the work of Thorn and Tsuchiya-Kanie \cite{Thorn:1984sn,Tsuchiya:1986}.   The structure of the Fock space is summarized in figure \ref{fig:FockStructure}, whose structure of inclusions should be contrasted with  that of figure \ref{fig:Verma}. In particular, while the dimension of each state appearing in this diagram matches that of the corresponding state in the Verma module diagram, the properties of the states are different. In the Fock space diagram, the vectors $v_0,u_k$ labeled by black dots represent the only singular vectors. In particular, $w_0$, which replaces the first singular vector in the Verma module diagram, is \textit{not} a singular vector. A useful example to keep in mind is the identity Fock space: after the twist the current algebra descendant $w_0=j_{-1}|0\rangle$ is no longer singular, but the highest weight state can be written $v_0=|0\rangle \; \propto\; L_{1}j_{-1}|0\rangle$. The vectors $v_{\pm k}$ depicted in red in figure \ref{fig:FockStructure} have the property that they become singular vectors if one quotients the Fock space by the $u_k$. Finally, the $w_k$ become singular if one quotients out by the $v_k$'s. This characterization in terms of successive quotients is depicted in figure \ref{fig:FockQuotient}. Upon performing all quotients, one arrives at a Hilbert space isomorphic to the irreducible representation of the Virasoro algebra. 
The physical interpretation of these states is most transparent in Felder's BRST formulation, and we postpone further discussion until section \ref{sec:Felder}.

\subsection{Dotsenko and Fateev's construction and the screening charges}\label{sec:DotsenkoFateev}
The discovery of the Feigin-Fuchs bosonic resolution made it clear that the degenerate representations of the Virasoro algebra can be embedded within the Hilbert space of the timelike linear dilaton, but said nothing about dynamics. Around the same time, it was independently discovered \cite{Polyakov:1984yq} that null-state decoupling conditions lead to ordinary differential equations for the minimal model four-point functions. These ODE's are generalizations of the hypergeometric equation, and their solutions have integral representations. Dotsenko and Fateev realized that these integrals could be represented in terms of integrated free field correlation functions, and this insight led to the development of a powerful formalism that computes any genus zero minimal model observable in terms of free field data. In this section we will briefly review this Coulomb gas formalism. For more complete reviews see \cite{Dotsenko:1986ca,dotsenko:cel-00092929,DiFrancesco:1997nk}.

Recall that the linear dilaton exponential operators $e^{i\alpha\phi(z)}$ have left-moving weights
\begin{equation}
    h=\frac{\alpha}{2}\left(\frac{\alpha}{2}+iQ\right) \; 
\end{equation}
and that this expression is invariant under the ``reflection" map
\begin{equation}
    \alpha \to 2\alpha_+ +2\alpha_--\alpha \; . 
\end{equation}
When the momentum selection rule is satisfied, the free field expectation value of these operators takes the form 
\begin{equation}
    \langle e^{i\alpha_1\phi(z_1)}\dots e^{i\alpha_n\phi(z_n)} \rangle=\prod_{i>j}(z_i-z_j)^{\alpha_i\alpha_j/2}  \; . 
\end{equation}
The vertex operators
\begin{equation}\label{eq:CGOps}
    V_{r,s}(z)=e^{i\alpha_{r,s}\phi(z)}\; , \hspace{1 in} \alpha_{r,s} = (1-r)\alpha_+ +(1-s)\alpha_- \; , \hspace{.25 in} r,s\in \mathbb{Z} 
\end{equation}
have the same conformal dimensions as the minimal model primaries, as do their reflection partners $V_{-r,-s}(z)$.
In the original application of the Coulomb gas formalism, one simply restricts attention to the set of operators \eqref{eq:CGOps} with values of $r,s$ appearing in the Kac table. However, when we construct Felder's BRST complex we will have to consider operators lying outside of this range, but still of the form \eqref{eq:CGOps}.

The genus zero selection rule 
\begin{equation}\label{eq:Select}
    \sum_i \alpha_i =2\alpha_++2\alpha_- \; 
\end{equation}
 sets most correlation functions of \textit{local} operators to zero.
There are however two $h=1$ operators in the model with momenta that lie outside of the Kac table:
\begin{equation}\label{eq:ScreeningOp}
 M_+(z)\equiv e^{2i\alpha_+\phi(z)}\; , \hspace{1 in}M_-(z)\equiv  e^{2i\alpha_-\phi(z)}. 
\end{equation}
Importantly, these exponentials have non-vanishing $U(1)$ charges and 
the nonlocal ``screening operators"
\begin{equation}
    S_{\pm}\equiv\oint dz M_{\pm}(z) 
\end{equation}
commute with the Virasoro algebra for suitably chosen closed contours. Since these operators carry charge, they can be used to satisfy the selection rule \eqref{eq:Select} without altering the conformal properties of a correlator.   The first step in producing the minimal model correlation function
\begin{equation}
    \langle M_{r_1,s_1}(z_1,\bar{z}_1)\dots M_{r_n,s_n}(z_n,\bar{z}_n)\rangle 
\end{equation}
is to determine the appropriate number of screening charges needed to satisfy the selection rule in the linear dilaton theory:
\begin{equation}
\langle V_{r_1,s_1}(z_1)\dots V_{r_n,s_n}(z_n)S_+^{n_+}S_-^{n_-}\rangle \; . 
\end{equation}
There is no principle which independently fixes the specific combination of contours in this holomorphic correlation function. Indeed, for each correlator there is a fixed number of independent contours, each corresponding to a particular conformal block. The correct combination is instead determined by combining the holomorphic correlator with its anti-holomorphic counterpart and requiring locality (absence of branch cuts) in the final answer. This prescription reproduces the full minimal model correlator up to normalization factors. 

The textbook example of an application of the Coulomb gas formalism is the calculation of the four-point correlation function
\begin{equation}
    \langle M_{r_1,s_1}(0)M_{1,2}(z,\bar{z})M_{r_3,s_4}(1)M_{r_4,s_4}(\infty)\rangle \; 
\end{equation}
with the three operators chosen so that a single screening charge is needed. Ignoring factors that do not produce interesting analytic structure when combined with the right moving correlator, we find
\begin{equation}
    \langle e^{i\alpha_{1}\phi(0)}e^{-i\alpha_{-}\phi(z)}e^{i\alpha_{3}\phi(1)}e^{i\alpha_{4}\phi(\infty)}\oint dw e^{2i\alpha_+\phi(w)}\rangle \;\propto \; 
    \oint dw w^{\alpha_1\alpha_+}(w-1)^{\alpha_3\alpha_+}(w-z)^{-\alpha_-\alpha_+} \; . 
\end{equation}
The integrand has branch points at $0,z,1,\infty$ and  the integral has two independent closed contours. Up to phase factors arising from crossing the cuts (which we ignore since we are blind to normalization at this stage), they can be shrunk to run from  $0$ to $z$ and from $1$ to $\infty$. We denote the two independent integrals
\begin{equation}
    I_i(z)=\oint_{C_i} dw w^a(w-1)^b(w-z)^c \; , 
\end{equation}
with the obvious identifications.

The full minimal model correlator must be proportional to a sum of products of the holomorphic and antiholomorphic correlators  
\begin{equation}
 \langle M_{r_1,s_1}(0)M_{1,2}(z,\bar{z})M_{r_3,s_3}(1)M_{r_4,s_4}(\infty) \rangle   \sim \sum X_{ij}I_i(z)\overline{I_j(z)} \; , 
\end{equation}
and the correct combination is determined by requiring the absence of branch cuts in the full correlator. Since the integrands of the $I_i(z)$ have branch points at $0,1$, and $\infty$, taking $z$ around a closed path encircling $0$ or $1$ results in a  monodromy transformation $M_{ij}$ that mixes the different solutions of the hypergeometric equation
\begin{equation}
    I_i(z)\to G_{ij}^{a}I_j(z)\; , \hspace{.5 in} a=0,1 \; .
\end{equation}
In the case at hand the monodromy around $0$ is a diagonal transformation, so that $X_{ij}$ must be a diagonal matrix. The monodromy around the point $1$ is not diagonal in this basis, but there is another basis of solutions $\tilde{I}_i(1-z)$ in which it is. The two bases are linearly related
\begin{equation}
    I_i(z)=\sum \alpha_{ij}\tilde{I}_j(1-z)
\end{equation}
and the locality requirement translates into the condition 
\begin{equation}
    \sum_i\alpha_{ij}X_i\alpha_{ik}=0 \; , \hspace{.5 in} j\neq k \; ,
\end{equation}
where $X_i$ denote the diagonal elements of $X_{ij}$. Once the $\alpha_{ij}$ are known, the $X_i$ are fixed (up to normalization) and the solution to the crossing constraint obtained.
Once the four-point function has been determined, it is factorized onto three-point functions to determine the fusion rules.
Extra work is required to determine the correct normalization: this is most easily done using the two-dimensional surface integrals we will discuss in section \ref{sec:PerturbationTheory}.

\section{Felder's BRST construction}\label{sec:Felder}
The screening charge prescription of Dotsenko and Fateev is capable of computing genus zero minimal model correlators. However, higher genus observables naively receive contributions from the spurious states that are present in the linear dilaton but not in the minimal model, and it is necessary to understand how to remove these contributions.  This problem was solved algebraically by Felder, who unearthed a hidden BRST structure implicit in the works of Feigin-Fuchs and Thorn.  We will begin by analyzing a very simple model where the BRST structure is extremely transparent. Felder's general construction is reviewed in section \ref{sec:FelderGeneral}.

\subsection{Warmup: the $(2,1)$ model and the $\eta\xi$ system }\label{sec:EtaXi}
In this subsection we will analyze what might be called the $(2,1)$ minimal model. According to section \ref{sec:Minimal_Models}, this system should have no local operators, and understanding how this statement is realised within the compact timelike linear dilaton theory will prove useful in the analysis of the more complicated models. In fact, Felder's BRST construction is extremely transparent in this model, and will be immediately recognized by those familiar with the quantization of the superstring. The standard description of the holomorphic part of the model uses a pair of anticommuting fields $\eta(z)$ and $\xi(z)$ with central charge $c=-2$ and an action of the form
\begin{equation}\label{eq:etaxiaction}
    S=\frac{1}{2\pi}\int d^2z \;  \eta \bar{\partial} \xi \; . 
\end{equation} 
The operator $\eta(z)$ is naturally a vector current with $h=1$, while $\xi(z)$ is an $h=0$ scalar. The Coulomb gas scalar can be viewed as a ``bosonization" of this system, but several peculiar properties of the Coulomb gas formalism have simple explanations within the $\eta\xi$ description.

Although our primary goal is to motivate Felder's BRST construction, there are several reasons to consider this model. There is an argument\footnote{Distler identifies the $\eta\xi$ system (viewed as the $(2,1)$ minimal model) combined with the Liouville field $\varphi$ as a bosonization of the $\beta\gamma$ ghosts, which combine with the $bc$ reparameterization ghosts to form the $\beta\gamma bc$ BRST multiplet of topological gravity. Certain global issues, including non-compactness of the Liouville field, the absence of the Liouville potential, and the proper treatment of the $\xi(z)$ zero mode seem puzzling in this derivation \cite{Gaiotto:2003yb}. As we will describe, the $(2,1)$ model is treated differently when viewed as a minimal model than it is in string theory.} due to Distler \cite{Distler:1989ax} that relates the $(2,1)$ model, coupled to Liouville, to topological gravity \cite{Witten:1989ig,Witten:1990hr} and the one-matrix model \cite{Brezin:1990rb,Douglas:1989ve,Gross:1989vs}. The $\eta\xi$ system also makes an appearance in the free field realization of the WZW models\cite{Distler:1989xv}, and most famously, in the
``bosonization" of the $\beta \gamma$ superconformal ghosts\cite{Friedan:1985ge}:
\begin{equation}\label{eq:betagamma}
    \beta=e^{\sqrt{2}\varphi}\partial \xi \; , \hspace{.5 in} \gamma=\eta e^{-\sqrt{2}\varphi}\; . 
\end{equation}
The action \eqref{eq:etaxiaction} has an obvious shift symmetry under which
\begin{equation}
    \xi \to \xi + \text{const.} \; , 
\end{equation}
as well as an anomalous ghost number current $j_g=-\eta\xi $ whose bosonization is simply the momentum current of the linear dilaton.
The ghost OPE 
\begin{equation}\label{eq:ghostOPE}
    \eta(z)\xi(w)\sim \frac{1}{z-w}\; , \hspace{.5 in} \eta(z)\eta(w)\sim 0 \; , \hspace{.5 in} \xi(z)\xi(w)\sim 0 \; 
\end{equation}
indicates that $j_B(z)=\eta(z)$ is the current for the $\xi(z)$ shift symmetry, and the corresponding charge is
\begin{equation}
    Q_B=\oint \frac{dz}{2\pi i}\eta(z) \; .
\end{equation}
Since $\eta$ is anticommuting, $Q_B^2=0$ and $Q_B$ is a \textit{scalar} BRST charge. Crucially, it commutes with the Virasoro algebra since the energy momentum tensor $T= \,\eta \partial \xi$ is shift invariant.
The BRST transformations are
\begin{align}
    \{ Q_B,\xi(z)\}=1 \; ,\\ 
    \{ Q_B,\eta(z)\}=0 \; . 
\end{align}
From the first of these equations we see that the identity operator is BRST exact, and $\xi(z)$ is not BRST closed. In particular, if we identify the dimension zero operator $\xi(z)$ as the ``reflection of the identity," then we see that neither putative identity operator is in the BRST cohomology.

It is helpful to translate some of these statements using the free field mode expansions
\begin{equation}
    \eta(z)=\sum_{n=-\infty}^{\infty}\frac{\eta_n}{z^{n+1}} \; , \hspace{.5 in} \xi(z)=\sum_{n=-\infty}^{\infty}\frac{\xi_n}{z^{n}} \; . 
\end{equation}
In this language the scalar BRST charge is the zero mode of $\eta$
\begin{equation}
    Q_{B}=\eta_0 \; . 
\end{equation}
The anticommutators of the modes follow directly from the ghost OPE \eqref{eq:ghostOPE}:
\begin{equation}
    \{ \eta_n, \xi_m \}=\delta_{n,-m} \; , \hspace{.5 in} \{\eta_n,\eta_m\}=\{\xi_n,\xi_m\}=0 \; . 
\end{equation}
Because the zero mode of $\xi$ does not appear in the $\beta \gamma$ model \eqref{eq:betagamma}, standard treatments distinguish between the ``large algebra," which contains $\xi_0$, and the ``small algebra" which does not. Indeed, the definition of the $SL(2,\mathbb{C})$ invariant vacuum requires
\begin{align}
    &\xi_n|0\rangle =0 \;\;\;\;\; n\geq 1 \; ,\\
    &\eta_n|0\rangle=0 \;\;\;\;\; n\geq 0 \; . 
\end{align}
Since $[L_0,\xi_0]=0$, when $\xi_0$ acts on the $SL(2,\mathbb{C})$ invariant vacuum, it generates a new state $\xi_0|0\rangle$ of the same energy. This is a state in the ``large" Hilbert space: it has the same energy, but carries different ghost charge. The vacuum is BRST invariant since $\eta_0|0\rangle=0$, and the states of the model are built using the usual fermionic Fock space construction. Since $\{\eta_0,\xi_m\}=\delta_{0,m}$, restricting to the small Hilbert space is the same as restricting to the kernel of the BRST charge $Q_B$. In other words, the small Hilbert space is the space of BRST invariant states, and passing to the kernel of $Q_B$ removes the ``reflected" operators from the spectrum. If we pass to the cohomology of $Q_B$ then we find, as expected, that there are no states.
Each potential state is constructed by acting on $|0\rangle$ with each fermionic mode at most once. Any state that includes $\xi_0$ is not BRST invariant, and any state without $\xi_0$ is BRST exact since
\begin{equation}
    |0\rangle=\{\eta_0, \xi_0\}|0\rangle = Q_B ( \xi_0|0\rangle) \; .
\end{equation}
The same statements hold in the Fock spaces built on different fillings of the Fermi sea. Felder's BRST construction is a more complicated (and bosonized) version of this story.

Given the $(2,1)$ model defined by \eqref{eq:etaxiaction}, how do we know that we are supposed to perform the BRST reduction? Apparently, it is a choice that we make in the definition of the model. When this system is used to bosonize the $\beta\gamma$ superconformal ghosts, we discard states that are not BRST invariant but keep BRST exact states.  The same model with no BRST projection at all actually describes the critical phase of dense polymers \cite{Duplantier:1987sh,Saleur:1991hk,Kausch:1995py} (obviously, this critical point exhibits many non-standard features). In this case, passing to the cohomology is a choice, and it determines the observables we allow ourselves to compute. For instance, we expect that if we perform a trace in the Hilbert space of the BRST quotiented model, we should get zero. Moreover, this trace should differ from the trace in the model without the BRST quotient. Since $\eta\xi$ are anticommuting variables, we need to pick a spin structure on the torus, and to describe the ``trivial" minimal model we should pick the one that gives zero. From the functional integral point of view, this will be the spin structure such that $\xi(z)$ has a zero mode. We are naturally led to the conclusion that the  path integral with periodic boundary conditions, 
\begin{equation}
    Z(\beta)=\text{Tr}_{R}(-1)^Fe^{-\beta H} \; , 
\end{equation}
is the quantity that actually computes the ``observables" in the $(2,1)$ model viewed as a trivial minimal model.
From the algebraic perspective, this quantity vanishes identically due to the exact degeneracy between states built on $|0\rangle$ and on $\xi_0|0\rangle$, which cancel in pairs. 

The appearance of a specific spin structure in this construction suggests that, in bosonizing the model to obtain the usual Coulomb gas description, we may need to introduce discrete gauge fields coupled to the spin structure through a topological term. In other words, what term do we add to the compact timelike dilaton action to compute zero with the functional integral?

In the treatment of the $\beta\gamma$ system, $\eta$ and $\xi$ are often further bosonized in terms of a scalar field $\phi(z)$ with a background charge $Q=\frac{i}{\sqrt{2}}$:
\begin{equation}
\xi(z)=e^{i\sqrt{2}\phi(z)}\;, \hspace{.5 in } \eta(z)=e^{-i\sqrt{2}\phi(z)} \; . 
\end{equation}
This description in terms of $\phi(z)$ is the analog of the Coulomb gas formulation of the $(2,1)$ minimal model, and we would like to use it to draw lessons for the physical models. We begin by reproducing the BRST discussion in these new variables. 
For reasons that will become clear in later sections, we will assume that the radius of the boson is $R=\sqrt{2}$. For a holomorphic vertex operator the OPE is
\begin{equation}
    e^{i\alpha_1\phi(z)}e^{i\alpha_2\phi(0)}\sim z^{\alpha_1\alpha_2/2}e^{i(\alpha_1+\alpha_2)\phi(0)}+\dots \;,
\end{equation}
the conformal weight is
\begin{equation}
    h=\frac{\alpha}{2}\left(\frac{\alpha}{2}-\alpha_+-\alpha_-\right) \; 
\end{equation}
and the reflection is  $\alpha \to -2iQ-\alpha$. Since the identity operator has $U(1)$ charge zero, we identify
\begin{equation}
 \xi(z)=   e^{i(2\alpha_++2\alpha_-)\phi(z)}=e^{i\sqrt{2}\phi(z)} \; . 
\end{equation}
 Since $\eta(z)$ should have opposite ghost charge, we identify
\begin{equation}
    \eta(z)= e^{2i\alpha_-\phi(z)}=e^{-i\sqrt{2}\phi(z)} \;, \hspace{.5 in} \tilde{\eta}(z)=e^{i2\sqrt{2}\phi(z)} \; .
\end{equation}
These identifications reproduce the correct form of the ghost OPE:
\begin{equation}
    \eta(z)\xi(0)\sim e^{-i\sqrt{2}\phi(z)}e^{i\sqrt{2}\phi(0)}\sim \frac{1}{z} \; , \hspace{.25 in}
    \eta(z)\eta(0)\sim ze^{-2\sqrt{2}i\phi(0)} +\dots \; , \hspace{.25 in}\xi(z)\xi(0)\sim ze^{2\sqrt{2}i\phi(0)}+\dots \; .
\end{equation}
In particular, the $\eta \eta$ and $\xi\xi$ OPEs are non-singular. It is also important to note that
\begin{equation}
    \eta(z)\tilde{\eta}(0)\sim \frac{1}{z^2}\left[ \xi(0)+zL_{-1}\xi(0)+\dots \right] 
\end{equation}
so that the level-one singular vector in the $\xi(z)$ Fock space is BRST exact:
\begin{equation}
[Q_B,\tilde{\eta}(0)]  = L_{-1}\xi(0) \; . 
\end{equation}
Also note that although $\tilde{\eta}(z)$ is not BRST closed, the non-local operator
\begin{equation}
  \oint \tilde{\eta}(z) dz= \oint e^{2i\alpha_+\phi(z)}dz 
\end{equation}
is BRST invariant since the variation of $\tilde{\eta}(z)$ is a total derivative.
We can briefly summarize the discussion with the statement that the BRST complex with cochain groups given by the Fock spaces
\begin{align}
\dots \xrightarrow{Q_B}    \widetilde{j}_B(z) \xrightarrow{Q_B}\widetilde{1}(z)\xrightarrow{Q_B}&1\xrightarrow{Q_B}j_B(z)\xrightarrow{Q_B} \dots =\notag\\
\dots \xrightarrow{Q_B}    e^{i2\sqrt{2}\phi(z)} \xrightarrow{Q_B}e^{i\sqrt{2}\phi(z)}\xrightarrow{Q_B}&1\xrightarrow{Q_B}e^{-i\sqrt{2}\phi(z)}\xrightarrow{Q_B} \dots 
\end{align}
is exact: there is no cohomology with respect to the differential $Q_B$. It is important to understand the structure of this BRST complex, because a more complicated version underlies Felder's construction. Each exponential operator to the left of the identity has a singular OPE with the BRST current and is not BRST closed. The $Q$-variations of these operators and their descendants prepare the singular vectors in the Fock spaces to their right. The exponential operators to the right of the identity all have non-singular OPE's with the BRST current and are therefore BRST closed. They are however also BRST exact. In order to understand this point, it is important to remember that the Fock spaces built on the exponential operators contain current algebra descendants in addition to Virasoro descendants. In particular, while the BRST charge commutes with the Virasoro algebra, it does not commute with individual elements of the current algebra (for instance, it carries $U(1)$ charge). For instance, since the identity operator is BRST exact, each of its Virasoro descendants is also $Q$-exact and therefore has trivial image in the Fock space built on $j_B=e^{2i\alpha_-\phi}$. However, the level one current algebra descendant in the identity Fock space is not BRST closed and its image is precisely the BRST current: 
\begin{equation}
    [Q_B,j(z)]=j_B(z) \; . 
\end{equation}
This crucial formula actually holds in all of the linear dilaton models we consider, and explains how the $U(1)$ currents of the linear dilaton are eliminated when reducing to the minimal model.

\subsection{Felder's BRST complex: the general case}\label{sec:FelderGeneral}
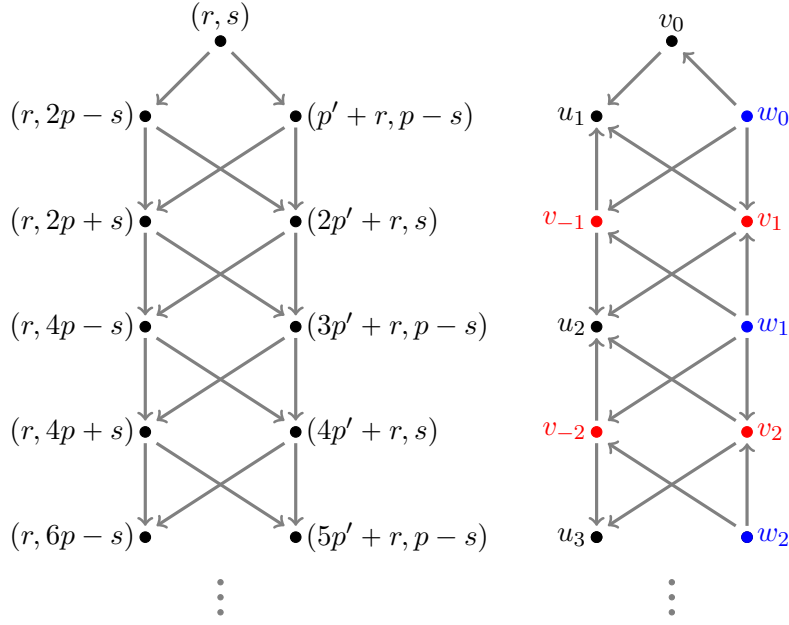
\begin{figure}[h]
\begin{tikzpicture}

\draw[->,gray, very thick] (.15,-.15) -- (.85,-.85);
\draw[->,gray, very thick] (-.15,-.15) -- (-.85,-.85);
\filldraw[black] (0,0) circle (2pt) node[anchor=south] { $(r,s)$};

\draw[->,gray, very thick] (1,-1.15) -- (1,-2.25);
\draw[->,gray, very thick] (-1,-1.15) -- (-1,-2.25);
\draw[->,gray, very thick] (.85,-1.15) -- (-.85,-2.25);
\draw[->,gray, very thick] (-.85,-1.15) -- (.85,-2.25);

\filldraw[black] (1,-1) circle (2pt) node[anchor=west] { $(p'+r,p-s)$};
\filldraw[black] (1,-2.4) circle (2pt) node[anchor=west] { $(2p'+r,s)$};

\filldraw[black] (-1,-1) circle (2pt) node[anchor=east] { $(r,2p-s)$};
\filldraw[black] (-1,-2.4) circle (2pt) node[anchor=east] { $(r,2p+s)$};

\draw[->,gray, very thick] (-1,-2.55) -- (-1,-3.65);
\draw[->,gray, very thick] (1,-2.55) -- (1,-3.65);

\draw[->,gray, very thick] (-.85,-2.55) -- (.85,-3.65);
\draw[->,gray, very thick] (.85,-2.55) -- (-.85,-3.65);

\filldraw[black] (-1,-3.8) circle (2pt) node[anchor=east] {$(r,4p-s)$ };
\filldraw[black] (1,-3.8) circle (2pt) node[anchor=west] {$(3p'+r,p-s)$ };

\filldraw[black] (-1,-5.2) circle (2pt) node[anchor=east] { $(r,4p+s)$};
\filldraw[black] (1,-5.2) circle (2pt) node[anchor=west] {$(4p'+r,s)$};

\draw[->,gray, very thick] (1,-5.35) -- (1,-6.45);
\draw[->,gray, very thick] (-1,-5.35) -- (-1,-6.45);

\draw[->,gray, very thick] (.85,-5.35) -- (-.85,-6.45);
\draw[->,gray, very thick] (-.85,-5.35) -- (.85,-6.45);
\filldraw[black] (-1,-6.6) circle (2pt) node[anchor=east] {$(r,6p-s)$ };
\filldraw[black] (1,-6.6) circle (2pt) node[anchor=west] {$(5p'+r,p-s)$};

\filldraw[gray] (0,-7.2) circle (1pt) node[anchor=south] {};
\filldraw[gray] (0,-7.4) circle (1pt) node[anchor=south] {};
\filldraw[gray] (0,-7.6) circle (1pt) node[anchor=south] {};

\draw[->,gray, very thick] (-.85,-3.95) -- (.85,-5.05);
\draw[<-,gray, very thick] (-.85,-5.05) -- (.85,-3.95);

\draw[->,gray, very thick] (-1,-3.95) -- (-1,-5.05);
\draw[->,gray, very thick] (1,-3.95) -- (1,-5.05);

\draw[<-,gray, very thick] (6.15,-.15) -- (6.85,-.85);
\draw[->,gray, very thick] (5.85,-.15) -- (5.15,-.85);
\filldraw[black] (6,0) circle (2pt) node[anchor=south] { $v_0$};

\draw[->,gray, very thick] (7,-1.15) -- (7,-2.25);
\draw[->,gray, very thick] (5,-2.25) -- (5,-1.15) ;
\draw[<-,gray, very thick] (5.15,-2.25) -- (6.85,-1.15);
\draw[<-,gray, very thick] (5.15,-1.15) -- (6.85,-2.25);

\filldraw[blue] (7,-1) circle (2pt) node[anchor=west] { $w_0$};
\filldraw[red] (7,-2.4) circle (2pt) node[anchor=west] { $v_1$};

\filldraw[black] (5,-1) circle (2pt) node[anchor=east] { $u_1$};
\filldraw[red] (5,-2.4) circle (2pt) node[anchor=east] { $v_{-1}$};

\draw[->,gray, very thick] (5,-2.55) -- (5,-3.65);
\draw[->,gray, very thick] (7,-3.65) -- (7,-2.55);

\draw[<-,gray, very thick] (5.15,-2.55) -- (6.85,-3.65);
\draw[->,gray, very thick] (6.85,-2.55) -- (5.15,-3.65);

\filldraw[black] (5,-3.8) circle (2pt) node[anchor=east] { $u_2$};
\filldraw[blue] (7,-3.8) circle (2pt) node[anchor=west] { $w_1$};

\draw[->,gray, very thick] (7,-3.95) -- (7,-5.05);
\draw[<-,gray, very thick] (5,-3.95) -- (5,-5.05);

\filldraw[red] (5,-5.2) circle (2pt) node[anchor=east] { $v_{-2}$};
\filldraw[red] (7,-5.2) circle (2pt) node[anchor=west] {$v_2$};

\draw[<-,gray, very thick] (7,-5.35) -- (7,-6.45);
\draw[->,gray, very thick] (5,-5.35) -- (5,-6.45);

\draw[->,gray, very thick] (6.85,-3.95) -- (5.15,-5.05);
\draw[->,gray, very thick] (6.85,-5.05) -- (5.15,-3.95);

\draw[->,gray, very thick] (6.85,-5.35) -- (5.15,-6.45);
\draw[<-,gray, very thick] (5.15,-5.35) -- (6.85,-6.45);
\filldraw[black] (5,-6.6) circle (2pt) node[anchor=east] { };
\filldraw[black] (7,-6.6) circle (2pt) node[anchor=west] {};

\filldraw[gray] (6,-7.2) circle (1pt) node[anchor=south] {};
\filldraw[gray] (6,-7.4) circle (1pt) node[anchor=south] {};
\filldraw[gray] (6,-7.6) circle (1pt) node[anchor=south] {};

\filldraw[black] (5,-6.6) circle (2pt) node[anchor=east] { $u_{3}$};
\filldraw[blue] (7,-6.6) circle (2pt) node[anchor=west] {$w_2$};

\end{tikzpicture}
\caption{Verma module structure versus Fock space structure.}\label{fig:VermavsFock}
\end{figure}

In section \ref{sec:Minimal_Models}, we saw that the degenerate Verma module associated to the primary of weight $h_{r,s}$ has two infinite classes of singular vectors. The first class occurred  with weights $h_{r_k,s_k}$ given by
\begin{equation}\label{eq:Type1}
  \hspace{1 in} r_k=r+kp' \; , \hspace{.25 in} s_k=(-1)^ks+[1-(-1)^k]\frac{p}{2} \; . 
\end{equation}
The other class was of the form $h_{r,s_k}$ with
\begin{equation}\label{eq:Type2}
 s_k=kp+(-1)^ks+[1-(-1)^k]\frac{p}{2} \; . 
\end{equation}
 The structure of the Fock space discussed in section \ref{sec:FeiginFuchs} was slightly more complicated. In particular, the full Fock space had fewer singular vectors than the reducible Verma module. However, upon performing successive quotients by submodules of singular vectors, the Fock space is reduced to the irreducible representation of the Virasoro algebra needed for the construction of the minimal model. When placed side by side, as in figure \ref{fig:VermavsFock}, the vectors in each diagram share the same conformal dimension. 

Felder's innovation was to reformulate the successive quotients of the Fock space by the submodules of singular vectors in the language of BRST.

To begin, Felder considers a sequence of cochain groups given by Fock spaces $F_{r_k,s_k}$ whose highest weight vectors have the dimensions $h_{r_k,s_k}$ appearing in figure \ref{fig:VermavsFock}:
\begin{equation}\label{eq:CoChainComplex}
\dots \xrightarrow{} F_{r,4p-s}\xrightarrow{}F_{r,2p+s}\xrightarrow{} F_{r,2p-s}  \xrightarrow{}  F_{r,s}\xrightarrow{}F_{p'+r,p-s}\xrightarrow{}F_{2p'+r,s}\xrightarrow{}F_{3p'+r,p-s} \xrightarrow{}\dots 
\end{equation}
 Note that the Fock spaces to the right of $F_{r,s}$ have momenta corresponding to \eqref{eq:Type1}, while those to the left have momenta given by \eqref{eq:Type2}. It had been known since the work of Thorn \cite{Thorn:1984sn} that multiple nested integrals of the $h=1$ operator $M_{-}(z)$ could be used to write the singular vectors in the Fock space $F_{r,s}$ in terms of exponential operators with different momenta. This led Felder to identify the operator
\begin{equation}\label{eq:HolBRST}
    Q_-=\oint \frac{dz}{2\pi i} M_-(z) 
\end{equation}
as the appropriate differential for the complex \eqref{eq:CoChainComplex}. For a closed contour, the operator $Q_-$ commutes with the Virasoro algebra. However, since the holomorphic operator $M_-(z)$ is not itself mutually local with all exponentials in the model, it is actually an appropriately chosen power of $Q_-$ that acts nilpotently on the individual Fock spaces. It is also precisely this power of $Q_-$ which was known to prepare singular vectors in the Fock space $F_{r,s}$. The actual BRST complex takes the form
\begin{equation}\label{eq:ComplexDifferential}
\xrightarrow{}F_{r,4p-s}\xrightarrow{Q_-^{p-s}}F_{r,2p+s}\xrightarrow{Q_-^s} F_{r,2p-s}  \xrightarrow{Q_-^{p-s}}  F_{r,s}\xrightarrow{Q_-^s}F_{p'+r,p-s}\xrightarrow{Q_-^{p-s}}F_{2p'+r,s}\xrightarrow{Q_-^s}F_{3p'+r,p-s}\xrightarrow{}
\end{equation}
Using the results of Feigin and Fuchs, Felder proved that the only nontrivial cohomology of this complex lies in $F_{r,s}$. Therefore, a trace in the cohomology of $F_{r,s}$ can be extended to a trace in the cohomology of the entire complex, which by the algebraic Lefshetz principle can be represented by an alternating trace in the full cochain groups themselves. This fact is the basis for Felder's simple rederivation of the Rocha-Caridi form of the minimal model characters. 
An equivalent complex can be constructed using appropriate powers of the BRST charge 
\begin{equation}
    Q_+=\oint \frac{dz}{2\pi i} M_+(z) \; . 
\end{equation}
We will illustrate the general idea with a few simple examples before describing the full action of the BRST charge within each Fock space.

\subsubsection*{Example: the identity Fock space}
If we specialize the complex \eqref{eq:ComplexDifferential} to the case $r=s=1$, it becomes
\begin{equation}
\dots \xrightarrow{}F_{1,2p-1}\xrightarrow{Q_-^{p-1}}    F_{1,1}\xrightarrow{Q_-}F_{1,-1} \xrightarrow{}\dots 
\end{equation}
The exponential operator acting as the highest weight state in $F_{1,-1}$ is simply $e^{2i\alpha_-\phi(z)}$, which is the BRST current $j_B(z)$. Now according to formula  \eqref{eq:VertexComm} we have
\begin{equation}
    [j_{-1},e^{2i\alpha_-\phi(z)}]=\frac{\alpha_-}{\sqrt{2}}\frac{1}{z}e^{2i\alpha_-\phi(z)} \; . 
\end{equation}
Integrating both sides of the equality on a contour surrounding the origin, we find
\begin{equation}
    [j_{-1},Q_-]= \frac{\alpha_-}{\sqrt{2}}e^{2i\alpha_-\phi(0)}\; \propto \; j_B(0) \; . 
\end{equation}
 This equation says that the $U(1)$ momentum current is not BRST closed, and that its $Q_-$ variation simply produces the BRST current $j_B(z)$:
\begin{equation}
    Q_-j_{-1}|0\rangle=[Q_-,j_{-1}]|0\rangle \;\;\propto \;\;j_B(0)|0\rangle \; . 
\end{equation}
Restoring position dependence, we have the operator equality $[Q_-,j(z)]=j_B(z)$. This is an extremely important equation, since the minimal model obviously does not contain any $U(1)$ momentum current. In this example, the identity operator in the Verma module has two singular descendants at levels one and $(p'-1)(p-1)$. In the Fock module, the singular descendant at level one is replaced by a nonsingular vector which is however not BRST closed. Meanwhile, the singular vector at level $(p'-1)(p-1)$ is $Q$-exact and is produced by the action of $Q_-^{p-1}$ on the highest weight exponential $e^{i(2-2p)\alpha_-\phi(z)}$ in $F_{1,2p-1}$. To see this, note that since $Q_-$ commutes with the Virasoro algebra, the dimension and charge of the $Q$-exact operator $[Q_-^{p-1},e^{i(2-2p)\alpha_-\phi(z)}]$ matches that of the singular descendant in $F_{1,1}$ which was known from the work of Feigin and Fuchs. Moreover, since the exponential is singular (highest weight), it's $Q$-variation will be singular provided that it is non-vanishing. One then verifies that this vector is nonzero by computing its inner product with another state using the Dotsenko-Fateev integrals. 

Another simple example demonstrates why the reflection of the identity operator does not introduce position dependence in BRST invariant correlation functions. The reflection of the identity  is in the Fock space $F_{p'-1,p-1}$, and the relevant segment of the BRST complex is
\begin{equation}
 F_{1,-1}  \xrightarrow{Q_+}  F_{p'-1,p-1} \; . 
\end{equation}
The action of the BRST charge is given by
\begin{align}
    [Q_+,e^{2i\alpha_-\phi(z)}]&=\oint \frac{dw}{2\pi i} (w-z)^{2\alpha_+\alpha_-}:e^{2i\alpha_+\phi(w)}e^{2i\alpha_-\phi(z)}:\notag\\
    &=\oint \frac{dw}{2\pi i} \frac{1}{(w-z)^2}[1+2i\alpha_+(w-z)\partial \phi+\dots]e^{i(2\alpha_++2\alpha_-)\phi(z)}\\
    &\propto \;\;\partial_z e^{i(2\alpha_++2\alpha_-)\phi(z)} \; .\notag 
\end{align}
 The derivative of the reflection of the identity is BRST exact, so $e^{i(2\alpha_++2\alpha_-)\phi(z)}$ is a topological operator.

Another tractable example is the construction of the leading singular vector in the Fock space $F_{1,p-1}$.
The relevant segment of the BRST complex is
\begin{equation}
 F_{1,p+1}  \xrightarrow{Q_-}  F_{1,p-1} \; . 
\end{equation}
The BRST charge should produce the null state at level $l=(p'-r)(p-s)=p'-1$. Its action on the exponential operator in $F_{1,p+1}$ is given by
\begin{align}
        [Q_-,V_{1,p+1}(0)]&=\oint dz z^{-\alpha_-^2p}:e^{2i\alpha_-\phi(z)}e^{-ip\alpha_-\phi(0)}:\\
        &=\oint dz \frac{1}{z^{p'}}[V_{1,p-1}+\dots + z^{p'-1}(L_{-(p'-1)}+\dots)V_{1,p-1}+\dots]\; , 
\end{align}
and the contour integral picks out the correct descendant at level $p'-1$.

\subsubsection*{The general mechanism}
\begin{figure}[h]
\begin{centering}
\begin{tikzpicture}

\draw[<-,gray, very thick] (.15,-.15) -- (.85,-.85);
\draw[->,gray, very thick] (-.15,-.15) -- (-.85,-.85);
\filldraw[black] (0,0) circle (2pt) node[anchor=south] { $v_0$};

\draw[->,gray, very thick] (1,-1.15) -- (1,-2.25);
\draw[->,gray, very thick] (-1,-2.25) -- (-1,-1.15) ;
\draw[<-,gray, very thick] (-.85,-2.25) -- (.85,-1.15);
\draw[<-,gray, very thick] (-.85,-1.15) -- (.85,-2.25);

\filldraw[black] (1,-1) circle (2pt) node[anchor=west] { $w_0$};
\filldraw[black] (1,-2.4) circle (2pt) node[anchor=west] { $v_1$};

\draw[->,red, very thick] (1.75,-1) -- (4.25,-1);
\draw[->,red, very thick] (1.75,-2.4) -- (2.75,-2.4);
\draw[->,red, very thick] (1.75,-3.8) -- (2.75,-3.8);
\draw[->,red, very thick] (1.75,-5.2) -- (2.75,-5.2);
\draw[->,red, very thick] (1.75,-6.6) -- (2.75,-6.6);

\filldraw[black] (4.75,-1) circle (2pt) node[anchor=south] { $v_0$};
\filldraw[black] (3.75,-2.4) circle (2pt) node[anchor=east] { $u_1$};
\filldraw[black] (3.75,-3.8) circle (2pt) node[anchor=east] { $v_{-1}$};
\filldraw[black] (3.75,-5.2) circle (2pt) node[anchor=east] { $u_{2}$};
\filldraw[black] (3.75,-6.6) circle (2pt) node[anchor=east] { $v_{-2}$};

\filldraw[black] (5.75,-2.4) circle (2pt) node[anchor=west] { $w_0$};
\filldraw[black] (5.75,-3.8) circle (2pt) node[anchor=west] { $v_{1}$};
\filldraw[black] (5.75,-5.2) circle (2pt) node[anchor=west] { $w_{1}$};
\filldraw[black] (5.75,-6.6) circle (2pt) node[anchor=west] { $v_{2}$};

\draw[->,red, very thick] (6.5,-2.4) -- (7.5,-2.4);
\draw[->,red, very thick] (6.5,-3.8) -- (7.5,-3.8);
\draw[->,red, very thick] (6.5,-5.2) -- (7.5,-5.2);
\draw[->,red, very thick] (6.5,-6.6) -- (7.5,-6.6);

\filldraw[black] (-4.75,-1) circle (2pt) node[anchor=south] { $v_0$};
\filldraw[black] (-5.75,-2.4) circle (2pt) node[anchor=east] { $u_1$};
\filldraw[black] (-5.75,-3.8) circle (2pt) node[anchor=east] { $v_{-1}$};
\filldraw[black] (-5.75,-5.2) circle (2pt) node[anchor=east] { $u_{2}$};
\filldraw[black] (-5.75,-6.6) circle (2pt) node[anchor=east] { $v_{-2}$};

\draw[->,red, very thick] (-7.75,-2.4) -- (-6.75,-2.4);
\draw[->,red, very thick] (-7.75,-3.8) -- (-6.75,-3.8);
\draw[->,red, very thick] (-7.75,-5.2) -- (-6.75,-5.2);
\draw[->,red, very thick] (-7.75,-6.6) -- (-6.75,-6.6);

\filldraw[black] (-3.75,-2.4) circle (2pt) node[anchor=west] { $w_0$};
\filldraw[black] (-3.75,-3.8) circle (2pt) node[anchor=west] { $v_{1}$};
\filldraw[black] (-3.75,-5.2) circle (2pt) node[anchor=west] { $w_{1}$};
\filldraw[black] (-3.75,-6.6) circle (2pt) node[anchor=west] { $v_{2}$};

\draw[->,red, very thick] (-4.5,-1) -- (-2,-1);
\draw[->,red, very thick] (-3,-2.4) -- (-2,-2.4);
\draw[->,red, very thick] (-3,-3.8) -- (-2,-3.8);
\draw[->,red, very thick] (-3,-5.2) -- (-2,-5.2);
\draw[->,red, very thick] (-3,-6.6) -- (-2,-6.6);

\filldraw[black] (-1,-1) circle (2pt) node[anchor=east] { $u_1$};
\filldraw[black] (-1,-2.4) circle (2pt) node[anchor=east] { $v_{-1}$};

\draw[->,gray, very thick] (-1,-2.55) -- (-1,-3.65);
\draw[->,gray, very thick] (1,-3.65) -- (1,-2.55);

\draw[<-,gray, very thick] (-.85,-2.55) -- (.85,-3.65);
\draw[->,gray, very thick] (.85,-2.55) -- (-.85,-3.65);

\filldraw[black] (-1,-3.8) circle (2pt) node[anchor=east] { $u_2$};
\filldraw[black] (1,-3.8) circle (2pt) node[anchor=west] { $w_1$};

\draw[->,gray, very thick] (1,-3.95) -- (1,-5.05);
\draw[<-,gray, very thick] (-1,-3.95) -- (-1,-5.05);

\filldraw[black] (-1,-5.2) circle (2pt) node[anchor=east] { $v_{-2}$};
\filldraw[black] (1,-5.2) circle (2pt) node[anchor=west] {$v_2$};

\draw[<-,gray, very thick] (1,-5.35) -- (1,-6.45);
\draw[->,gray, very thick] (-1,-5.35) -- (-1,-6.45);

\draw[->,gray, very thick] (.85,-3.95) -- (-.85,-5.05);
\draw[->,gray, very thick] (.85,-5.05) -- (-.85,-3.95);

\draw[->,gray, very thick] (.85,-5.35) -- (-.85,-6.45);
\draw[<-,gray, very thick] (-.85,-5.35) -- (.85,-6.45);
\filldraw[black] (-1,-6.6) circle (2pt) node[anchor=east] { };
\filldraw[black] (1,-6.6) circle (2pt) node[anchor=west] {};

\filldraw[gray] (0,-7.2) circle (1pt) node[anchor=south] {};
\filldraw[gray] (0,-7.4) circle (1pt) node[anchor=south] {};
\filldraw[gray] (0,-7.6) circle (1pt) node[anchor=south] {};

\filldraw[black] (-1,-6.6) circle (2pt) node[anchor=east] { $u_{3}$};
\filldraw[black] (1,-6.6) circle (2pt) node[anchor=west] {$w_2$};

\draw[->,gray, very thick] (-3.75,-5.35) -- (-3.75,-6.45);
\draw[<-,gray, very thick] (-3.75,-3.95) -- (-3.75,-5.05);
\draw[->,gray, very thick] (-3.75,-2.55) -- (-3.75,-3.65);

\draw[<-,gray, very thick] (-5.75,-5.35) -- (-5.75,-6.45);
\draw[->,gray, very thick] (-5.75,-3.95) -- (-5.75,-5.05);
\draw[<-,gray, very thick] (-5.75,-2.55) -- (-5.75,-3.65);

\draw[->,gray, very thick] (5.75,-5.35) -- (5.75,-6.45);
\draw[<-,gray, very thick] (5.75,-3.95) -- (5.75,-5.05);
\draw[->,gray, very thick] (5.75,-2.55) -- (5.75,-3.65);

\draw[<-,gray, very thick] (3.75,-5.35) -- (3.75,-6.45);
\draw[->,gray, very thick] (3.75,-3.95) -- (3.75,-5.05);
\draw[<-,gray, very thick] (3.75,-2.55) -- (3.75,-3.65);

\draw[<-,gray, very thick] (4.9,-1.15) -- (5.6,-2.25);
\draw[->,gray, very thick] (4.6,-1.15) -- (3.9,-2.25);
\draw[->,gray, very thick] (5.6,-2.55) -- (3.9,-3.65);
\draw[->,gray, very thick] (5.6,-3.65) -- (3.9,-2.55);
\draw[->,gray, very thick] (5.6,-5.05) -- (3.9,-3.95);
\draw[->,gray, very thick] (5.6,-6.45) -- (3.9,-5.35);
\draw[->,gray, very thick] (5.6,-3.95) -- (3.9,-5.05);
\draw[->,gray, very thick] (5.6,-5.35) -- (3.9,-6.45);

\draw[->,gray, very thick] (-4.9,-1.15) -- (-5.6,-2.25);
\draw[<-,gray, very thick] (-4.6,-1.15) -- (-3.8,-2.25);

\draw[->,gray, very thick] (-3.9,-2.55) -- (-5.6,-3.65);
\draw[->,gray, very thick] (-3.9,-3.95) -- (-5.6,-5.05);
\draw[->,gray, very thick] (-3.9,-5.35) -- (-5.6,-6.45);

\draw[->,gray, very thick] (-3.9,-3.65) -- (-5.6,-2.55);
\draw[->,gray, very thick] (-3.9,-5.05) -- (-5.6,-3.95);
\draw[->,gray, very thick] (-3.9,-6.45) -- (-5.6,-5.35);

\node[draw] at (0,1) {$F_{r,s}$};
\node[draw] at (-4.75,0) {$F_{r,2p-s}$};
\node[draw] at (4.75,0) {$F_{r,-s}$};

\end{tikzpicture}
\end{centering}
\caption{BRST structure of Fock modules.}\label{fig:BRSTmaps}
\end{figure}
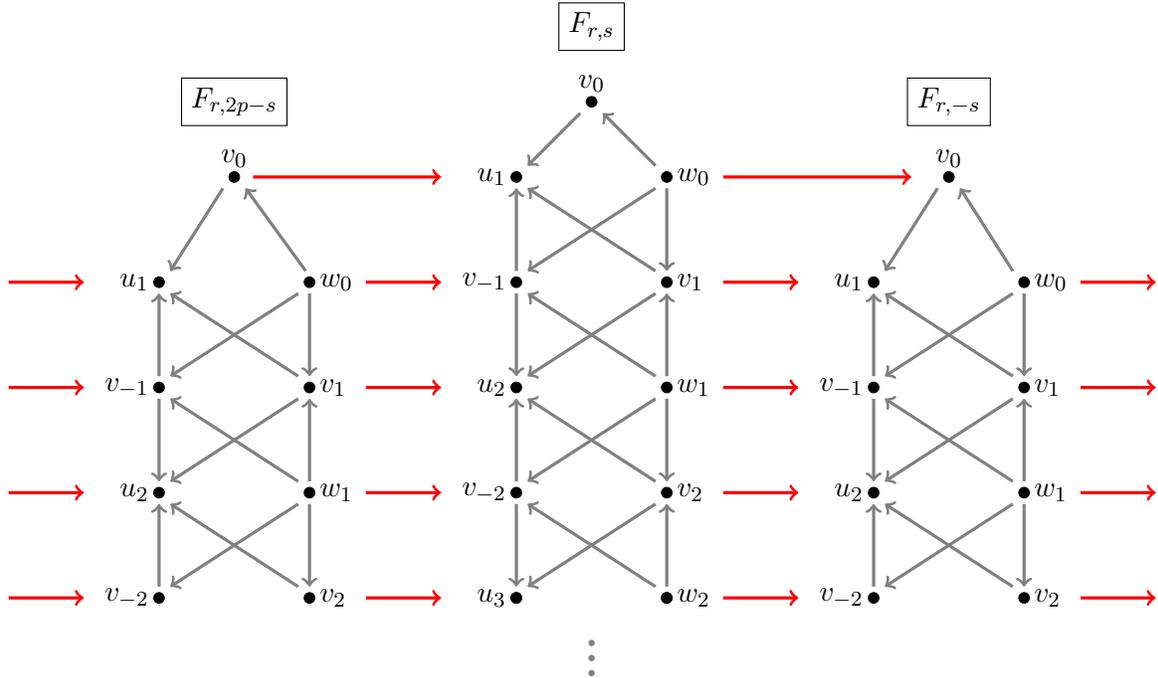

The action of the BRST charge in a general Fock space $F_{r,s}$ corresponding to a minimal model primary is shown in figure \ref{fig:BRSTmaps}.
The vector $w_0$ (the analog of the singular vector at level $l=rs$ in the Verma module), while not singular, is not $Q$-closed and is mapped to the highest weight state in $F_{r,-s}$. On the other hand, the vector $u_1$ is singular and corresponds to the null descendant at level $(p'-r)(p-s)$. It is the image of the highest weight state in $F_{r,2p-s}$ under the action of the BRST charge.   Outside of $F_{r,s}$ there is no cohomology: the other exponential operators appearing in the complex are all either BRST exact or not BRST closed. In general, the states appearing in the right column are not BRST closed and are mapped into the vectors $u_i,v_{k<1}$ in the Fock space to the right. The vectors in the left-hand columns are correspondingly BRST-exact. Demonstrating that the BRST charge acts nilpotently and produces non-vanishing singular vectors is nontrivial: for details see section 4 of \cite{Felder:1988zp}. Resolutions of the degenerate Verma modules located outside of the Kac table were obtained in \cite{Bouwknegt:1991mv}.

\subsubsection*{The special case $p'=2$}
For the special case of the $(2,p)$ minimal models, the entries of the Kac table all have $r=1$. Felder's BRST complex with $Q_+$ as the differential takes the form
\begin{equation}
 \dots\xrightarrow{Q_+^r}F_* \xrightarrow{Q_+^{p'-r}}  F_{r,s}\xrightarrow{Q_+^r}F_*\xrightarrow{Q_+^{p'-r}}\dots 
\end{equation}
For $p'=2$ and $r=1$ this becomes
\begin{equation}
 \dots\xrightarrow{Q_+}F_* \xrightarrow{Q_+}  F_{1,s}\xrightarrow{Q_+}F_*\xrightarrow{Q_+}\dots 
\end{equation}
and no nested contours are necessary. This is because the BRST current is mutually local with all of the minimal model exponentials $V_{1,s}(z)$:
\begin{equation}
    J_+(z)V_{1,s}(z)\sim z^{s-1}V_{-1,s}(0) +\dots 
\end{equation}
As we will see, this simplification is preserved in the full quantum field theory. 

\subsection{$SL(2,\mathbb{R})$ quantum Hamiltonian reduction, Felder's BRST, and $\eta\xi$}
In the introduction we mentioned that there exists yet another construction related to the minimal models, due to Bershadsky and Ooguri \cite{Bershadsky:1989mf}, that makes use of $SL(2,\mathbb{R})$ quantum Hamiltonian reduction. In fact, there is an argument that relates this construction to Felder's BRST complex, although it requires invoking yet another free field realization.

Bershadsky and Ooguri begin by considering the $SL(2,\mathbb{R})_k$ WZW model at level $k+2=p/p'$ with the twisted energy momentum tensor $T(z)\to T(z)-\partial J_3(z)$. The central charge of this twisted energy momentum tensor nearly reproduces that of the corresponding minimal model
\begin{equation}
    c_{SL(2)}-2=\frac{3k}{k+2}-6k-2=c_{p,p'} \;, 
\end{equation}
but the systems are clearly not the same. The basic idea of the construction is to associate the $-2$ appearing in this formula with the $\eta\xi$ system discussed in section \ref{sec:EtaXi}. The first step in reducing the model involves fixing lightcone gauge through the constraint $J_-(z)=1$.
 In order for this equation to make any sense, the scaling dimension of $J_-(z)$ should be zero, motivating the twist of the Sugawara stress tensor 
\begin{equation}
    T(z)\to T(z)-\partial J_3(z) \; . 
\end{equation}
 The imposition of the constraint in the path integral is accompanied by the introduction of the $\eta \xi$ BRST multiplet, and the BRST charge is
\begin{equation}
    Q_{\text{BRST}}=\oint \frac{dz}{2\pi i}(J_-(z)-1)\eta(z) \; . 
\end{equation}
Bershadsky and Ooguri argue that the cohomology of this charge in the combined ghost-$SL(2,\mathbb{R})_k$ Hilbert space reproduces the irreducible representations of the Virasoro algebra.  In order to connect with Felder's BRST and the Coulomb gas formalism,
 one replaces the $SL(2,\mathbb{R})_k$ WZW model with its Wakimoto free field resolution \cite{Wakimoto:1986gf}. The relationship between this free field representation and the WZW model itself involves a separate BRST quotient, as demonstrated by Bernard and Felder \cite{Bernard:1989iy}. We denote this second BRST charge as $Q_{SL_2}$.
 
The Wakimoto realization bosonizes the current $J_3(z)$ in terms of $\partial \phi(z)$ and the scalar/vector $bc$ ghosts. Bershadsky and Ooguri demonstrate that passing to the $Q_{\text{BRST}}$ cohomology of  the combined  $\phi bc -\eta \xi$ system simply removes all of the ghosts, leaving only the Fock spaces of the scalar $\phi(z)$ which is then interpreted as the Feigin-Fuchs field.
The BRST charge $Q_{SL_2}$ associated to the Wakimoto realization is then shown to be $Q_{\text{BRST}}$-equivalent to Felder's Coulomb gas BRST operator $Q_{CG}$. Therefore, taking the cohomology in the $\phi bc-\eta \xi$ system with respect to both $Q_{\text{BRST}}$ and $Q_{SL_2}$  is equivalent to Felder's BRST quotient of the Feigin-Fuchs Fock space using $Q_{CG}$. This in turn establishes that passing to the $Q_{\text{BRST}}$ cohomology of the $SL(2,\mathbb{R})_k-\eta\xi$ system produces the irreducible representation of the Virasoro algebra.

Although this analysis is primarily kinematic, it does seem to provide a moral reason for the existence of Felder's BRST construction. It would be interesting to establish the precise connection between this model and the quantum field theory that we describe in the next two sections at the level of correlation functions.

\section{Lagrangian formulation: kinematics}\label{sec:LinDil}

The combined results of Feigin-Fuchs, Dotsenko-Fateev, and Felder strongly suggest that there is a physical relationship between some variation of the timelike linear dilaton theory and the minimal models. The correct quantum field theory should consolidate the ``representation-by-representation" analysis of Feigin-Fuchs and Felder, and should \textit{explain} the computational rules invented by Dotsenko-Fateev. The goal of the rest of the paper is to embed all of these ingredients in an explicitly local Lagrangian quantum field theory, paying close attention to global issues which have often been ignored. Morally, we would like to perform a sequence of standard, well-defined quantum field theory operations (in this case gauging, BRST projection, and deformation by a marginal operator) that turns the timelike linear dilaton into the minimal model. 

The systems that we will ultimately consider have actions of the form
\begin{equation}\label{eq:LinDilAct}
    S[\phi,g]=\frac{1}{4\pi}\int \sqrt{g}\,d^2x \left[g^{ab}\partial_a\phi \partial_b \phi +Q\phi R(g)+4\pi \mu^+ e^{2 i\alpha_+ \phi} +4\pi \mu^- e^{2 i\alpha_- \phi}\right] \; .
\end{equation}
We will view the last two terms as marginal deformations of the timelike linear dilaton, and treat them in perturbation theory. An important point to note is that both deformations are charged under the $U(1)$ symmetry of the model. This is essential, since none of the minimal models support continuous symmetries. More importantly, if we deform by a single marginal operator, then only a single order in perturbation theory can contribute to any physical observable. This exact calculability appears to be lost if one turns on the second deformation, since $p'\alpha_+ + p\alpha_-=0$ and higher orders in perturbation theory naively contribute.\footnote{This puzzle already arises in the holomorphic Coulomb gas construction with line integral screening charges. Authors always choose the minimal number of screening charges to compute a given correlator, and we are not aware of any justification for this choice. It seems plausible that the perturbation series does terminate at leading order due to the BRST structure inherent in the model, but we have been unable to verify this. In spacelike Liouville theory, the nonperturbative DOZZ formula does exhibit a double lattice $mb +n b^{-1}$ of poles, which is suggestive of the double deformation.} We will have more to say about this in the following section.

The analysis of this model splits up naturally into two parts. The Feigin-Fuchs resolution and Felder's BRST analysis are purely kinematical, and can be studied before deforming away from the free theory.
So we begin by studying a single scalar $\phi$ with the action
\begin{equation}\label{eq:Free}
    S[\phi, g]=\frac{1}{4\pi}\int \sqrt{g}d^2x \left[g^{ab}\partial_a\phi \partial_b \phi +Q\phi R(g)\right] \; 
\end{equation}
and a background charge $Q=i(\alpha_++\alpha_-)$ chosen to match the central charge of the minimal model
\begin{equation}
    c=1+6Q^2=1-6\frac{(p-p')^2}{pp'} \; . 
\end{equation}
Interpreted naively, the system has a number of pathologies. The partition function vanishes on all surfaces except the torus due to the background charge asymmetry. The spectrum is continuous, and many states have zero or negative norm. We would like a model with a normalizable $SL(2,\mathbb{C})$ invariant groundstate and a discrete spectrum, but the background charge term renders the shift symmetry $\phi(x)\to \phi(x)+\text{const}.$ anomalous on  a curved surface and seems to prevent us from gauging the discrete subgroup $\phi\to \phi +2\pi R$. It is at this point that we encounter the first qualitative difference between the $Q\in \mathbb{R}$ spacelike case used in Liouville theory and the $Q\in i\mathbb{R}$ timelike case relevant for the minimal models. Indeed, although the action itself is not invariant under any constant shift of $\phi(x)$, the exponential of the action is still invariant provided
\begin{equation}
    2\pi R |Q| \chi \in 2\pi\mathbb{Z} \; . 
\end{equation}
Since the minimal models make sense on surfaces with boundary (for which $\chi \in \mathbb{Z}$ can be odd), the radius must be an integer\footnote{If we only require the model to make sense on closed surfaces, we could quotient by a half-integer multiple of$\frac{\sqrt{pp'}}{p-p'}$.}  multiple of $\frac{\sqrt{pp'}}{p-p'}$. In order to retain the Dotsenko-Fateev screening operators $e^{2i\alpha_{\pm}\phi}$ for generic $p,p'$, this integer is fixed so that the radius is $R=\sqrt{pp'}$. We therefore identify
\begin{equation}
    \phi(x) \sim \phi(x) + 2\pi \sqrt{pp'} \; . 
\end{equation}
Note that this is the radius that appears naturally in the formula \eqref{eq:Partition} for the torus partition function. 

The compactness of the timelike linear dilaton has important consequences for our construction. It immediately cuts down the number of momentum states in the model, making the identification of the ``minimal model exponentials" more natural. It also introduces a new set of operators with nontrivial winding. These operators are never discussed in the Coulomb gas formalism, but they will play an important role in our construction. Finally, the compactness of the zero mode of $\phi(x)$ has a dramatic effect on the correlators of the deformed theory \eqref{eq:LinDilAct}, producing a stark contrast with the spacelike Liouville theory. 
When performing the path integral, we can separate out the zero-mode $\phi(x)=\phi_0+\Phi(x)$ and integrate over $\phi_0$ and $\Phi(x)$ separately. Then on a surface of constant curvature the un-normalized correlation functions in the free theory take the form
\begin{equation}
    \langle \prod_i e^{i\alpha_i\phi(x_i)} \rangle=\int D\phi_0 D\Phi e^{-\phi_0 Q \chi + i\phi_0\sum_i \alpha_i }e^{-S_{\text{free}}[\Phi,g]} \prod_i e^{i\alpha_i\Phi(x_i)} \; . 
\end{equation}
Since the zero-mode enters linearly in the action, it acts as a Lagrange multiplier enforcing the constraint
\begin{equation}
    \sum_i \alpha_i= -iQ\chi \; . 
\end{equation}
The strategy to calculate correlators of the deformed model \eqref{eq:LinDilAct} in section \ref{sec:PerturbationTheory} treats the deformation as a perturbation. The typical calculation involves a free field integral of the form
\begin{equation}
    \int D\phi_0 D\Phi e^{-\phi_0 Q \chi + i\phi_0\sum_i \alpha_i }e^{-S_{\text{free}}[\Phi,g]} \prod_i e^{i\alpha_i\Phi(x_i)} \prod_{j=1}^{n_+}\int d^2x_j e^{2i\alpha_+\phi(x_j)}\prod_{k=1}^{n_-}\int d^2y_k e^{2i\alpha_-\phi(y_k)}\; , 
\end{equation}
and the selection rule for non-vanishing correlators becomes 
\begin{equation}
     \sum_i \alpha_i + 2n_+\alpha_+ +2n_-\alpha_-= -iQ\chi \; . 
\end{equation}
This is the same prescription used by Goulian and Li \cite{Goulian:1990qr} to calculate resonance correlators in spacelike Liouville theory. In spacelike Liouville most correlators are off resonance, and their technique only works for combinations of exponential operators whose momenta can be exactly screened by some number of the screening charges. For those observables, the zero mode of $\phi(x)$ is completely undamped in the path integral, and integrating over this noncompact direction in field space yields a divergent answer. Analytic continuation away from the Goulian-Li momenta yields the DOZZ formula, which exhibits poles at the points where the zero-mode integral diverges.
In contrast, the interesting observables in our version of ``timelike Liouville theory" are precisely the resonance correlators which can be screened by integer numbers of screening charges, since these correspond to the minimal model correlation functions. The answers we obtain are finite because the zero-mode is compact. One can think of the pole as being replaced by $2\pi R$, which diverges in the decompactification limit.

\subsection{States and local operators }
Performing the quotient $\phi(x)\sim \phi(x)+2\pi R$ projects onto operators with momenta of the form
\begin{equation}
    V_n(x)=e^{i\frac{n}{R}\phi(x)} \; , \hspace{.5 in} n\in \mathbb{Z} \; .
\end{equation}
The improvement term in the energy momentum tensor shifts the conformal weights of these exponentials
\begin{equation}
    h=\bar{h}=\frac{n}{2R}\left(\frac{n}{2R}+iQ\right) \; 
\end{equation}
but does not spoil their transformation properties (as it does for the $U(1)$ current). 
Since the scalar field is compact, there is a second $U(1)$ symmetry of the model generated by the current $j_{\text{w}}(x)=d\phi(x)$. In the non-compact theory this current is exact and there are no local charged operators. In the compact theory, $\phi(x)$ is no longer gauge invariant and winding operators appear in the spectrum. The dyonic operators charged under both $U(1)$'s take the form
\begin{equation}
    V_{n,w}(x)=e^{ik_L\phi_L(x)+ik_R\phi_R(x)}  \equiv \exp\left[i\left(\frac{n}{R}+wR\right)\phi_L(x)+i\left(\frac{n}{R}-wR\right)\phi_R(x)\right]\; 
\end{equation}
up to cocycle factors that we will suppress. The dimension and spin of this operator is given by
\begin{equation}\label{spec}
\Delta=\frac{n(n-2(p-p'))}{2pp'}+\frac{pp'w^2}{2} \; , \hspace{1 in}    S=w(n-(p-p')) \; . 
\end{equation}
These formulas follow immediately from the left and right conformal weights 
\begin{equation}\label{eq:Dimensions}
    h=\frac14 \left(\frac{n}{R}-\alpha_+-\alpha_-+wR\right)^2+\frac14Q^2\;,
    \hspace{.25 in}
    \bar{h}=\frac14 \left(\frac{n}{R}-\alpha_+-\alpha_--wR\right)^2+\frac14Q^2 \;. 
\end{equation}
Comparing with the spectrum of the diagonal minimal models singles out two sets of operators satisfying
\begin{equation}
    \frac{n^{\pm}_{r,s}}{R}\equiv(1\pm r)\alpha_+ + (1\pm s)\alpha_- \; , \hspace{.25 in} w=0\; . 
\end{equation}
This parametrization is redundant since
\begin{equation}
    n^{\pm}_{r,s}=(1\pm r)p -(1\pm s)p'=n^{\pm}_{r+p',s+p} \; . 
\end{equation} 
It is important to note that, for fixed $r$ and $s$, there are two physically distinct vertex operators which could be identified with the minimal model primary. Both operators have the same scaling dimension, but differ in their $U(1)$ charges.
This degeneracy of the spectrum is the remnant of the $\phi\to -\phi$ symmetry of the model, which is broken by the $\phi R$ coupling. Instead, the model is invariant under the substitution $\phi \to -\phi$ and $Q\to -Q$, which is equivalent to switching $p$ and $p'$.

The timelike linear dilaton still has a state-operator correspondence. However, because $\partial \phi$ is not a primary operator, there is an inhomogeneous relationship between the charges of states on the cylinder and of operators on the plane. The formula relating the charge of the state $\hat{n}$ to the momentum $n$ of the operator is
\begin{equation}\label{eq:StatesCylinder}
    n=\hat{n} +p-p' \; , 
\end{equation}
and in terms of this variable we have
\begin{equation}
    h=\frac{1}{4}\left(\frac{\hat{n}}{R}+wR\right)^2+\frac{1}{4}Q^2 \; , \hspace{.5 in} \bar{h}=\frac{1}{4}\left(\frac{\hat{n}}{R}-wR\right)^2+\frac{1}{4}Q^2 \; . 
\end{equation}

\subsection{Marginal deformations}\label{subsec:Marginal}
 Regardless of whether or not the scalar field is taken to be compact, the timelike linear dilaton has two $\Delta=2, \; S=0$ marginal  operators given by
\begin{equation}
   M_\pm(x)= e^{2i\alpha_{\pm}\phi(x)} \; . 
\end{equation}
 In the compact model, these operators are not charged under the $U(1)$ winding symmetry. Deforming away from the compact linear dilaton by either of these operators explicitly breaks the $U(1)$ momentum symmetry of the model but preserves the winding symmetry. Given a conformal field theory with one or more marginal operators, one can consider infinitesimal deformations of the form
 \begin{equation}\label{eq:Deformation}
    \delta S = \int \sqrt{g}d^2x \left[\mu^+ M_+(x) + \mu^- M_-(x)\right]\; . 
\end{equation}
 If the operators are exactly marginal (which is certainly not guaranteed and is in fact quite rare), then the deformation parameters $\mu^{\pm}$ provide local coordinates on the conformal manifold. In these coordinates the Zamolodchikov metric has components
\begin{equation}
    g_{++}=g_{--}=0 \; , \hspace{.5 in}    g_{+-}=(x-y)^4\langle e^{2i\alpha_+ \phi(x)} e^{2i\alpha_- \phi(y)}\rangle =1  \; 
\end{equation}
in our example at leading order. Note that the diagonal entries vanish due to the selection rule \eqref{eq:Select}.

Now to leading order in conformal perturbation theory, exact marginality requires the vanishing of the beta function for each deformation 
\begin{equation}
    \beta^k=\sum C_{ij}^k \mu^i\mu^j = 0 \; , 
\end{equation}
where $i,j,k\in \{+,-\}$. It is easy to verify that the 3-point functions of the marginal deformations all vanish due to the selection rule \eqref{eq:Select}
\begin{equation}
    C_{++}^+ =C_{--}^-=C_{+-}^+=C_{+-}^-=0 \; , 
\end{equation}
so that to first order in perturbation theory the resulting system is still conformally invariant as was to be expected. This alone does not guarantee exact conformal invariance, and in principle one should continue to higher orders in perturbation theory to be sure that no scale enters the model. The relation 
\begin{equation}
    p'\alpha_+ +p\alpha_-=0 
\end{equation}
implies that the correlation function
\begin{equation}
    \langle M_+(x_1)\dots M_+(x_{1+p'})M_-(y_1)\dots M_-(y_{1+p})\rangle 
\end{equation}
requires no screening and does not vanish. Since all non-trivial minimal models have $p+p'+2\geq 9$, the first dangerous correlation function is a nine-point function, but the possibility to gap out the system remains.\footnote{There is a second set of marginal operators with fractional winding
\begin{equation}
    n=p-p' \; , \hspace{.5 in}  w=\pm \frac{p+p'}{pp'} \; , 
\end{equation}
which we denote $W_{\pm}(x)$. Ordinarily we would discard these operators, since they are not mutually local with the minimal model exponentials. However, as we will describe later, Felder's BRST current also has fractional winding, so there might be an interesting role for these operators. 
 These deformations would break the winding symmetry of the system, but since both $W_{\pm}(x)$ have the same $U(1)$ momentum charge, multiple orders of perturbation could not contribute to a single observable due to the selection rule \eqref{eq:Select} when $\mu^+=\mu^-=0$.} Note that if we deform by a single marginal operator, this potential problem disappears.

We can also make a trivial check that the scaling dimensions of the minimal model exponentials do not change to first order in perturbation theory. The relevant formula
\begin{equation}
    \delta h_{r,s}=\delta\bar{h}_{r,s}=-\sum c_{\{r,s\},\{r,s\},i}\mu^i=0
\end{equation}
is easily verified, again due to the selection rule. Therefore, the scaling dimensions of the ``minimal model exponentials" do not change to leading order, which is required but was not guaranteed.

The above analysis suggests that the deformation \eqref{eq:Deformation} will produce scale-invariant correlation functions for a set of primary operators whose conformal dimensions match those of the minimal models. This might seem puzzling at first glance, since there is certainly no continuous family of minimal models labeled by $\mu^{\pm}$ for fixed $(p,p')$. The situation is analogous to that of spacelike Liouville theory. There, the cosmological constant term appears to be a marginal deformation and yet there is a single theory for fixed $Q$. In section \ref{sec:PerturbationTheory} we demonstrate that the dependence of physical correlators on $\mu^{\pm}$ is trivial for canonically normalized operators.

\subsection{The reflection identification  }\label{sec:Reflection}
The spectrum \eqref{spec} of the compact linear dilaton is doubly degenerate. Each operator $V_{n,w}(x)$ of a given dimension and spin has a ``reflection partner" $\widetilde{V}_{\tilde{n},\tilde{w}}(x)$ with the quantum numbers 
\begin{equation}\label{eq:Reflectnw}
    \widetilde{n}= 2(p-p')-n, \hspace{1 in} \widetilde{w}= -w \; . 
\end{equation}
 As discussed previously, this is the $Q$-shifted version of the the $n\to -n, w\to -w$ symmetry of the free compact boson. 
The double degeneracy means that there are actually two candidate operators $V_{n_{r,s},0}(x)$ and $V_{n_{-r,-s},0}(x)$ which could be identified with the corresponding minimal model primary.  
 
 A similar degeneracy (with $w=0$) occurs for the non-compact spacelike linear dilaton with a real background charge. When this model is deformed by the Liouville exponential, the spectrum is effectively halved: in the full non-perturbative Liouville theory, there is an operator identification $V_p(x)=R(p)V_{-p}(x)$. This reflection amplitude $R(p)$ is related to the quantum mechanics of the zero mode scattering off of the unbounded Liouville potential.

In seeking a relationship between the deformed compact timelike linear dilaton and the minimal model, one might hope for a similar operator identification 
\begin{equation}\label{eq:Ident}
    V_{n_{r,s},0}(x)\stackrel{?}{=}R(r,s)V_{n_{-r,-s},0}(x) \; . 
\end{equation}
The two operators certainly cannot be identified before deforming the compact dilaton by the marginal operators $M_{\pm}(x)$: generic non-vanishing correlation functions involving $V(x)$ will vanish upon substitution of $\widetilde{V}(x)$ due to the selection rule \eqref{eq:Select}. The identification is only made possible by the deformation, which allows one to trade $V(x)$ and $\widetilde{V}(x)$ inside correlation functions while balancing the resulting charge-asymmetry with additional screening operators. 

There is an even more important distinction to be made between the operator identification in Liouville theory and the corresponding identification in our model. The operator identification  \eqref{eq:Ident}, and its analog in Liouville theory, is only possible because the continuous shift symmetry of the boson is explicitly broken by the perturbation. If this were not the case, the operators being identified would transform in inequivalent representations of a global symmetry group of the theory, which is impossible. 
However, the deformation \eqref{eq:Deformation} does not break the $U(1)$ winding symmetry of the compact dilaton. This means that there can be no analog of \eqref{eq:Ident} for operators with winding in the theory defined by \eqref{eq:LinDilAct}. Indeed, substituting $\widetilde{V}_{\tilde{n},\tilde{w}}(x)$ for $V_{n,w}(x)$ in a non-vanishing correlator would introduce an asymmetry in the global winding charge, which could not be screened by the perturbation. This subtlety does not exist in Liouville theory (where the scalar is non-compact) and is irrelevant for the identification of the minimal model primaries \eqref{eq:Ident}, all of which are winding singlets. However, in the next section we will encounter winding operators (BRST currents) that play a crucial role in reducing the theory \eqref{eq:LinDilAct} to the minimal model, and for which the identification cannot hold.

Since the reflection identification cannot be made before the deformation of the compact dilaton, it certainly makes sense to perturb with both $M_+(x)$ and $M_-(x)$ simultaneously. Whether or not this is the correct prescription seems less clear, and we will postpone this question until section \ref{sec:PerturbationTheory}.

\subsection{BRST currents}\label{sec:BRSTCurrents}
In the holomorphic Coulomb gas formalism of section \ref{Sec:CoulombGas}, we encountered special $h=1$ operators which seemed to play several roles. In particular, the operators $e^{2i\alpha_{\pm}\phi(z)}$ were used to construct Felder's BRST operators as well as the screening charges used in the calculation of correlation functions. In the full theory, these holomorphic operators must be paired with their right-moving counterparts to form genuine local operators. As we saw in section \ref{subsec:Marginal}, the holomorphic screening operators of the Coulomb gas formalism are naturally completed into  $(1,1)$ marginal operators $M_{\pm}(x)$. In this section we will try to identify Felder's BRST currents in the spectrum of the compact dilaton.

There are two potential sets of $(1,0)$ operators in the compact linear dilaton theory. Their momenta and winding, as well as the left and right momenta, are summarized in the following table:
\begin{center}
\begin{tabular}{ |c|c|c|c|c|c| } 
 \hline
  & $n$ &$\hat{n}$& $w$&$k_L$&$k_R$ \\ \hline
$J_+(x)$ & $p$ &$p'$& $\frac{1}{p'}$&$2\alpha_+$&0 \\ \hline
 $\widetilde{J}_+(x)$ & $p-2p'$ &$-p'$& $-\frac{1}{p'}$&$2\alpha_-$&$2\alpha_++2\alpha_-$ \\ \hline
  $J_-(x)$ & $-p'$ &$-p$&$-\frac{1}{p}$&$2\alpha_-$&$0$ \\ \hline
   $\widetilde{J}_-(x)$ & $2p-p'$  &$p$& $\frac{1}{p}$&$2\alpha_+$&$2\alpha_++2\alpha_-$ \\ 
 \hline
\end{tabular}
\end{center}
$J_i(x)$ and $\widetilde{J}_i(x)$ are related by the reflection transform \eqref{eq:Reflectnw}, but are not identified as operators. Similarly, there are two potential sets of $(0,1)$ currents:
\begin{center}
\begin{tabular}{ |c|c|c|c|c|c| } 
 \hline
  & $n$ &$\hat{n}$& $w$&$k_L$&$k_R$ \\ \hline
    $K_+(x)$ & $p$ &$p'$& $-\frac{1}{p'}$&$0$&$2\alpha_+$ \\ \hline
     $\widetilde{K}_+(x)$ &$p-2p'$ &$-p'$& $\frac{1}{p'}$&$2\alpha_++2\alpha_-$&$2\alpha_-$ \\ \hline
      $K_-(x)$ & $-p'$ &$-p$& $\frac{1}{p}$&$0$&$2\alpha_-$ \\ \hline
       $\widetilde{K}_-(x)$ & $2p-p'$ &$p$& $-\frac{1}{p}$&$2\alpha_++2\alpha_-$&$2\alpha_+$ \\ \hline
\end{tabular}
\end{center}
The first important point to note is that each of these operators carries winding. This could have been predicted on general grounds: the null descendants of the non-holomorphic minimal model primary $M_{r,s}(x)$ generically carry spin. Therefore, the exponential operators surrounding the minimal model exponential in the BRST complex must also carry spin (since the BRST charge commutes with both Virasoro algebras), and in order for an exponential operator to carry spin, it must have nonzero winding. The more troubling fact is that the winding carried by the BRST currents is fractional.
We can define the \textit{scalar} BRST charge operators 
\begin{equation}
    Q_{\pm}=\oint J_{\pm}(z)dz\; , \hspace{.4 in} \bar{Q}_{\pm}=\oint K_{\pm}(\bar{z})d\bar{z}\; , \hspace{.4 in } W_{\pm}=\oint \widetilde{J}_{\pm}(z)dz \; , \hspace{.4 in} \bar{W}_{\pm}=\oint \widetilde{K}_{\pm}(\bar{z})d\bar{z} \; .
\end{equation}
Each charge formally commutes with both copies of the Virasoro algebra, but carries non-trivial momentum and winding. The fact that the BRST charges carry winding has an important consequence: the insertion of a charge in any correlation function consisting only of operators of the form $V_{n_{r,s},0}(x)$ (the minimal model exponentials) vanishes trivially due to winding conservation. A second important consequence is that the full BRST complex will now involve operators with non-trivial winding.

\subsubsection*{Mutual locality}
In an ordinary interpretation of the compact boson CFT, one does not permit operators with fractional winding since they are not mutually local with respect to the momentum operators. However, Felder's holomorphic BRST construction produced a charge which was not mutually local with respect to all of the exponential vertex operators needed for the construction of the minimal model. Rather, different combinations of multiple line integrals with nested contours were needed for each separate Fock space.

The OPE of two operators with momentum and winding takes the form
\begin{equation}
 e^{ik_L\phi_L(z)+ik_R\phi_R(\bar{z})}e^{ip_L\phi_L(0)+ip_R\phi_R(0)}\sim   z^{k_Lp_L/2}\bar{z}^{k_Rp_R/2}e^{i(k_L+p_L)\phi_L(0)+i(k_R+p_R)\phi_R(0)} +\dots 
\end{equation}
Sending $z\to e^{2\pi i}z$ one obtains the condition for the mutual locality of two operators:
\begin{equation}
    k_Lp_L-k_Rp_R\in 2\mathbb{Z} \; . 
\end{equation}
 Since the BRST currents $J_{\pm}$ (resp. $K_{\pm}$) have $k_R=0$ (resp. $k_L=0$) while the minimal model exponentials have $p_L=p_R=\frac{n_{r,s}}{R}$, it is clear that the lack of mutual locality in Felder's construction is not removed by considering the full theory. 

Therefore, it may be too much to ask for the BRST currents $J_{\pm}(x)$ and $K_{\pm}(x)$ to be mutually local with respect to all of the operators in the compact timelike linear dilaton model: we seem to have to tolerate a non-standard spectrum in order to reproduce the minimal model. Since the majority of the local operators of the system (including the BRST currents) do not belong to the physical subsector, it seems enough to require mutual locality among the minimal model exponentials (which is guaranteed). 
Nevertheless, the construction does not seem completely natural and we would welcome a simpler explanation.

\subsubsection*{The special case $p'=2$}
As we noted in section \ref{sec:Felder}, Felder's holomorphic BRST complex appears particularly natural when $p'=2$. In that case, the BRST current $J_+(z)$ is mutually local with all of the minimal model exponentials $V_{1,s}(z)$, and only a single power of the BRST charge $Q_+$ is needed in the complex (rather than multiple nested contours). This simplification is maintained in our full quantum field theory. Although the BRST current $J_+(x)$ has winding $w=1/2$, all of the minimal model exponentials (as well as the marginal operators) have even momenta since $n_{1,s}=(s-1)p'=2(s-1)$. This suggests that the more natural radius for this subset of models might be $R'=R/2$, although the discrete shift-symmetry of the scalar would be lost on a surface of odd Euler character. Note that for this case
\begin{equation}
    R'=\frac{1}{2}\sqrt{pp'}=\alpha_+ \; . 
\end{equation}
Indeed, a naive candidate for the completion of the holomorphic minimal model exponentials with
\begin{equation}
    k_L=(1-r)\alpha_+ +(1-s)\alpha_- 
\end{equation}
would have been to interpret $\alpha_+$ as the radius, $(s-1)$ as the momentum and $(1-r)$ as the winding. This is incorrect since the resulting operator in the full theory would have spin, except for the case $p'=2$ when $r=1$.

\subsection{Full BRST complex }\label{Sec:BRSTComplex}
The minimal model primaries $M_{r,s}(z,\bar{z})$ are simply the products of their holomorphic and antiholomorphic counterparts, and they have their first singular vectors at the levels
\begin{equation}\label{eq:Null}
    (h_{r,s}+rs, \bar{h}_{r,s}),\hspace{.25 in}(h_{r,s}+(p'-r)(p-s), \bar{h}_{r,s}), \hspace{.25 in}(h_{r,s}, \bar{h}_{r,s}+rs), \hspace{.25 in}(h_{r,s}, \bar{h}_{r,s}+(p'-r)(p-s)) \; . 
\end{equation}
Similarly, the minimal model exponentials in the linear dilaton model holomorphically factorize into products of left and right-moving exponentials
\begin{equation}
    V_{n_{r,s},0}(z,\bar{z})=e^{i\frac{n_{r,s}}{R}\phi_L(z)}e^{i\frac{n_{r,s}}{R}\phi_R(\bar{z})} \; .
\end{equation}
 Since $J_-(z)=e^{2i\alpha_-\phi_L(z)}$ is purely holomorphic and $K_-(\bar{z})=e^{2i\alpha_-\phi_R(\bar{z})}$ is purely antiholomorphic, Felder's holomorphic BRST construction can be applied separately to the left-moving and the right-moving part of each operator.
Combining this with the momentum and winding charges of the BRST currents, it is possible to map out the full BRST complex. For instance, the four Fock spaces directly adjacent to the minimal model Fock space $F_{r,s}$ in the double complex are
\begin{equation*}
\xymatrix{
  &{\color{red} _{-\frac{(p-s)}{p}}F_{r-p',0} \ar[d]^{\bar{Q}_-^{p-s}} } &  \\
{\color{red} _{\frac{(p-s)}{p}}F_{r-p',0}} \ar[r]^{\;\;\;\; Q_-^{p-s}}& {\color{blue}_0F_{r,s} }\ar[r]^{Q_-^s} \ar[d]^{\bar{Q}^s_-} & {\color{red} _{-\frac{s}{p}}F_{r,0} }\\
 &{\color{red} _{\frac{s}{p}}F_{r,0} } & 
}
\end{equation*}
Here $_{w}F_{r,s}$ denotes the Fock space with momentum $n_{r,s}$ and winding $w$. As expected, the highest weight states in each of these Fock spaces carries winding since the descendants \eqref{eq:Null} all carry spin. As a result, these Fock spaces carry different momenta than those appearing in Felder's BRST complex: Felder labeled the Fock spaces with the left-moving momentum $k_L$ while we label them with momentum and winding charge. Expanding the complex out further helps to exhibit the patterns more clearly. The result is depicted in figure \ref{fig:FullBRST}.

\begin{figure}[h]
\begin{equation*}\scalebox{0.85}{
\xymatrix{
{\color{blue}_0F_{r-4p',-s}}\ar[r]^{Q^{p-s}}\ar[d]^{\bar{Q}^{p-s}}&{\color{red}_{-1+\frac{s}{p}}F_{r-3p',0}}\ar[r]^{Q^s}\ar[d]^{\bar{Q}^{p-s}}&{\color{blue}_{-1}F_{r-3p',-s}}\ar[r]^{Q^{p-s}}\ar[d]^{\bar{Q}^{p-s}}&{\color{red}_{-2+\frac{s}{p}}F_{r-2p',0}}\ar[r]^{Q^s}\ar[d]^{\bar{Q}^{p-s}}&{\color{blue}_{-2}F_{r-2p',-s}}\ar[r]^{Q^{p-s}}\ar[d]^{\bar{Q}^{p-s}}&{\color{red}_{-3+\frac{s}{p}}F_{r-p',0}}\ar[r]^{Q^s}\ar[d]^{\bar{Q}^{p-s}}&{\color{blue}_{-3}F_{r-p',-s}}\ar[d]^{\bar{Q}^{p-s}}\\
{\color{red}_{1-\frac{s}{p}}F_{r-3p',0}}\ar[r]^{Q^{p-s}}\ar[d]^{\bar{Q}^{s}}&{\color{blue}_0F_{r-2p',s}}\ar[r]^{Q^{s}}\ar[d]^{\bar{Q}^{s}}& {\color{red}_{-\frac{s}{p}}F_{r-2p',0}}\ar[r]^{Q^{p-s}}\ar[d]^{\bar{Q}^{s}} &{\color{blue} _{-1}F_{r-p',s}} \ar[d]^{\bar{Q}^{s}} \ar[r]^{Q^s} &{\color{red} _{-1-\frac{s}{p}}F_{r-p',0}}\ar[r]^{Q^{p-s}}\ar[d]^{\bar{Q}^{s}}&{\color{blue}_{-2}F_{r,s}} \ar[d]^{\bar{Q}^{s}}\ar[r]^{Q^{s}}&{\color{red}_{-2-\frac{s}{p}}F_{r,0}}\ar[d]^{\bar{Q}^{s}}\\
{\color{blue}_1F_{r-3p',-s}}\ar[r]^{Q^{p-s}}\ar[d]^{\bar{Q}^{p-s}}& {\color{red}_{\frac{s}{p}}F_{r-2p',0}}\ar[r]^{Q^{s}}\ar[d]^{\bar{Q}^{p-s}}&{\color{blue}_0F_{r-2p',-s}}\ar[r]^{Q^{p-s}}\ar[d]^{\bar{Q}^{p-s}} &{\color{red} _{-1+\frac{s}{p}}F_{r-p',0} } \ar[d]^{\bar{Q}^{p-s}} \ar[r]^{Q^s}  &{\color{blue} _{-1}F_{r-p',-s}}\ar[d]^{\bar{Q}^{p-s}}\ar[r]^{Q^{p-s}} &{\color{red}_{-2+\frac{s}{p}}F_{r,0}}\ar[d]^{\bar{Q}^{p-s}}\ar[r]^{Q^{s}}&{\color{blue}_{-2}F_{r,-s}}\ar[d]^{\bar{Q}^{p-s}}\\
{\color{red}_{2-\frac{s}{p}}F_{r-2p',0}}\ar[r]^{Q^{p-s}}\ar[d]^{\bar{Q}^{s}}&{\color{blue}_{1}F_{r-p',s}}\ar[r]^{Q^{s}}\ar[d]^{\bar{Q}^{s}}&{\color{red} _{1-\frac{s}{p}}F_{r-p',0}} \ar[r]^{Q^{p-s}}\ar[d]^{\bar{Q}^{s}}& {\color{blue}_0F_{r,s}} \ar[r]^{Q^s} \ar[d]^{\bar{Q}^s} & {\color{red}_{-\frac{s}{p}}F_{r,0}}\ar[r]^{Q^{p-s}}\ar[d]^{\bar{Q}^{s}}&{\color{blue}_{-1}F_{r+p',s}}\ar[d]^{\bar{Q}^{s}}\ar[r]^{Q^{s}}&{\color{red}_{-1-\frac{s}{p}}F_{r+p',0}}\ar[d]^{\bar{Q}^{s}}\\
{\color{blue}_{2}F_{r-2p',-s}}\ar[r]^{Q^{p-s}}\ar[d]^{\bar{Q}^{p-s}}&{\color{red}_{1+\frac{s}{p}}F_{r-p',0}}\ar[r]^{Q^{s}}\ar[d]^{\bar{Q}^{p-s}}&{\color{blue}_{1}F_{r-p',-s}}\ar[r]^{Q^{p-s}}\ar[d]^{\bar{Q}^{p-s}} &{\color{red} _{\frac{s}{p}}F_{r,0} }\ar[r]^{Q^s} \ar[d]^{\bar{Q}^{p-s}}&{\color{blue} _{0}F_{r,-s}}\ar[r]^{Q^{p-s}}\ar[d]^{\bar{Q}^{p-s}}&{\color{red}_{-1+\frac{s}{p}}F_{r+p',0}}\ar[d]^{\bar{Q}^{p-s}}\ar[r]^{Q^{s}}&{\color{blue}_{-1}F_{r+p',-s}}\ar[d]^{\bar{Q}^{p-s}}\\
{\color{red}_{3-\frac{s}{p}}F_{r-p',0}}\ar[r]^{Q^{p-s}}\ar[d]^{\bar{Q}^{s}}&{\color{blue}_{2}F_{r,s}}\ar[r]^{Q^{s}}\ar[d]^{\bar{Q}^{s}}&{\color{red}_{2-\frac{s}{p}}F_{r,0}}\ar[r]^{Q^{p-s}}\ar[d]^{\bar{Q}^{s}} &{\color{blue} _{1}F_{r+p',s}} \ar[r]^{Q^s}\ar[d]^{\bar{Q}^{s}} &{\color{red} _{1-\frac{s}{p}}F_{r+p',0}}\ar[r]^{Q^{p-s}}\ar[d]^{\bar{Q}^{s}}&{\color{blue}_{0}F_{r+2p',s}}\ar[d]^{\bar{Q}^{s}}\ar[r]^{Q^{s}}&{\color{red}_{-\frac{s}{p}}F_{r+2p',0}}\ar[d]^{\bar{Q}^{s}}\\
{\color{blue}_3F_{r-p',-s}}\ar[r]^{Q^{p-s}}&{\color{red}_{2+\frac{s}{p}}F_{r,0}}\ar[r]^{Q^{s}}&{\color{blue}_{2}F_{r,-s}}\ar[r]^{Q^{p-s}}&{\color{red}_{1+\frac{s}{p}}F_{r+p',0}}\ar[r]^{Q^{s}}&{\color{blue}_1F_{r+p',-s}}\ar[r]^{Q^{p-s}}&{\color{red}_{\frac{s}{p}}F_{r+2p',0}}\ar[r]^{Q^{s}}&{\color{blue}_{0}F_{r+2p',-s}}\\
}}
\end{equation*}
\caption{Structure of the full BRST complex.}\label{fig:FullBRST}
\end{figure}
Since all of the cohomology is concentrated in $_0F_{r,s}$, the terms in blue will contribute to the torus partition function with positive signs, while those in red will enter with negative signs in order to cancel states that are not present in the minimal model. In particular, note that all of the blue terms actually have integer winding, while the red ``ghost states" are built on vertex operators with bad fractional winding which nonetheless have integer spins. This structure is related to the expression \eqref{eq:Partition} for the minimal model partition function in terms of $c=1$ partition functions at different radii: terms contributing with a positive sign naturally live at the radius $R=\sqrt{pp'}$, while the ghost states that contribute with a negative sign seem to live at the radius $R=\frac{1}{p}\sqrt{pp'}$. Next we would like to verify that this set of signs appearing in the BRST complex matches those that arise in the minimal model character formula.

\subsection{Torus partition function}
In this section we will show that the ``supertrace" in the full Hilbert space of the compact time-like linear dilaton produces precisely the trace in the Hilbert space of the minimal model. The Rocha-Caridi form of the holomorphic character for the minimal model is
\begin{equation}
    \chi_{r,s}=K_{r,s}^{(p,p')}(q)- K_{r,-s}^{(p,p')}(q) \; , \hspace{1 in}
      K_{r,s}^{(p,p')}(q)=\frac{1}{\eta(q)}\sum_{j\in \mathbb{Z}}q^{(\frac{-pr+p's}{R}-2jR)^2/4} \; , 
\end{equation}
where $R=\sqrt{pp'}$. We would like to express the quantity $\chi_{r,s}\bar{\chi}_{r,s}$ as a sum over momenta and winding states in order to compare with the BRST complex in section \ref{Sec:BRSTComplex}. It will prove useful to have an expression for the trace in the linear dilaton system when restricted to momenta and winding charges in the sets $n\in S_1$ and $w\in S_2$. For an operator with $k_L=\frac{n}{R} + wR$, we have 
\begin{align}
h-\frac{c-1}{24}&=\frac{1}{4}\left(k_L+iQ\right)^2 + \frac{Q^2}{4}-\frac{1+6Q^2-1}{24}\notag\\
&=\frac{1}{4}\left(k_L+iQ\right)^2 \; . 
\end{align}
Recalling the relation \eqref{eq:StatesCylinder} between the $U(1)$ charges of operators on the plane and states on the cylinder, we find a simple formula for the trace over operators with $n\in S_1$ and $w\in S_2$: 
\begin{equation}
    \text{Tr}_{\substack{n\in S_1 \\w\in S_2}}q^{L_0-c/24}\bar{q}^{\bar{L}_0-c/24}=|\eta(q)|^{-2}\sum_{\substack{n\in S_1 \\w\in S_2}} q^{(\frac{\hat{n}}{R}+wR)^2/4}\bar{q}^{(\frac{\hat{n}}{R}-wR)^2/4} \; , \hspace{.5 in} n=\hat{n}+(p-p') \;. 
\end{equation}
The BRST complex of figure \ref{fig:FullBRST} indicates that the partition function will be a sum of terms of this type, with relative signs determined by $S_1$ and $S_2$. Make the definition
\begin{equation}
     \hat{n}_{r,s}=-rp+sp'=n_{r,s}-(p-p') \; , 
\end{equation}
and consider the first term in the product $\chi_{r,s}\bar{\chi}_{r,s}$, which is
\begin{equation}
   K_{r,s}^{(p,p')}(q)K_{r,s}^{(p,p')}(\bar{q}) =|\eta(q)|^{-2}\sum_{j,j'}q^{(\frac{\hat{n}_{r,s}}{R}-2jR)^2/4}\bar{q}^{(\frac{\hat{n}_{r,s}}{R}-2j'R)^2/4}\;. 
\end{equation}
In order to relate this to the BRST complex, we would like to equate
\begin{equation}
    \left(\frac{\hat{n}}{R}+wR\right)^2=\left(\frac{\hat{n}_{r,s}}{R}-2jR\right)^2 \; , \hspace{.5 in}
    \left(\frac{\hat{n}}{R}-wR\right)^2=\left(\frac{\hat{n}_{r,s}}{R}-2j'R\right)^2 \; .
\end{equation}
There are four solutions to this set of equations (each solution relating to one of the four possible BRST currents encountered in section \ref{sec:BRSTCurrents} that could be used to construct the BRST complex). For the complex based on $Q_-$, the relevant solution is
\begin{equation}
    \hat{n}=\hat{n}_{r,s}-(j+j')R^2 \; , \hspace{.25 in} w=j'-j \; . 
\end{equation}
 We write
\begin{equation}
   K_{r,s}^{(p,p')}(q)K_{r,s}^{(p,p')}(\bar{q}) =|\eta(q)|^{-2}\sum_{\substack{n\in n_{r,s}+pp'\mathbb{Z}\\w\in \mathbb{Z}}}q^{(\frac{\hat{n}}{R}+wR)^2/4}\bar{q}^{(\frac{\hat{n}}{R}-wR)^2/4} 
\end{equation}
where importantly both integers in the sum have the same parity. The second term in the product
\begin{equation}
    -K_{r,s}^{(p,p')}(q)K_{r,-s}^{(p,p')}(\bar{q})=-|\eta(q)|^{-2}\sum_{j,j'}q^{(\frac{\hat{n}_{r,s}}{R}-2jR)^2/4}\bar{q}^{(\frac{\hat{n}_{r,-s}}{R}-2j'R)^2/4}
\end{equation}
has a similar solution with
\begin{equation}
    \hat{n}=\hat{n}_{r,0}-(j+j')R^2\; , \hspace{ .5 in} w=\frac{s}{p}+j'-j \;, 
\end{equation}
and can be written
\begin{equation}
    -K_{r,s}^{(p,p')}(q)K_{r,-s}^{(p,p')}(\bar{q}) =-|\eta(q)|^{-2}\sum_{\substack{n\in n_{r,0}+pp'\mathbb{Z}\\w\in \frac{s}{p}+\mathbb{Z}}}q^{(\frac{\hat{n}}{R}+wR)^2/4}\bar{q}^{(\frac{\hat{n}}{R}-wR)^2/4} \; . 
\end{equation}
Again, the two integers summed over are restricted to be simultaneously even or odd. The third term
\begin{equation}
    -K_{r,-s}^{(p,p')}(q)K_{r,s}^{(p,p')}(\bar{q})=-|\eta(q)|^{-2}\sum_{j,j'}q^{(\frac{\hat{n}_{r,-s}}{R}-2jR)^2/4}\bar{q}^{(\frac{\hat{n}_{r,s}}{R}-2j'R)^2/4} 
\end{equation}
can be represented with momenta and winding given by
\begin{equation}
       \hat{n}=\hat{n}_{r,0}-(j+j')R^2\; , \hspace{ .5 in} w=-\frac{s}{p}+j'-j \; . 
\end{equation}
With this representation it takes the form
\begin{equation}
    -K_{r,-s}^{(p,p')}(q)K_{r,s}^{(p,p')}(\bar{q})=
    -|\eta(q)|^{-2}\sum_{\substack{n\in n_{r,0}+pp'\mathbb{Z}\\w\in -\frac{s}{p}+\mathbb{Z}}}q^{(\frac{\hat{n}}{R}+wR)^2/4}\bar{q}^{(\frac{\hat{n}}{R}-wR)^2/4} \;  
\end{equation}
with both integers of identical parity. Finally, the fourth term is easily obtained from the first term and is given by
\begin{equation}
    K_{r,-s}^{(p,p')}(q)K_{r,-s}^{(p,p')}(\bar{q})=|\eta(q)|^{-2}\sum_{\substack{n\in n_{r,-s}+pp'\mathbb{Z}\\w\in \mathbb{Z}}}q^{(\frac{\hat{n}}{R}+wR)^2/4}\bar{q}^{(\frac{\hat{n}}{R}-wR)^2/4} \; 
\end{equation}
with the same condition on the sum. The final expression for the character is
\begin{align}
    |\eta(q)|^{2}    \chi_{r,s}\bar{\chi}_{r,s}&=\sum_{\substack{n\in n_{r,s}+pp'\mathbb{Z}\\w\in \mathbb{Z}}}q^{(\frac{\hat{n}}{R}+wR)^2/4}\bar{q}^{(\frac{\hat{n}}{R}-wR)^2/4}
    +\sum_{\substack{n\in n_{r,-s}+pp'\mathbb{Z}\\w\in \mathbb{Z}}}q^{(\frac{\hat{n}}{R}+wR)^2/4}\bar{q}^{(\frac{\hat{n}}{R}-wR)^2/4} \notag\\
    &-\sum_{\substack{n\in n_{r,0}+pp'\mathbb{Z}\\w\in \frac{s}{p}+\mathbb{Z}}}q^{(\frac{\hat{n}}{R}+wR)^2/4}\bar{q}^{(\frac{\hat{n}}{R}-wR)^2/4}
    -\sum_{\substack{n\in n_{r,0}+pp'\mathbb{Z}\\w\in -\frac{s}{p}+\mathbb{Z}}}q^{(\frac{\hat{n}}{R}+wR)^2/4}\bar{q}^{(\frac{\hat{n}}{R}-wR)^2/4} \; , 
\end{align}
with the two integers always of the same parity. The ranges of momenta and winding match the BRST complex in  figure \ref{fig:FullBRST}, which reproduces the signs in the partition function. Note that in order to reproduce the partition function of the minimal model, we do not separately sum over the reflection states. Doing so would produce a factor of two in the partition function.

\section{Lagrangian formulation: dynamics }\label{sec:PerturbationTheory}
In section \ref{sec:LinDil} we saw that the spectrum of the compact timelike linear dilaton, after an appropriate BRST quotient, reproduces the spectrum of the corresponding minimal model. This is a statement purely within the realm of representation theory, and says noting about the dynamics of either system. The results of Dotsenko and Fateev reviewed in section \ref{Sec:CoulombGas} indicate that we must somehow incorporate the ``screening charges" if we want the dynamics to match. In this section we will try to reproduce the Coulomb gas calculations using conformal perturbation theory for the compact linear dilaton deformed by the marginal operators discussed in section \ref{subsec:Marginal}:
\begin{equation}
    \delta S = \int \sqrt{g}d^2x \left[\mu^+ M_+(x) + \mu^- M_-(x)\right] \; . 
\end{equation}
A construction along these lines was originally suggested in an appendix of the second paper by Dotsenko and Fateev\cite{Dotsenko:1984ad,Dotsenko:1986ca}, where they also noted that infinitely many orders in perturbation theory might contribute (and gap out the system) when both deformations are turned on. We do not have a fully satisfactory resolution of this puzzle, and feel that it is an important question that requires further clarification. In much of what follows, we will restrict attention to the $(2,p)$ timelike linear dilaton. These systems have the special property that a representative of each correlation function of physical operators can be screened using only $M_-(x)$ (this is no longer the case when $p'\neq 2$). Even within this restricted class of models, we will encounter subtleties when attempting to reproduce the truncation of the OPE expected in the minimal models. Nevertheless, the ultimate conclusion is that, for this class of models, the BRST quotiented Lagrangian quantum field theory really does describe the minimal model. 

\subsection{Converting surface integrals into the Dotsenko-Fateev line integrals}\label{sec:Integrals}
The main advantage of a Lagrangian formulation of the Coulomb gas formalism is that it produces manifestly local correlation functions. The abstract axioms of conformal field theory like associativity of the OPE and modular covariance, which are constraining when applied to non-Lagrangian systems, are more or less guaranteed by the local path integral formalism. The quantities computed in conformal perturbation theory and identified with minimal model observables take the schematic form
\begin{equation*}
    \langle M_{r_1,s_1}(x_1)\dots M_{r_n,s_n}(x_n) \rangle \sim  \langle e^{i\alpha_{r_1,s_1}\phi(x_1)}\dots e^{i\alpha_{r_n,s_n}\phi(x_n)}\prod_{i=1}^{n_+} \int e^{2i\alpha_+\phi(w_i)}d^2w_i \prod_{j=1}^{n_-} \int e^{2i\alpha_-\phi(w_j)}d^2w_j \rangle \; ,
\end{equation*}
where the number of surface integrals is determined by the selection rule $\sum \alpha_i =-iQ\chi$ just as in the Coulomb gas formalism. There is no choice of contours to be made, and one can compute the three-point functions directly rather than resorting to monodromy constraints on holomorphic four-point functions. The method should be especially powerful when applied to computations on higher genus surfaces, where producing modular covariant answers using the holomorphic line integral prescription is more difficult.

Before calculating any observables, it is important to first understand how the two dimensional surface integrals of conformal perturbation theory are capable of reproducing the Dotsekno-Fateev line integral prescription.
 Conformal perturbation theory requires regularization, since the integrals 
 \begin{equation}\label{eq:IntExample}
    \langle e^{i\alpha_{r_1,s_1}\phi(x_1)}\dots e^{i\alpha_{r_n,s_n}\phi(x_n)}\prod_{i=1}^{n_+} \int d^2w_i e^{2i\alpha_+\phi(w_i)} \prod_{j=1}^{n_-} \int d^2w_j e^{2i\alpha_-\phi(w_j)} \rangle  
 \end{equation}
generically diverge for physical values of the momenta $\alpha_i$ due to short-distance singularities in the operator product expansion.
 The scheme that will be adopted first factorizes the surface integrals into products of Dotsenko-Fateev-type line integrals, and defines all expressions by a suitable analytic continuation from regions of parameter space where the integrals are convergent. This prescription for defining integrals of the type \eqref{eq:IntExample} is familiar in perturbative string theory\cite{Witten:2013pra}, and the relationship between the surface integrals and the Dotsenko-Fateev line integrals is nothing more than the KLT double copy \cite{Kawai:1985xq} in a different setting.
\subsubsection*{Toy model}
In order to illustrate the general procedure in a concrete setting, we will evaluate the surface integral 
\begin{equation}\label{eq:ToyModel}
 R(a,b)=   \int d^2z |z|^{2a}|z-1|^{2b} 
\end{equation}
by reducing it to a product of line integrals, following closely the discussion in \cite{Dotsenko:1986ca}. This integral is simpler than the generic Coulomb gas integral, but would arise when computing a correlator
\begin{equation}
    \langle e^{i\alpha_1\phi(0)}e^{i\alpha_2\phi(1)}\int e^{2i\alpha_+\phi(z)}d^2z \rangle = \int d^2z |z|^{2\alpha_+\alpha_1}|1-z|^{2\alpha_+\alpha_2} \; 
\end{equation}
requiring a single screening charge. The first step in evaluating the integral \eqref{eq:ToyModel} is to take the Euclidean coordinate $z=x+i\tau$ and Wick rotate to Lorentzian signature $\tau=ie^{-2i\varepsilon}t$. The integral becomes
\begin{equation}
 R(a,b)=   i\int dtdx [x^2-t^2e^{-4i\varepsilon}]^{a}[(x-1)^2-t^2e^{-4i\varepsilon}]^{b} \; . 
\end{equation}
Taking lightcone coordinates $x_\pm=x\pm t$ and rearranging terms puts the integral in a factorized form  
\begin{align}\label{eq:Factorized}
    R(a,b)&= \frac{i}{2}\int_{-\infty}^{\infty} dx_+    (x_+-i\varepsilon(x_+-x_-))^a  (x_+-1-i\varepsilon(x_+-x_-))^b \notag\\
    &\hspace{.5 in} \times\int_{-\infty}^{\infty} dx_- (x_-+i\varepsilon(x_+-x_-))^a(x_--1+i\varepsilon(x_+-x_-))^b \; .
\end{align}
Assuming for the moment that $a,b$ are such that the integrals converge for large $x_{\pm}$, each integrand has two potential singularities, which are generically branch points. The $x_+$ integration cycle can be separated into three pieces,
which we denote by $C_1=(-\infty,0)$, $C_2=(0,1)$, and $C_3=(1,\infty)$. 
When $x_+\in C_1$, $\text{Im}[i\varepsilon x_+]<0$ and $\text{Im}[i\varepsilon(x_+-1)]<0$. In this case, both singularities in the $x_-$ integrand  lie in the same half plane, so we can deform the  contour to infinity (assuming convergence) and the integral vanishes.
Likewise when $x_+\in C_3$, $\text{Im}[i\varepsilon x_+]>0$ and $\text{Im}[i\varepsilon(x_+-1)]>0$ so the $x_-$ contour can be deformed to infinity and the integral vanishes.

\begin{figure}
\begin{tikzpicture}
[
nodewitharrow/.style 2 args={                
            decoration={             
                        markings,   
                        mark=at position {#1} with { 
                                    \arrow{>},
                                    \node[transform shape,above] {#2};
                        }
            },
            postaction={decorate}
}
]

\filldraw[black] (-4,0) circle (2pt) node[anchor=east] { $0$};
\filldraw[black] (4,0) circle (2pt) node[anchor=east] { $1$};
\draw [smooth,gray,very thick,domain=10:350] plot ({(2*cos(\x)-4)}, {2*sin(\x)});
\draw [smooth,gray,very thick,domain=55:395] plot ({(1.5*cos(\x)-4)}, {1.5*sin(\x)});

\draw [smooth,gray,very thick,domain=167:485] plot ({(1.5*cos(\x)+4)}, {1.5*sin(\x)});
\draw [smooth,gray,very thick,domain=190:515] plot ({(2*cos(\x)+4)}, {2*sin(\x)});

  \draw[nodewitharrow={0.5}{},gray, very thick] (-4+2*.98480775301,2*.17364817766) -- (4-1.5*.96480775301,2*.17364817766);
  \draw[nodewitharrow={0.5}{},gray, very thick] (4-2*.98480775301,-2*.17364817766) -- (-4+2*.98480775301,-2*.17364817766);
  
   \draw[nodewitharrow={0.5}{},gray, very thick] (4-1.5*.57357643635,1.5*.81915204428) -- (-4+1.5*.57357643635,1.5*.81915204428);
   \draw[nodewitharrow={0.5}{},gray, very thick] (-4+1.5*.81915204428,1.5*.57357643635) -- (4-2*.91 ,1.5*.57357643635);

\end{tikzpicture}
\caption{Pochhammer contour.}\label{fig:Pochhammer}
\end{figure}
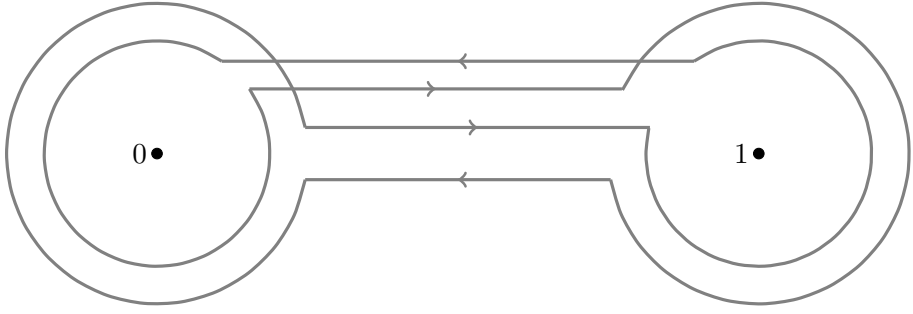

The nonzero contribution to the integral \eqref{eq:Factorized} arises when $x_+\in C_2$, in which case $\Im{[i\varepsilon x_+]}>0$ while $\Im{[i\varepsilon(x_+-1)]}<0$. In this case the $x_-$ integration contour is trapped and cannot be deformed to infinity. Instead, we rotate the $x_-$ contour around the singularity at $x_-=1$ to obtain
\begin{align}
   \begin{tikzpicture}
   [nodewitharrow/.style 2 args={                
            decoration={             
                        markings,   
                        mark=at position {#1} with { 
                                    \arrow{>},
                                    \node[transform shape,above] {#2};
                        }
            },
            postaction={decorate}
    }, baseline={([yshift=-18pt]current bounding box.north)}
    ]
    \filldraw[black] (-4,0) circle (2pt) node[anchor=east] { $1$};
    \draw [smooth,gray,very thick,domain=10:350] plot ({(.5*cos(\x)-4)}, {.5*sin(\x)});
    \draw[nodewitharrow={0.5}{},gray, very thick] (-1,.5*.17364817766) -- (-4+.5*.98480775301,.5*.17364817766);
    \draw[nodewitharrow={0.5}{},gray, very thick] (-4+.5*.98480775301,-.5*.17364817766) -- (-1,-.5*.17364817766);
\end{tikzpicture}
   &=\hspace{.15 in} \int_{1}^\infty x_-^a (x_--1)^bdx_--\int_{1}^\infty x_-^a (e^{-2\pi i}(x_--1))^bdx_-  \notag\\ &=2i\sin (\pi b) e^{-i\pi b}  \int_{1}^\infty x_-^a (x_--1)^bdx_- \; . 
\end{align}
Restoring the $x_+$ integration, the full integral now reads
\begin{equation}
R(a,b)
=-\sin (\pi b)   \int_0^1 x_+^a(1-x_+)^bdx_+ \int_{1}^\infty x_-^a (x_--1)^b dx_-\; .
\end{equation}
Changing variables $x_-\to 1/x_-$ in the second integral expresses $R(a,b)$ as a product of Euler Beta functions
\begin{equation}
    R(a,b)=- \sin (\pi b) \int_0^1 dx_+x_+^a(1-x_+)^b  \int_{0}^1 dx_- x_-^{-2-a-b}(1-x_-)^b \; , 
\end{equation}
and the final answer is
\begin{equation}
    R(a,b)=-\sin(\pi b) \frac{\Gamma(1+a)\Gamma(1+b)}{\Gamma(2+a+b)}\frac{\Gamma(-1-a-b)\Gamma(1+b)}{\Gamma(-a)} \; . 
    \end{equation}
It will be useful to recall the geometric picture for the analytic structure of this formula\cite{Witten:2013pra,Hanson:2006zc}.  The naive integral expression for the Euler Beta function
\begin{equation}
    B(a,b)=\int_0^1 t^{a-1}(1-t)^{b-1}dt
\end{equation}
does not converge for all possible values of its arguments. When the values of $a,b$ are such that the singularities at $t=0,1$ become unintegrable, the function must be defined by analytic continuation. The most transparent definition makes use of the the Pochhammer contour (P.C.), which is depicted in figure \ref{fig:Pochhammer}. This contour zig-zags back and forth between $t=0$ and $t=1$ on different sheets of the Riemann surface associated to the different branches of the multivalued integrand. Keeping track of the phase shifts as the contour circles around the singular points, one obtains the formula
\begin{equation}
    (1-e^{2\pi i a})(1-e^{2\pi i b})B(a,b)=\int_{P.C.}t^{a-1}(1-t)^{b-1}dt \; . 
\end{equation}
Since the Pochhammer contour is compact, the integral on the right hand side of the formula is manifestly finite for all values of $a,b$. It vanishes when either of the singular points becomes regular and the contour becomes contractible, which occurs when $a-1$ or $b-1$ is a non-negative integer.  The Beta function therefore has poles only when $a$ or $b$ is a non-positive integer. Similarly, when $-(a+b)$ is a non-negative integer, the point at infinity becomes regular and the contour can be shrunk to zero on the other side of the Riemann sphere. Therefore, if $a,b\notin \mathbb{Z}_{\leq 0}$ but $-(a+b)$ is a non-negative integer then the Beta function vanishes. This example is important because it illustrates the mechanism by which certain OPE coefficients vanish in the Coulomb gas formalism. The Coulomb gas integrals have strictly positive integrands but are generically divergent. The regularization scheme defines them using analytic continuation in a manner similar to the definition of the Beta function in terms of the Pochhammer contour and thus allows for zeros.

\subsubsection*{Surface integral identities}
The conversion of more complicated surface integrals into products of Dotsenko-Fateev line integrals follows the same basic steps described above. Dotsenko and Fateev developed the technique to evaluate the integrals arising in the computation of four-point functions, but to analyze three-point functions we only need the integrals 
\begin{align}\label{eq:1charge}
   &J_m(a,b;c)\equiv \frac{1}{m!}\int \prod_1^{m}\frac{i}{2}dz_id\bar{z}_i\prod_{1}^m |z_i|^{2a}|1-z_i|^{2b}\prod_{i<j}^m |z_i-z_j|^{4c}\\
   &=\pi^m\left(\frac{\Gamma(1-c)}{\Gamma(c)}\right)^m\prod_{k=1}^m\frac{\Gamma(kc)}{\Gamma(1-kc)}\times \prod_{k=0}^{m-1}\frac{\Gamma(1+a+kc)\Gamma(1+b+kc)\Gamma(-1-a-b-(m-1+k)c)}{\Gamma(-a-kc)\Gamma(-b-kc)\Gamma(2+a+b+(m-1+k)c)}\notag \;  
\end{align}
and 
\begin{align}\label{eq:SurfInt}
    &\int \prod_{i=1}^{n_+}d^2z_i\int\prod_{j=1}^{n_-}d^2w_j
     \prod_{i=1}^{n_+}|z_i|^{2a'}  |1-z_i|^{2b'}\prod_{j=1}^{n_-}|w_j|^{2a}|1-w_j|^{2b}                   \prod_{i<k}^{}|z_i-z_k|^{4c'}\prod_{j<l}^{}|w_j-w_l|^{4c}\prod_{i,j}|z_i-w_j|^{-4}\notag\\
    &=\frac{\pi^{n_++n_-}n_+!n_-!}{c^{4n_+n_-}}\left(\frac{\Gamma(1-c')}{\Gamma(c')}\right)^{n_+}\left(\frac{\Gamma(1-c)}{\Gamma(c)}\right)^{n_-}
    \times
    \prod_{k=1}^{n_+}\frac{\Gamma(kc'-n_-)}{\Gamma(1-kc'+n_-)}\prod_{k=1}^{n_-}\frac{\Gamma(kc)}{\Gamma(1-kc)}\notag\\
    &\times \prod_{k=0}^{n_+-1}\frac{\Gamma(1-n_-+a'+kc')\Gamma(1-n_-+b'+kc')\Gamma(-1+n_--a'-b'-(n_+-1+k)c')}{\Gamma(n_--a'-kc')\Gamma(n_--b'-kc')\Gamma(2-n_-+a'+b'+[n_+-1+k]c')}\notag\\
    &\times\prod_{k=0}^{n_--1}\frac{\Gamma(1+a+kc)\Gamma(1+b+kc)\Gamma(-1+2n_+-a-b-(n_--1+k)c)}{\Gamma(-a-kc)\Gamma(-b-kc)\Gamma(2-2n_++a+b+[n_--1+k]c)} \; . 
\end{align}
Both of these formulas first appeared in the appendix of \cite{Dotsenko:1985hi}.

\subsection{Reflection amplitudes}
Having seen that the surface integrals of marginal perturbations are capable of being factorized into products of Dotsenko-Fateev line integrals, we can begin to reproduce some of the Coulomb gas results using conformal perturbation theory. Correlation functions in the deformed theory are calculated using the correlation functions of the undeformed theory with an exponential insertion of the deformation
\begin{equation}
    \langle \mathcal{O}_1(x_1)\cdots \mathcal{O}_n(x_n)\rangle_{\text{deformed}} = \langle \mathcal{O}_1(x_1)\cdots \mathcal{O}_n(x_n) e^{-\int d^2x \mu^+M_+(x)-\int d^2x \mu^-M_-(x)}\rangle_{\text{undeformed}} \; . 
\end{equation}
In what follows we will drop the labels on the correlation functions since the meaning will always be clear. Conformal perturbation theory requires some scheme for regulating the integrals appearing on the right hand side of this equation.  The scheme adopted here is to convert the surface integrals into Dotsenko-Fateev line integrals as in section \ref{sec:Integrals}, and to define these integrals by analytic continuation from regions in parameter space where convergence is guaranteed. In the models that we consider, the observables that we would like to compute typically vanish in the unperturbed theory. Moreover, provided that we only deform the model by the marginal operator $M_{-}(x)$, a single term in the expansion of the exponential will contribute to any given correlator due to the selection rule on the momenta.
When the selection rule $\sum \alpha_i=2\alpha_++2\alpha_- $ is satisfied, the unperturbed compact timelike linear dilaton correlator takes the form
\begin{equation}
    \langle e^{i\alpha_1\phi(x_1)}\dots e^{i\alpha_n\phi(x_n)} \rangle=\prod_{i>j}|x_i-x_j|^{\alpha_i\alpha_j}  \; . 
\end{equation}
\subsubsection*{Reflection map }
We saw in section \ref{sec:FelderGeneral} and in section \ref{Sec:BRSTComplex} that the level-one descendant of the reflection partner of the identity operator is BRST exact:
\begin{equation}
    \partial \tilde{1}(x)= [Q, \cdot] \; .
\end{equation}
 This equation has several important consequences. First, it means that inserting the reflection of the identity inside a correlation function of BRST invariant operators does not introduce any position dependence: the reflection of the identity is a topological local operator. Now in the free theory without a background charge, the reflection map simply sends the momentum $n\to -n$, which one could view as a map
 \begin{equation*}
     V_{\alpha}(x)\to 1(x)V_{-\alpha}(x) \; .
 \end{equation*}
The generalization to the model with a background charge is immediate:
 \begin{equation*}
     V_{\alpha}(x)\to \tilde{1}(x)V_{-\alpha}(x) \; .
 \end{equation*}
Because the reflection of the identity is essentially topological, no special care is needed in defining this composite operator. In the undeformed theory, the operator and its reflection cannot be identified,
 since replacing an operator by its reflection introduces a charge asymmetry into the correlation function which cannot be screened.

\subsubsection*{Reflection coefficient}

We would like to work with a basis of operators $\mathcal{O}_i(x)$ with canonically normalized two point functions 
\begin{equation}
    \langle \mathcal{O}_i(x_1)\mathcal{O}_j(x_2)\rangle = \frac{\delta_{ij}}{(x_1-x_2)^{2\Delta_i}} \; .
\end{equation}
 As we will see, there are important subtleties with the operator normalizations in this model which are tied in with (but do not fully resolve) the apparent violations of truncation that we will encounter in section \ref{sec:ThreePoint}.

In the undeformed linear dilaton theory, the non-vanishing two point function involves an operator and its reflection partner:
\begin{equation}
    \langle{V_{\alpha}(x_1)V_{2\alpha_++2\alpha_--\alpha}(x_2)}\rangle =\frac{1}{(x_1-x_2)^{\alpha(\alpha+2iQ)}} \; . 
\end{equation}
The coefficient in this formula is finite because the zero mode of the dilaton is compact (we absorb a factor of the radius into the normalization of these operators). In the deformed theory, it is possible to screen the charge asymmetry in the two point function so that the correlator
\begin{equation}\label{eq:ReflectionScreened}
   \frac{(\mu^+)^{n_+}(\mu^-)^{n_-}}{n_+!n_-!} \langle V_{r,s}(x_1)V_{r,s}(x_2)\tilde{1}(x_3)\prod_{i=1}^{n_+}\int d^2y_i e^{2i\alpha_+\phi(y_i)}\prod_{j=1}^{n_-}\int d^2w_j e^{2i\alpha_-\phi(w_j)}\rangle\equiv \frac{R(r,s)}{(x_1-x_2)^{\alpha_{r,s}(\alpha_{r,s}+2iQ)}} 
\end{equation}
is generically non-vanishing for some choice of positive integers $n_{\pm}$. Note that the definition of the reflection coefficient involves the insertion of one factor of the reflection of the identity\cite{dotsenko:cel-00092929}.

 These considerations suggest that the correctly normalized minimal model correlation functions will be produced if one identifies 
\begin{equation}\label{eq:Normalization}
    V_{r,s}(x)=\sqrt{R(r,s)}M_{r,s}(x) \; . 
\end{equation}
Similarly one would like to identify 
\begin{equation}
    \sqrt{R(r,s)}V_{-r,-s}(x)=M_{r,s}(x) \; , 
\end{equation}
provided that the reflection coefficient is finite and non-vanishing. 
These integrals were originally calculated by Dotsenko and Fateev and are given in eq. \eqref{eq:Reflection} up to factors of $\mu^{\pm}$. For the $(2,p)$ models, they take a particularly simple form \cite{dotsenko:cel-00092929}
\begin{equation}
    R(1,s)=(\mu^-)^{s-1}\prod_{j=1}^{s-1} \frac{\Gamma(1-j\alpha_-^2)\Gamma(-1+(1+j)\alpha_-^2)}{\Gamma(j\alpha_-^2)\Gamma(2-(1+j)\alpha_-^2)} \; .
\end{equation}
 An important point to note is that some of these coefficients are zero, which complicates the identifications \eqref{eq:Normalization}. The first zero is encountered at the border of the Kac table at $s=p$, and in fact all $R(1,kp)$ with $k\in \mathbb{N}$ vanish.
 
 In spacelike Liouville theory, the cosmological constant $\mu$ is not an actual parameter of the theory. Similarly, we do not expect any interesting $\mu^{\pm}$ dependence in the deformed, BRST quotiented timelike linear dilaton theory given the uniqueness of the minimal models.  In order to screen the correlator \eqref{eq:ReflectionScreened}, one needs
\begin{equation}
    (2n_+-2r+2)\alpha_+ +(2n_--2s+2)\alpha_-=0 \; 
\end{equation}
so that the number of screening charges is given by\footnote{This derivation assumes that only one order in perturbation theory contributes. The exactness of the reflection amplitude (needed in order to match the minimal model data) seems to be evidence that perturbation theory does truncate at the first nontrivial order. }
\begin{equation}
    n_+=r-1, \hspace{.25 in } n_-=s-1 \; . 
\end{equation}
Therefore $R(r,s)\sim (\mu^+)^{r-1}(\mu^-)^{s-1}$. If we instead consider a $k$-point function (again with an insertion of the reflection of the identity), the selection rule requires
\begin{equation}
    (k-\sum r_i+2n_+)\alpha_+ +(k-\sum s_i +2n_-)\alpha_-=0\; . 
\end{equation}
The correlator therefore scales like
\begin{equation}
    \langle \prod_{i=1}^k M_{r_i,s_i}(x_i)\rangle \sim (\mu^+)^{-\frac12 \sum (r_i-1)}(\mu^-)^{-\frac12 \sum (s_i-1)}(\mu^+)^{\frac12[-k+\sum r_i]}(\mu^-)^{\frac12[-k+\sum s_i]} \; \sim (\mu_+\mu_-)^0 
\end{equation}
and there is no interesting dependence on the deformation parameter.

\subsection{Zero-mode quantum mechanics }
In spacelike Liouville theory, significant insight is gained by studying the dynamics of the zero-mode. In particular, the analysis ``explains" the halving of the physical spectrum induced by the Liouville potential, and provides a way to calculate the reflection amplitude. We would like to perform the same analysis on our model. Putting the model on the cylinder and considering the region of field space in which $\phi(t,\sigma)$ does not depend on the spatial coordinate $\sigma$, we obtain an effective quantum mechanics for the zero mode.
This leads us to study the spectrum and eigenfunctions of the Hamiltonian
\begin{equation}
    H=-\frac{d^2}{dx^2} + e^{2i\alpha_-x} \; 
\end{equation}
with periodic boundary conditions $\psi(x+2\pi R)=\psi(x)$ for the eigenfunctions. This Hamiltonian is obviously not Hermitian since the potential is complex. However, since $V(x)^*=V(-x)$, the operator is $PT$-symmetric \cite{Bender:1998uc} and may still make sense as a quantum mechanical system. 
If we make the change of variables $w=\alpha_+e^{i\alpha_-x}$ then the Hamiltonian becomes 
\begin{equation}
    H=\alpha_-^2\left(w^2 \frac{d^2}{dw^2}+w\frac{d}{dw}\right) + \alpha_-^2w^2 \; . 
\end{equation}
The eigenfunctions with energy $E$ are solutions of the Bessel equation
\begin{equation}
    \left[w^2 \frac{d^2}{dw^2}+w\frac{d}{dw} + w^2 -E\alpha_+^2\right] \psi(w)=0 \; . 
\end{equation}
For arbitrary (non-integer) $\sqrt{E\alpha_+^2}$, the two linearly independent solutions to this equation are
\begin{equation}
    J_{\pm\sqrt{E\alpha_+^2}}\left(\alpha_+e^{i\alpha_-x}\right) \; . 
\end{equation}
Although the solution appears automatically periodic, caution is required since the argument is complex. The Bessel function $J_\nu(w)$ has a branch cut beginning at $w=0$ and extending to infinity, so sending $x\to x+2\pi R$ moves the argument onto the next sheet. Requiring absence of the branch cut requires that the order $\sqrt{E\alpha_+^2}$ be an integer, so we may write $E=n^2\alpha_-^2$ with $n\in \mathbb{Z}$. The solution becomes
\begin{equation}
    J_{\pm n}(\alpha_+e^{i\alpha_-x}) \; . 
\end{equation}
When the order is an integer, the two solutions satisfy a ``reflection identification"
\begin{equation}
    J_{-n}(w)=(-1)^nJ_n(w) \; 
\end{equation}
and are no longer linearly independent. In this case the second linearly independent solution is the Bessel function of the second kind. This function always has a branch point at $w=0$ so we discard it.

For integer order, the Bessel function of the first kind is an entire function with a Taylor series
\begin{equation}\label{eq:BesselPlaneWave}
    J_n(w)=\sum_{j=0} \frac{(-1)^j}{j!\Gamma(j+n+1)}\left(\frac{w}{2}\right)^{n+2j} \; . 
\end{equation}
Recalling that $w=\alpha_+e^{i\alpha_-x}$, we see that the wavefunction has support on an \textit{infinite} number of plane waves. This should be contrasted with the case of spacelike Liouville theory, where the scattering states only have support on a given momentum and its reflection. This is surely at the heart of the drastic reduction in the number of independent states in the full model.
The shift in momentum between terms in \eqref{eq:BesselPlaneWave} is of the form $\Delta n =2jp'$. Returning to figure \ref{fig:FullBRST}, one sees that the exponentials that appear in the BRST complex based on $F_{r,s}$ all either contain a shift of $r$ by a multiple of $p'$ or shift of $s$. In either case, the momentum $n$ shifts by a multiple of $p'$ as in the Bessel function.

\subsection{Three-point functions and truncation of the OPE }\label{sec:ThreePoint}
In this subsection we will attempt to reproduce the minimal model fusion rules using the compact timelike linear dilaton deformed by both marginal operators. The OPE  we would like to reproduce is
\begin{equation}\label{eq:FusionSec6}
    M_{r_1,s_1}\times M_{r_2,s_2}=\sum_{r_3=1+|r_1-r_2|}^{r_{\text{max}}} \sum_{s_3=1+|s_1-s_2|}^{s_{\text{max}}} M_{r_3,s_3} \; , 
\end{equation}
with $r_1+r_2+r_3$ and $s_1+s_2+s_3$ odd and
\begin{align}
    r_{\text{max}}&=\min (r_1+r_2-1,2p'-1-r_1-r_2)\; , \notag\\
    s_{\text{max}}&=\min (s_1+s_2-1,2p-1-s_1-s_2) \; .\label{eq:FusionBound} 
\end{align} 
We will refer to the upper bounds $2p'-1-r_1-r_2$ and $2p-1-s_1-s_2$ as truncation from above. Our notation will distinguish between the minimal model operator $M_{r,s}(x)$, its linear dilaton avatar $V_{r,s}(x)$ and the reflection partner $\tilde{V}_{r,s}(x)=V_{-r,-s}(x)$.

\subsubsection*{A quick but misleading derivation of truncation}
There is a formal argument \cite{AlvarezGaume:1989vk,AlvarezGaume:1991rm} which  reproduces some of the fusion rules of the minimal models from the deformed timelike linear dilaton. The quantity that we would like to calculate in the minimal model is the three-point function of primary operators within the Kac table
\begin{equation}
    \langle M_{r_1,s_1}M_{r_2,s_2}M_{r_3,s_3}\rangle \; . 
\end{equation}
In fact, for the purposes of this section we are really only interested in whether or not the coefficient vanishes.
Within the linear dilaton theory, there are eight distinct ways of calculating this coefficient, depending on which and how many reflection operators we choose to insert. They are
\begin{align}
    &\langle V_{r_1,s_1}V_{r_2,s_2}V_{r_3,s_3}\rangle \; , \hspace{.1 in}
    \langle V_{-r_1,-s_1}V_{r_2,s_2}V_{r_3,s_3}\rangle\; , \hspace{.1 in}
    \langle V_{r_1,s_1}V_{-r_2,-s_2}V_{r_3,s_3}\rangle\; , \hspace{.1 in}
    \langle V_{r_1,s_1}V_{r_2,s_2}V_{-r_3,-s_3}\rangle \; ,\\
    &\langle V_{-r_1,-s_1}V_{-r_2,-s_2}V_{r_3,s_3}\rangle\; , \hspace{.025 in}
    \langle V_{-r_1,-s_1}V_{r_2,s_2}V_{-r_3,-s_3}\rangle\; , \hspace{.025 in}
    \langle V_{r_1,s_1}V_{-r_2,-s_2}V_{-r_3,-s_3}\rangle\; , \hspace{.025 in}
    \langle V_{-r_1,-s_1}V_{-r_2,-s_2}V_{-r_3,-s_3}\rangle \; . \notag
\end{align}
For the moment, we will allow ourselves to use both marginal perturbations. In order to obtain a non-zero coefficient using the representative $\langle V_{-r_1,-s_1}V_{r_2,s_2}V_{r_3,s_3}\rangle$ we must have
\begin{equation}\label{eq:FusionRep1}
    (1+r_1-r_2-r_3+2n_+)\alpha_+ +(1+s_1-s_2-s_3 +2n_-)\alpha_-=0 
\end{equation}
for some number $n_+$ and $n_-$ of the two screening charges. Similarly, to screen the representative $\langle V_{r_1,s_1}V_{-r_2,-s_2}V_{r_3,s_3}\rangle$ one needs
\begin{equation}\label{eq:FusionRep2}
    (1-r_1+r_2-r_3+2m_+)\alpha_+ +(1-s_1+s_2-s_3 +2m_-)\alpha_-=0 
\end{equation}
for a different number $m_{\pm}$ of screening charges. Finally, in order to obtain a nonzero coefficient from the representative $\langle V_{r_1,s_1}V_{r_2,s_2}V_{-r_3,-s_3}\rangle$ one must have
\begin{equation}\label{eq:FusionRep3}
    (1-r_1-r_2+r_3+2l_+)\alpha_+ +(1-s_1-s_2+s_3 +2l_-)\alpha_-=0 \; . 
\end{equation}
Imagine for the moment that each term in parentheses must vanish identically in order to satisfy each equality (this would be the case if $\alpha_{\pm}$ were not rationally related). Then it is easy to see that each equality requires
\begin{equation}
    r_1+r_2+r_3 \hspace{.15 in}\text{odd} \; , \hspace{.5 in} s_1+s_2+s_3 \hspace{.15 in} \text{odd}\;. 
\end{equation}
This is the first ingredient in the fusion rules \eqref{eq:FusionSec6}. Now since $n_{\pm},m_{\pm},l_{\pm}$ are \textit{non-negative} integers, we also encounter three sets of inequalities. Equation \eqref{eq:FusionRep1} requires 
\begin{equation}\label{eq:lowbound1}
    1+r_1-r_2-r_3\leq0 \; , \hspace{.5 in }1+s_1-s_2-s_3 \leq 0 \; . 
\end{equation}
Similarly, equation \eqref{eq:FusionRep2} necessitates
\begin{equation}\label{eq:lowbound2}
    1-r_1+r_2-r_3 \leq 0 \; , \hspace{.5 in} 1-s_1+s_2-s_3 \leq 0 
\end{equation}
and equation \eqref{eq:FusionRep3} requires
\begin{equation}\label{eq:upbound1}
    1-r_1-r_2+r_3 \leq 0 \; , \hspace{.5 in} 1-s_1-s_2+s_3\leq 0 \; .
\end{equation}
Now if one requires that each set of inequalities be satisfied simultaneously (so that all three representatives separately yield a nonzero answer) one finds that
\begin{equation}
    r_3 \in [1+|r_1-r_2|, r_1+r_2-1] \; , \hspace{.5 in} s_3 \in [1+|s_1-s_2|, s_1+s_2-1] \; . 
\end{equation}
The lower bound comes from equations \eqref{eq:lowbound1}-\eqref{eq:lowbound2} while the upper bounds come from equation \eqref{eq:upbound1}. From this analysis we can conclude at most that
\begin{equation}\label{eq:HalfFusion}
    M_{r_1,s_1}\times M_{r_2,s_2} = \sum_{r_3=1+|r_1-r_2|}^{r_1+r_2-1} \sum_{s_3=1+|s_1-s_2|}^{s_1+s_2-1} M_{r_3,s_3} \; . 
\end{equation}
We have thus reproduced part of the fusion rule \eqref{eq:FusionBound}, but have not demonstrated truncation from above. 

The logic of this argument appears to be rather delicate.
Following the same steps as above, the representative $\langle V_{-r_1,-s_1}V_{-r_2,-s_2}V_{r_3,s_3}\rangle$ leads to an apparent inequality
\begin{equation}\label{eq:lowBad1}
 1+r_1+r_2-r_3 \leq 0   \; , \hspace{.5 in} 1+s_1+s_2-s_3 \leq 0 \; .
\end{equation}
This would seem to imply a bound $1+r_1+r_2 \leq r_3 $, which is certainly not obeyed in the minimal models. This is our first example of a naively problematic correlation function involving two reflection operators. More serious examples will be encountered later.
Similarly, the representative $\langle V_{-r_1,-s_1}V_{r_2,s_2}V_{-r_3,-s_3}\rangle$
leads to upper bounds on $(r_3,s_3)$ which are far too strong
\begin{equation}\label{eq:upBad1}
 1+r_1 -r_2 +r_3   \leq 0 \; , \hspace{.5 in} 1+s_1 -s_2 +s_3   \leq 0 \; . 
\end{equation}
The same goes for the representative $\langle V_{r_1,s_1}V_{-r_2,-s_2}V_{-r_3,-s_3}\rangle$:
\begin{equation}\label{eq:upBad2}
     1-r_1+r_2+r_3   \leq 0 \; , \hspace{.5 in} 1-s_1+s_2 +s_3   \leq 0 \; . 
\end{equation}
 Meanwhile, according to this logic the correlator $\langle V_{r_1,s_1}V_{r_2,s_2}V_{r_3,s_3}\rangle$ gives no constraints while the correlator $\langle V_{-r_1,-s_1}V_{-r_2,-s_2}V_{-r_3,-s_3}\rangle$ can never be screened.
There is clearly an issue in the treatment of correlators with multiple reflection operator insertions.

In demonstrating that the minimal model OPE closes on a finite number of primaries, it is very important that $\alpha_{\pm}$ are rationally related and $h_{r,s}=h_{p'-r,p-s}$.
Similarly, any attempt to fix the inequalities \eqref{eq:lowBad1}-\eqref{eq:upBad2} must somehow work
 $p,p'$ into the formulas. Now for rational $\alpha_{\pm}^2$, the identity $V_{p'-r,p-s}=V_{-r,-s}$ is tautological and results from the redundancy in the parametrization of the momentum $n_{r,s}=(1-r)p-(1-s)p'$. The formal manipulation that reproduces the minimal model fusion rules applies the logic of eqns. \eqref{eq:FusionRep1}-\eqref{eq:upbound1} to the correlator $\langle V_{p'-r_1,p-s_1}V_{p'-r_2,p-s_2}V_{r_3,s_3}\rangle$ at an irrational point, and then sets $V_{p'-r,p-s}=V_{-r,-s}=V_{r,s}$ at the end of the calculation. To be explicit, the representative $\langle V_{-(p'-r_1),-(p-s_1)}V_{p'-r_2,p-s_2}V_{r_3,s_3}\rangle$ leads to the inequalities
\begin{equation}
 1-r_1+r_2-r_3 \leq 0   \; , \hspace{.5 in} 1-s_1+s_2-s_3 \leq 0 \; ,
\end{equation}
while the representative $\langle V_{p'-r_1,p-s_1}V_{-(p'-r_2),-(p-s_2)}V_{r_3,s_3}\rangle$ requires
\begin{equation}
 1+r_1-r_2-r_3 \leq 0   \; , \hspace{.5 in} 1+s_1-s_2-s_3 \leq 0 \; .
\end{equation}
The truncation from below is therefore unaffected. However, the ability to screen the correlation function $\langle V_{p'-r_1,p-s_1}V_{p'-r_2,p-s_2}V_{-r_3,-s_3}\rangle$ leads to the requirement that
\begin{equation}
 1-(p'-r_1)-(p'-r_2)+r_3 \leq 0   \; , \hspace{.5 in} 1-(p-s_1)-(p-s_2)+s_3 \leq 0   \; , 
\end{equation}
 and we  conclude that
\begin{equation}
    M_{p'-r_1,p-s_1}\times M_{p'-r_2,p-s_2} = \sum_{r_3=1+|r_1-r_2|}^{2p'-1-r_1-r_2} \sum_{s_3=1+|s_1-s_2|}^{2p-1-s_1-s_2} M_{r_3,s_3} \; . 
\end{equation}
If we now go to the rational point, set $M_{p'-r,p-s}=M_{r,s}$, and require consistency with \eqref{eq:HalfFusion}, we obtain the full fusion rule \eqref{eq:FusionSec6} with truncation from above.
Note that for this entire argument to work, we \textit{must} make the reflection identification.

\subsection{Apparent failures of truncation and subtleties with the reflection identification}
Although the above derivation ultimately produces the desired answer, it is unsatisfying in several respects and requires justification. Requiring each term in \eqref{eq:FusionRep1} to vanish separately is not justified since $p'\alpha_++p\alpha_-=0$ and there can be cancellation between the two terms. This same basic issue is encountered in obtaining truncation from above, where one must obtain the fusion rules for $V_{p'-r,p-s}$ before taking $\alpha_{\pm}^2$ rational. The ultimate justification for these manipulations probably relies on the fact that, in order to even define the Coulomb gas integrals, one must analytically continue the momenta from regions where the integrals converge. Consideration of irrational quantities is seemingly inevitable even though the definition of the theory is purely rational.
However, it is important that we encountered problems with the naive interpretation of correlation functions involving multiple reflection operators,
and the use of analytic continuation also implies that the ability to screen a correlator does not guarantee that it is non-vanishing, as the simple Pochhammer contour example of section \ref{sec:Integrals} shows.

The best way to illustrate the subtleties in the calculation of the three-point functions is with examples. Perhaps the simplest OPE to check is
\begin{equation}
    M_{1,2}\times M_{1,2}= M_{1,1}+M_{1,3} \; . 
\end{equation}
In the Yang-Lee model, this would be the only fusion rule to verify since $M_{1,2}=M_{1,5-2}$. The first point to note is that we can represent the correlator $\langle M_{1,2}M_{1,2}M_{1,3} \rangle$ using a single reflection operator
\begin{equation}
    \langle V_{1,2}(x_1)V_{1,2}(x_2)V_{-1,-3}(x_3)\rangle
    =\langle e^{-i\alpha_-\phi(x_1)}e^{-i\alpha_-\phi(x_2)}e^{i(2\alpha_++4\alpha_-)\phi(x_3)}\rangle\; . 
\end{equation}
 This correlator does not require screening, so we conclude that the OPE coefficient will not vanish as expected. We could also choose to represent the OPE coefficient using a different combination of operators that requires a single screening charge:
\begin{equation}
    \langle V_{1,2}(x_1)V_{-1,-2}(x_2)V_{1,3}(x_3)\int d^2w e^{2i\alpha_-\phi(w)}\rangle \; .
\end{equation}
In this case, the integral that we need to do is
\begin{align}
    &\langle e^{-i\alpha_-\phi(x_1)}e^{i(2\alpha_++3\alpha_-)\phi(x_2)}e^{-2i\alpha_-\phi(x_3)}\int d^2w e^{2i\alpha_-\phi(w)}\rangle\notag\\
    &= |x_1-x_2|^{2-3\alpha_-^2}|x_1-x_3|^{2\alpha_-^2}|x_2-x_3|^{4-6\alpha_-^2} \int d^2w |x_1-w|^{-2\alpha_-^2}|x_2-w|^{-4+6\alpha_-^2}|x_3-w|^{-4\alpha_-^2} \; . 
\end{align}
In general this integral needs to be defined by analytic continuation from a region of parameter space where it converges. Sending $x_1=0,x_2=1,x_3\to \infty$ the integral takes the form \eqref{eq:1charge}. Evaluating the Dotsenko-Fateev expression, one obtains a finite limit with nonsingular terms multiplied by
$\lim_{\varepsilon\to 0} \frac{\Gamma(\varepsilon)}{\Gamma(\varepsilon)}$.
The two calculations agree and the OPE coefficient does not vanish. Similarly, the representation
\begin{align}
    &\langle V_{1,2}(x_1)V_{1,2}(x_2)V_{1,3}(x_3)\int d^2y e^{2i\alpha_+\phi(y)}\prod_{i=1}^3\int d^2w_i e^{2i\alpha_-\phi(w_i)}\rangle\\
    &= \langle e^{-i\alpha_-\phi(x_1)}e^{-i\alpha_-\phi(x_2)}e^{-2i\alpha_-\phi(x_3)}\int d^2y e^{2i\alpha_+\phi(y)}\int d^2w_1 e^{2i\alpha_-\phi(w_1)}\int d^2w_2 e^{2i\alpha_-\phi(w_2)}\int d^2w_3 e^{2i\alpha_-\phi(w_3)}\rangle\notag 
\end{align}
also yields a non-vanishing answer, although a finite limit of a different ratio of divergent gamma functions appears. 

Next we try to verify truncation in a few simple examples. The minimal model fusion rule requires the following three-point function to vanish:
\begin{equation}
    \langle M_{1,2}(x_1)M_{1,2}(x_2)M_{3,1}(x_3)\rangle=0 \; . 
\end{equation}
One representative of this correlation function, requiring a single screening charge, is
\begin{equation}
    \langle V_{1,2}(x_1)V_{-1,-2}(x_2)V_{3,1}(x_3)\int d^2y e^{2i\alpha_+\phi(y)}\rangle
    =    \langle e^{-i\alpha_-\phi(x_1)}e^{i(2\alpha_++3\alpha_-)\phi(x_2)}e^{-2i\alpha_+\phi(x_3)}\int d^2y e^{2i\alpha_+\phi(y)}\rangle \; . 
\end{equation}
This correlator satisfies the neutrality condition (can be screened), but should vanish due to the fusion rules. Applying \eqref{eq:1charge}, one finds non-singular terms multiplied by a vanishing ratio of gamma functions
\begin{equation}
    \lim_{\varepsilon \to 0}\frac{\Gamma(\varepsilon)}{\Gamma(\varepsilon)\Gamma(-1+\varepsilon)}=0 \; . 
\end{equation}
 We could try to calculate this three-point function without using any reflection operators. The relevant correlator is
\begin{equation}
    \langle e^{-i\alpha_-\phi(x_1)}e^{-i\alpha_-\phi(x_2)}e^{-2i\alpha_+\phi(x_3)} \int d^2y_1e^{2i\alpha_+\phi(y_1)}\int d^2y_2e^{2i\alpha_+\phi(y_2)}\int d^2w_1e^{2i\alpha_-\phi(w_1)}\int d^2w_2e^{2i\alpha_-\phi(w_2)}\rangle \; . 
\end{equation}
Unsurprisingly, one finds non-singular terms multiplied by a different vanishing ratio of gamma functions
\begin{align}
\lim_{\varepsilon \to 0}\frac{\Gamma(\varepsilon)^2}{\Gamma(\varepsilon)^2\Gamma(-2+\varepsilon)}=0 \; .
\end{align}
Truncation in the $s$-direction is similarly verified. For instance, the three-point function $\langle M_{1,2}M_{1,2}M_{1,5}\rangle$ should vanish, but it has linear dilaton representatives that can be screened. One representative is
\begin{align}
  & \langle e^{-i\alpha_-\phi(x_1)}e^{-i\alpha_-\phi(x_2)}e^{-4\alpha_-\phi(x_3)}\int d^2y e^{2i\alpha_+\phi(y)}\prod_{i=1}^4\int d^2w_i e^{2i\alpha_-\phi(w_i)}\rangle \; . 
\end{align}
Evaluating this integral in a region of convergence and continuing to the physical values, one obtains a non-singular prefactor multiplied by a vanishing combination of gamma functions
\begin{align}
    &\lim_{\varepsilon\to0}\frac{\Gamma(-2+\varepsilon)^2\Gamma(-4+\varepsilon)}{\Gamma(\varepsilon)^4}=0 \; . 
\end{align}
 Now let's consider a slightly more complicated fusion rule. We will take $p,p'$ large so that we do not need to worry about truncation from above, and consider the OPE
\begin{equation}
    M_{2,3}\times M_{2,3}=\sum_{r=1,3}\sum_{s=1,3,5}M_{r,s} \; . 
\end{equation}
The prediction from the minimal model is that any linear-dilaton representative of the correlator $\langle M_{2,3}M_{2,3}M_{5,9}\rangle$
should vanish. There is a simple representative that requires a single screening charge
\begin{align}
    &\langle V_{-2,-3}(x_1)V_{-2,-3}(x_2)V_{5,9}(x_3)\int d^2w e^{2i\alpha_-\phi(w)}\rangle \notag\\
    &=\langle e^{i(3\alpha_++4\alpha_-)\phi(x_1)}e^{i(3\alpha_++4\alpha_-)\phi(x_2)}e^{i(-4\alpha_+-8\alpha_-)\phi(x_3)}\int d^2w e^{2i\alpha_-\phi(w)}\rangle \;. 
\end{align}
Oddly, after analytic continuation and application of \eqref{eq:SurfInt} one obtains finite non-zero terms multiplied by a non-vanishing ratio of gamma functions
\begin{equation}
    \lim_{\varepsilon \to 0} \frac{\Gamma(\varepsilon)}{\Gamma(\varepsilon)}\neq 0 \; ,
\end{equation}
yielding an apparent violation of truncation. To test the seriousness of this violation, we can do the analogous calculation without the use of any reflection operators. The answer also does not vanish:
\begin{equation}
    \langle V_{2,3}(x_1)V_{2,3}(x_2)V_{5,9}(x_3)\prod_{i=1}^{4}\int d^2y_i e^{2i\alpha_+(y_i)}\prod_{i=1}^{7}\int d^2w_i e^{2i\alpha_-\phi(w_i)}\rangle \sim  \lim_{\varepsilon \to 0}\frac{\Gamma(-4+\varepsilon)^2}{\Gamma(-1+\varepsilon)^2} \; . 
\end{equation}
Unfortunately, this violation of fusion is the first among many. One seems to encounter the identical problem whenever a representative correlation function with two reflection operators can be screened. Low-lying examples include
$\langle M_{2,3}M_{3,4}M_{6,8}\rangle$ and $\langle M_{2,3}M_{2,3}M_{7,9}\rangle$. Perhaps the simplest example that illustrates the problem is the correlator $\langle M_{1,2}M_{1,2}M_{3,5} \rangle$. This three-point function should vanish in any minimal model irrespective of the values of $p,p'$. However, it has a representative involving two reflection operators
\begin{equation}\label{eq:Dangerous}
\langle V_{-1,-2}(x_1)V_{-1,-2}(x_2)V_{3,5}(x_3) \rangle=    \langle e^{(2i\ap + 3i\am)\phi(x_1)}e^{(2i\ap + 3i\am)\phi(x_2)}e^{(-2i\ap -4i\am)\phi(x_3)}\rangle 
\end{equation}
that does not even need to be screened and cannot vanish. Similar examples abound.

\subsection{Discussion and resolution for $(2,p)$ models}
The apparent failure of truncation in certain analytically continued Coulomb gas integrals has been noted before (see for example section 6.4 of \cite{DiFrancesco:1993cyw}). Although this appears discouraging, we have already encountered many instances where a naive interpretation of the timelike linear dilaton seems to conflict with the minimal model, so we need to be more careful in our interpretation of the results of calculations.

In the rest of this section we will restrict attention to the $(2,p)$ models deformed by the single marginal operator $e^{2i\alpha_-\phi(x)}$. This will allow us to avoid the question of the truncation of perturbation theory, and simplifies the analysis of the apparent violations of fusion. 

In these models, there are $\frac{p-1}{2}$ fundamental BRST invariant operators, which we label by $V_{1,s}$ with $s=1,2,\dots \frac{p-1}{2}$. The operators with $s=\frac{p+1}{2},\dots p-1$ are the reflections of these fundamental operators. Note that the reflection operation $V_{1,s}\to V_{1,p-s}$ flips the parity of the second Kac label since $p$ is by assumption odd. Because $p'=2$, any momentum operator in the theory can be written as $V_{0,t}$ or $V_{1,t}$ for some integer $t$, depending on the parity of the momentum $n$. The parameter $t$ is of course unbounded in both directions.   

\subsubsection*{Yang-Lee}
We are going to begin by studying the violation of truncation in the simplest possible case, the $(2,5)$ Yang-Lee edge singularity.  Since we do not deform the linear dilaton by $e^{2i\alpha_+\phi(x)}$,
the selection rule on the momenta is
\begin{align}
    \sum n_i -4n_- =2p-2p'&=6 \; . 
\end{align}
Here $n_-$ is the number of screening charge insertions, and it is important that its coefficient is negative. It will be helpful to switch from the redundant $r,s$ label to the momentum label $n$ which is unambiguous. In table \ref{table:YangLee} we list the momenta for low-lying values of $r,s$.

\begin{table}[h]
\centering
\begin{tabular}{ |c|c|c|c|c|c|c|c|c|c|c| } 
 \hline
  $n$& $s=-3$ &$s=-2$& $s=-1$&$s=0$&$s=1$&$s=2$&$s=3$&$s=4$&$s=5$&$s=6$ \\ \hline
    $r=0$ & $-3$ &$-1$& $1$&$3$&$5$&$7$&$9$&$11$&$13$ &$15$\\ \hline
     $r=1$ &$-8$ &$-6$& $-4$&$-2\cellcolor{lightgray}$&$0\cellcolor{teal}$&$2\cellcolor{cyan}$&$4 \cellcolor{cyan}$&$6\cellcolor{teal}$&$8\cellcolor{lightgray}$&$10$ \\ \hline
\end{tabular}
\caption{Values of the momentum $n$ of some operators in the $(2,5)$ compact linear dilaton. The correct fusion range for $M_{1,2}\times M_{1,2}$ is shaded in blue and green. The gray shaded boxes correspond to the operator $V_{3,5}$ which seemingly violates the fusion rule.}\label{table:YangLee}
\end{table}

The reflection map is simply $n\to 6-n$, and reflection pairs are shaded with the same color.
In the previous subsection, we encountered a troublesome three-point correlator $\langle V_{-1,-2}V_{-1,-2}V_{3,5} \rangle$ that seemed to violate fusion. In this model $V_{3,5}=V_{1,0}$, and we see that the correlator $\langle V_{-1,-2}V_{-1,-2}V_{3,5} \rangle$ does not vanish since $\sum n=4+4-2=6$.
Given the BRST structure of the model, the natural question to ask is whether or not this three-point function can contribute to correlation functions of BRST invariant operators. 

Consider all the representatives of the 4-point function for identical operators $V(x)\equiv V_{1,2}(x)$ and the reflection partner $\tilde{V}(x)\equiv V_{-1,-2}(x)$. They are
\begin{equation}
 \langle V(x_1)V(x_2)V(x_3)V(x_4)\rangle \, , \hspace{.25 in} \langle \tilde{V}(x_1)\tilde{V}(x_2)\tilde{V}(x_3)\tilde{V}(x_4) \rangle \; , \hspace{.25 in} \langle V(x_1)V(x_2)\tilde{V}(x_3)\tilde{V}(x_4) \rangle \; , 
 \end{equation}
 and
 \begin{equation}
\langle V(x_1)V(x_2)V(x_3)\tilde{V}(x_4) \rangle \; , \hspace{.25 in} \langle V(x_1)\tilde{V}(x_2)\tilde{V}(x_3)\tilde{V}(x_4) \rangle \; .
\end{equation}
The combined momentum charges of the operators in each correlation function on the top line are all equal to $0$ mod $4$ so these correlators cannot be screened. The combined charges of the operators  on the bottom line are both $2$ mod $4$ and these four-point functions can be screened. 

Consider first the correlation function $\langle V(x_1)V(x_2)V(x_3)\tilde{V}(x_4) \rangle$. When we evaluate this by factorizing onto three-point functions, we need to keep in mind that $|\mathcal{O}\rangle^\dagger=\langle \tilde{\mathcal{O}}|$ so that the insertion of the identity takes the form
\begin{equation}
\langle V(x_1)V(x_2)V(x_3)\tilde{V}(x_4) \rangle \sim \sum_{\mathcal{O}} \;  \langle V(x_1)V(x_2) |\mathcal{O}\rangle \langle \tilde{\mathcal{O}}| V(x_3)\tilde{V}(x_4)\rangle \; . 
\end{equation}
In order for an operator $\mathcal{O}$ to contribute via exchange, two separate three-point functions must be nonvanishing: we need $\langle V(x_1)V(x_2)\mathcal{O}(x_3)\rangle \neq 0$ and $\langle V(x_1)\tilde{V}(x_2)\tilde{\mathcal{O}}(x_3)\rangle\neq 0$. In order to screen both three-point functions, one must be able to solve the simultaneous equations
\begin{equation}
    4+n-4n_-=6 \; , \hspace{1 in} 6+(6-n)-4m_-=6 \; .
\end{equation}
 The only shared solutions are $n=2$ and $n=6$, which are precisely the operators expected to show up in the Yang-Lee OPE. The contribution from $|\tilde{\mathcal{O}}\rangle \langle \mathcal{O}|$ just picks out the reflections of these two operators, $n=0$ and $n=4$. The point of this example is to show that it is possible to have a non-vanishing three-point function outside of the fusion range which does not contribute to the four-point function.
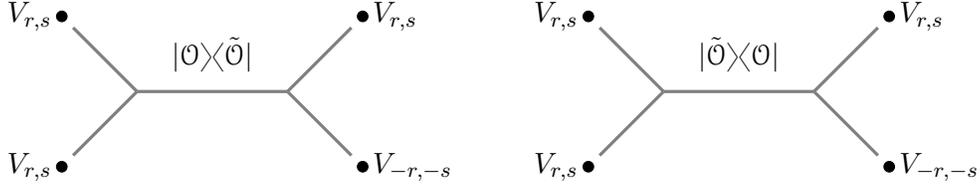
\begin{figure}[h]
\begin{tikzpicture}

\filldraw[black] (-6,5) circle (2pt) node[anchor=east] { $V_{r,s}$};
\filldraw[black] (-6,3) circle (2pt) node[anchor=east] { $V_{r,s}$};
\filldraw[black] (-2,5) circle (2pt) node[anchor=west] { $V_{r,s}$};
\filldraw[black] (-2,3) circle (2pt) node[anchor=west] { $V_{-r,-s}$};
\draw[-,gray, very thick] (-5.85,4.85) -- (-5,4);
\draw[-,gray, very thick] (-5.85,3.15) -- (-5,4);
\draw[-,gray, very thick] (-5,4) -- (-3,4);
\draw[-,gray, very thick] (-3,4) -- (-2.15,4.85);
\draw[-,gray, very thick] (-3,4) -- (-2.15,3.15);

\filldraw[black] (1,5) circle (2pt) node[anchor=east] { $V_{r,s}$};
\filldraw[black] (1,3) circle (2pt) node[anchor=east] { $V_{r,s}$};
\filldraw[black] (5,5) circle (2pt) node[anchor=west] { $V_{r,s}$};
\filldraw[black] (5,3) circle (2pt) node[anchor=west] { $V_{-r,-s}$};
\draw[-,gray, very thick] (1.15,4.85) -- (2,4);
\draw[-,gray, very thick] (1.15,3.15) -- (2,4);
\draw[-,gray, very thick] (2,4) -- (4,4);
\draw[-,gray, very thick] (4,4) -- (4.85,4.85);
\draw[-,gray, very thick] (4,4) -- (4.85,3.15);

\node[] at (-4,4.5) {$|\mathcal{O}\rangle\langle \tilde{\mathcal{O}}|$};
\node[] at (3,4.5) {$|\tilde{\mathcal{O}}\rangle\langle \mathcal{O}|$};
\end{tikzpicture}
\caption{Potential exchange for an operator $\mathcal{O}$ or its reflection $\tilde{\mathcal{O}}$ violating fusion. A \textit{pair} of distinct three-point functions must be non-vanishing simultaneously if the operator is to contribute to the four-point function.}
\end{figure}

Now consider factorizing the representative $\langle\tilde{V}(x_1)\tilde{V}(x_2)\tilde{V}(x_3)V(x_4)\rangle$ on three-point functions. In order to contribute, an operator $\mathcal{O}(x)$ must satisfy $\langle \tilde{V}(x_1)\tilde{V}(x_2)\mathcal{O}(x_3) \rangle \neq 0$ and $\langle V(x_1)\tilde{V}(x_2)\tilde{\mathcal{O}}(x_3)\rangle \neq 0$.
The simultaneous equations are
\begin{equation}
    8+n-4n_-=6 \; , \hspace{1 in} 6+(6-n)-4m_-=6 \; . 
\end{equation}
There is now an extra shared solution at $n=-2$ corresponding to the coefficient $\langle \tilde{V}_{1,2}\tilde{V}_{1,2} V_{3,5} \rangle$ that we encountered previously. In order to determine whether or not this term contributes, we need to evaluate  $\langle V(x_1)\tilde{V}(x_2)\tilde{\mathcal{O}}(x_3)\rangle$ explicitly, since considering factorization on $\langle \tilde{V}(x_1)\tilde{V}(x_2)\tilde{\mathcal{O}}(x_3) \rangle$ and $\langle V(x_1)\tilde{V}(x_2)\mathcal{O}(x_3)\rangle$ allows possible contributions from the reflected operators at $n=0,4,8$ and does not solve the problem. This correlation function requires two screening operators, but one can check that it does vanish for the correct values of the parameters:
\begin{equation}\label{eq:Vanishing}
    \langle e^{-i\am\phi(x_1)}e^{(2i\ap + 3i\am)\phi(x_2)}e^{(4i\ap +6i\am)\phi(x_3)}\prod_{i=1}^2\int d^2w_i e^{2i\alpha_-\phi(w_i)}\rangle \sim \lim_{\varepsilon\to 0}\frac{1}{\Gamma(\varepsilon)}=0 \; .
\end{equation}
 So although the reflected representative of the four-point function naively allows for contributions from outside of the fusion range, only the expected operators actually contribute. This points to a possible mechanism for the deformed linear dilaton to reproduce the correct minimal model correlation functions while botching some three-point functions: the linear dilaton does not have to set every representative of a vanishing fusion coefficient to zero, but only certain combinations.

The non-obvious vanishing of the integral \eqref{eq:Vanishing} was necessary in order for the dangerous three-point function \eqref{eq:Dangerous} to have no effect on the correlator $\langle\tilde{V}(x_1)\tilde{V}(x_2)\tilde{V}(x_3)V(x_4)\rangle$. It is tempting to look for an explanation of this mechanism in terms of the BRST structure of the model. Indeed, at least in the $(2,5)$ model the operator $V_{3,5}$ is not in the BRST cohomology and calculating the three-point function \eqref{eq:Dangerous} is akin to calculating a non-gauge-invariant quantity: the value does not matter since it will not contribute to physical BRST invariant  observables. This argument seems too fast, since we could find a model in which $V_{3,5}$ is in the cohomology while \eqref{eq:Dangerous} would still represent an apparent violation of truncation. 
Indeed, we now turn to an example in which an apparent failure of truncation is not remedied solely by the vanishing of a reflection partner three-point function.

\subsubsection*{The $(2,7)$ model}
Next we would like to verify the fusion rules in the $(2,7)$ model. The reflection map sends $n\to 10 -n$ and the selection rule is $\sum n_i -4n_-=10$. There are three fusion rules in this model to verify
\begin{align}
    M_{1,2}\times M_{1,2} &=1 + M_{1,3}\; ,\label{eq:27F1}\\
    M_{1,2}\times M_{1,3}&=M_{1,2}+M_{1,4}=M_{1,2}+M_{1,3} \; ,\label{eq:27F2}\\
    M_{1,3}\times M_{1,3}&=1  +M_{1,3}+M_{1,5}=1 +M_{1,2} +M_{1,3} \; . \label{eq:27F3} 
\end{align}
In table \ref{table:27} we list the linear dilaton momenta corresponding to these operators. 

\begin{table}[h]
\centering
\begin{tabular}{ |c|c|c|c|c|c|c|c|c|c|c| } 
 \hline
  $n$&  $s=-2$& $s=-1$&$s=0$&$s=1$&$s=2$&$s=3$&$s=4$&$s=5$&$s=6$&$s=7$ \\ \hline
    $r=0$  &$1$& $3$&$5$&$7$&$9$&$11$&$13$&$15$ &$17$& $19$\\ \hline
     $r=1$  &$-6$& $-4$&$-2\cellcolor{lightgray}$&$0\cellcolor{teal}$&$2\cellcolor{cyan}$&$4 \cellcolor{green}$&$6\cellcolor{green}$&$8\cellcolor{cyan}$&$10\cellcolor{teal}$&$12\cellcolor{lightgray}$ \\ \hline
\end{tabular}
\caption{Values of the momentum $n$ of some operators in the $(2,7)$ compact linear dilaton. The correct truncated fusion range for the model is shaded in blue and green.}\label{table:27}
\end{table}

We will consider first the four-point functions with a single reflection operator insertion. In order to check the first fusion rule \eqref{eq:27F1} we consider
\begin{equation}
\langle V_{1,2}(x_1)V_{1,2}(x_2)V_{1,2}(x_3)\tilde{V}_{1,2}(x_4) \rangle \sim \sum_{\mathcal{O}}   \; \langle V_{1,2}(x_1)V_{1,2}(x_2) |\mathcal{O}\rangle \langle \tilde{\mathcal{O}}| V_{1,2}(x_3)\tilde{V}_{1,2}(x_4)\rangle \; . 
\end{equation}
In order for an operator to contribute to this correlator one must have $\langle V_{1,2}(x_1)V_{1,2}(x_2)\mathcal{O}_{r,s}(x_3)\rangle \neq 0$ and $\langle V_{1,2}(x_1)\tilde{V}_{1,2}(x_2)\tilde{\mathcal{O}}_{r,s}(x_3)\rangle\neq 0$. This leads to two constraints
\begin{equation}
    n_{r,s}=6+4n_- \; , \hspace{.5 in}     n_{r,s}=10-4m_- \; . 
\end{equation}
The simultaneous solutions are $n_{r,s}=6$ and $n_{r,s}=10$ corresponding to $M_{1,3}$ and $M_{1,1}$ as expected.
Next we check the third fusion rule \eqref{eq:27F3} for $M_{1,3}\times M_{1,3}$ using the correlator $\langle V_{1,3}(x_1)V_{1,3}(x_2)V_{1,3}(x_3)\tilde{V}_{1,3}(x_4) \rangle$. In order for an operator $\mathcal{O}_{r,s}$ to contribute, it must satisfy the constraints
\begin{equation}
    n_{r,s}=2+4n_- \; , \hspace{.5 in} n_{r,s}=10-4m_- \; . 
\end{equation}
The simultaneous solutions are $n_{r,s}=2,6,10$, corresponding to the operators $M_{1,2}, M_{1,3}$ and $1$. In order to verify  the last fusion rule \eqref{eq:27F2}, we consider the correlator $\langle V_{1,2}(x_1)V_{1,3}(x_2)V_{1,2}(x_3)\tilde{V}_{1,3}(x_4)\rangle$. In order for an operator $\mathcal{O}_{r,s}$ to contribute, it must be the case that
\begin{equation}
    n_{r,s}=4+4n_- \; , \hspace{.5 in} n_{r,s}=8-4m_- \; . 
\end{equation}
The simultaneous solutions are $n_{r,s}=4$ and $n_{r,s}=8$, corresponding to the expected operators $M_{1,3}$ and $M_{1,2}$. At this point the general mechanism for obtaining truncation from the $\langle VVV\tilde{V}\rangle$ correlator should be clear: the truncation from below arises from the $\langle VV\mathcal{O}_{r,s}\rangle$ three-point function, where $n_{r,s}$ enters the selection rule with a positive sign. Likewise, the truncation from above is enforced by the non-vanishing of the three-point function $\langle V\tilde{V}\tilde{\mathcal{O}}_{r,s}\rangle$, in which $n_{r,s}$ enters the selection rule with a negative sign. 

Now we would like to repeat the above analysis for the four-point function with three reflection operator insertions, denoted schematically $\langle\tilde{V}\tilde{V}\tilde{V} V\rangle$. One might imagine that this correlator leads to identical constraints, but we already saw that this is not the case in the $(2,5)$ model. For instance, we can revisit the $M_{1,2}\times M_{1,2}$ fusion rules using the correlation function $\langle\tilde{V}_{1,2}(x_1)\tilde{V}_{1,2}(x_2)\tilde{V}_{1,2}(x_3)V_{1,2}(x_4)\rangle$. Factorizing on an operator $\mathcal{O}_{r,s}$, one encounters a pair of constraints from the three-point functions $\langle \tilde{V}_{1,2}(x_1)\tilde{V}_{1,2}(x_2)\mathcal{O}_{rs}(x_3)\rangle$ and $\langle \tilde{V}_{1,2}(x_1)V_{1,2}(x_2)\tilde{\mathcal{O}}_{rs}(x_3)\rangle$. They are
\begin{equation}
    n_{r,s}=-6+4n_- \; , \hspace{.5 in} n_{r,s}=10-4m_- \; . 
\end{equation}
There are five solutions $n=-6,-2,2,6,10$ to this set of equations, but the fusion rules predict only $n=6$ and $n=10$. In particular, for the case $n=-6$ the three-point function $\langle\tilde{V}_{1,2}(x_1)\tilde{V}_{1,2}(x_2)\mathcal{O}_{n=-6}(x_3)\rangle$ does not even need to be screened and does not vanish. In order to determine whether or not this apparent violation of fusion actually contributes to the four-point function, we need to check if $\langle \tilde{V}_{1,2}(x_1)V_{1,2}(x_2)\tilde{\mathcal{O}}_{n=-6}(x_3)\rangle$ vanishes.
The integral is of the usual form, but now one finds a finite coefficient multiplied by a nonvanishing ratio of divergent Gamma functions
\begin{equation}
    \langle e^{(2i\alpha_++3\alpha_-)\phi(x_1)}e^{-i\alpha_-\phi(x_2)}e^{-8\alpha_-\phi(x_3)}\prod_{i=1}^4\int d^2w_i e^{2i\alpha_-\phi(w_i)}\rangle \sim \lim_{\varepsilon \to 0} \frac{\Gamma(-1+\varepsilon)}{\Gamma(\varepsilon)} \; . 
\end{equation}
 This calculation seems to represent a genuine violation of the fusion rules (in particular, the reflection coefficient for this operator is finite). Note that, in contrast with the four-point function with a single reflection operator insertion, \textit{both} three-point functions in this example contain two reflection operators. The root of the problem seems to be that the reflection operators have larger momenta and contribute to the selection rule in the wrong direction, loosening the bounds. Apparently, the nonzero contribution from this operator is cancelled by other apparent violations of fusion in order to reproduce the correct four-point function.
As we will see in the next section, one never encounters this subtlety if one restricts attention to the four-point function with a single reflection operator.

\subsubsection*{General case $(2,p)$}
In this section we will derive the $(2,p)$ minimal model fusion rules
\begin{equation}\label{eq:Sec6Fusion}
    M_{1,s_1}\times M_{1,s_2}=\sum_{s_3=1+|s_1-s_2|}^{\min(s_1+s_2-1,2p-s_1-s_2-1)} M_{1,s_3} 
\end{equation}
from the timelike linear dilaton deformed by the marginal operator $M_-(x)$. We will only make use of the four-point function with a single reflection operator insertion.
There are two types of OPE to consider: that of an operator with itself, and the OPE between two distinct operators in the fundamental range $s\in [1,\frac{p-1}{2}]$. To treat the first case, we consider the correlator $\langle V_{1,s}(x_1)V_{1,s}(x_2)V_{1,s}(x_3)\tilde{V}_{1,s}(x_4)\rangle$
factorized onto a product of three-point functions:
\begin{equation}\label{eq:OPEsame}
    \langle V_{1,s}(x_1)V_{1,s}(x_2)|\mathcal{O}\rangle \times \langle \tilde{\mathcal{O}}|V_{1,s}(x_3)\tilde{V}_{1,s}(x_4)\rangle \; . 
\end{equation}
In order to contribute, the momentum $n$ of the operator $\mathcal{O}$ must satisfy the relations
\begin{equation}\label{eq:OPEconstraintSame}
    n=2p-4s+4n_- \; ,\hspace{.5 in} n=2p-4-4m_- \; . 
\end{equation}
Taking the difference of these two equations one finds $s-1=n_-+m_-$. There are $s$ distinct combinations of positive integers $(n_-,m_-)$ satisfying this equation. Similarly, taking the sum of \eqref{eq:OPEconstraintSame}, one finds
\begin{equation}
    n=2p-2s+2n_--2m_--2 \; .
\end{equation}
For the allowed partitions of $s-1=n_-+m_-$, the quantity $(n_--m_--1)$ spans the range $[-s,s-2]$ in increments of 2 so that the allowed values of $n$ span the range $[2p-4s,2p-4]$ in increments of 4. These are are precisely the momenta allowed by the fusion rules \eqref{eq:Sec6Fusion}. The upper bound $2p-4=2p-2p'$ is simply the reflection of the identity, and the decreases of momenta in increments of $4$ pick out a single representative of each allowed operator in the OPE. 

In order to check the OPE between two distinct operators $M_{1,s_1}(x)$ and $M_{1,s_2}(x)$ we consider the correlation function $\langle V_{1,s_1}(x_1)V_{1,s_2}(x_2)V_{1,s_1}(x_3)\tilde{V}_{1,s_2}(x_4)\rangle$
factorized onto a product of three-point functions:
\begin{equation}\label{eq:OPEdiff}
    \langle V_{1,s_1}(x_1)V_{1,s_2}(x_2)|\mathcal{O}\rangle \times \langle \tilde{\mathcal{O}}|V_{1,s_1}(x_3)\tilde{V}_{1,s_2}(x_4)\rangle \; . 
\end{equation}
Without loss of generality we will assume $s_2>s_1$. In order to contribute, the momentum $n$ of the operator $\mathcal{O}$ must satisfy
\begin{equation}
    n=2p-2s_1-2s_2+4n_- \; , \hspace{.5 in} n=2p+2s_1-2s_2-4-4m_-\;.
\end{equation}
Taking the difference of these two equations, one finds $s_1-1=n_-+m_-$. There are $s_1$ such partitions. Similarly, taking the sum yields the equation
\begin{equation}
    n=2p-2s_2+2n_--2m_--2 \; . 
\end{equation}
For the allowed partitions of $s_1-1=n_-+m_-$, the quantity $(n_--m_--1)$ spans the range $[-s_1,s_1-2]$ in increments of 2 so that the allowed values of $n$ span the range $[2p-2s_1-2s_2,2p-4+2s_1-2s_2]$ in increments of 4. These are are precisely the momenta allowed by the fusion rules \eqref{eq:Sec6Fusion}.

As discussed earlier, if we were willing to work at an irrational point (so that there can be no compensation between $\alpha_+$ and $\alpha_-$ in the selection rule) and then continue to the rational point, then there would be a shortcut to obtaining the fusion rules. In this scenario, at least for the $(2,p)$ models deformed by the single marginal deformation $M_-(x)$, it is not possible to screen a correlation function with more than a single reflection operator. Using this fact one could read off the fusion rules directly from the three-point function without resorting to the four-point function.
As an example, note that the correlator $\langle M_{1,s_1}M_{1,s_2}M_{1,s_1+s_2-1+n}\rangle$ should vanish for $n>0$. If we represent it  using a single reflection operator
\begin{equation}
    \langle e^{i(1-s_1)\am \phi(x_1)}e^{i(1-s_2)\am \phi(x_2)}e^{[2i\ap +i(s_1+s_2+n)\am] \phi(x_3)}\rangle 
\end{equation}
we find that the three-point function cannot be screened. In particular, since all of the problematic three-point functions that we encountered in previous sections involved multiple reflection operator insertions, we avoid the subtle puzzles and their resolutions.

\subsection{Comments on truncation of perturbation theory and general $(p,p')$ models}
In this section we will investigate truncation in the general $(p,p')$ compact timelike linear dilaton models. These cases involve puzzles not present in the $(2,p)$ models.

\subsubsection*{Ising model}
The Ising model corresponds to $(p,p')=(4,3)$. There are three Virasoro primaries in the model
\begin{equation}
    M_{1,1}=M_{2,3}\equiv 1\;, \hspace{.5 in} M_{1,2}=M_{2,2}\equiv \sigma \; , \hspace{.5 in}  M_{1,3}=M_{2,1}\equiv \varepsilon \; . 
\end{equation}
Their fusion rules are
\begin{equation}
    M_{12}\times M_{12}=M_{11}+M_{13}\;, \hspace{.5 in} M_{13}\times M_{13}=M_{11}\;, \hspace{.5 in}  M_{12}\times M_{13}=M_{12} \; . 
\end{equation}
For all of the unitary models, the selection rule on momenta is given by
\begin{equation}
    \sum n +2n_+p -2n_-p'= 2 \; . 
\end{equation}
In table \ref{table:Ising} we list the momenta of the linear dilaton exponentials corresponding to Ising model primaries. Our convention will be to denote the operator with the smallest value of $n$ as $V(x)$, and its reflection partner as $\tilde{V}(x)$.

Consider first the $\sigma \times \sigma$ OPE. If we only deform the linear dilaton theory by the operator $M_-(x)$, then the only representative of the four-point function $\langle \sigma\sigma\sigma\sigma\rangle$ that can be screened is 
\begin{equation}
    \langle V_{1,2}(x_1)V_{1,2}(x_2)V_{1,2}(x_3)V_{2,2}(x_4)\int d^2w e^{2i\alpha_-\phi(w)}\rangle \; .
\end{equation}
Factorizing the correlator on three-point functions, one finds conditions on the exchanged operator $\mathcal{O}_n$:
\begin{equation}
    n=-4+6n_- \; , \hspace{.5 in} n=2-6m_- \; . 
\end{equation}

\begin{table}[h]
\centering
\begin{tabular}{ |c|c|c|c|c|c|c|c|c|c|c| } 
 \hline
  $n$&  $s=-2$& $s=-1$&$s=0$&$s=1$&$s=2$&$s=3$&$s=4$&$s=5$&$s=6$&$s=7$ \\ \hline
    $r=0$  &$-5$& $-2$&$1$&$4$&$7$&$10$&$13$&$16$ &$19$& $22$\\ \hline
     $r=1$  &$-9$& $-6$&$-3$&$0\cellcolor{teal}$&$3\cellcolor{cyan}$&$6 \cellcolor{green}$&$9$&$12$&$15$&$18$ \\ \hline
     $r=2$  &$-13$& $-10$&$-7$&$-4\cellcolor{green}$&$-1\cellcolor{cyan}$&$2 \cellcolor{teal}$&$5$&$8$&$11$&$14$ \\ \hline
\end{tabular}
\caption{Values of the momentum $n$ of some operators in the $(4,3)$ compact linear dilaton. }\label{table:Ising}
\end{table}

The simultaneous solutions are $n=-4$  and $n=2$, as expected from exchange of $\varepsilon$ and the identity. Similarly, one can check that the only representative of the four-point function $\langle\varepsilon\varepsilon\varepsilon\varepsilon \rangle $ that can be screened is 
\begin{equation}
    \langle V_{1,3}(x_1) V_{1,3}(x_2)V_{1,3}(x_3)V_{2,1}(x_4)\int d^2w_1 e^{2i\alpha_-\phi(w_1)}\int d^2w_2 e^{2i\alpha_-\phi(w_2)}  \rangle \; . 
\end{equation}
Factorizing onto three-point functions, the constraints for exchange are
\begin{equation}
    n=-10+6n_- \; , \hspace{.5 in} n=2-6m_- \; . 
\end{equation}
The simultaneous solutions are $n=2,-4,-10$, while only $n=2$ is expected from the fusion rules. This appears to be the same phenomena encountered in the $(2,p)$ models when considering the four-point function with three reflection operator insertions (in this case $V_{2,1}$ has the lesser momentum and the reflection operator is $V_{1,3}$): delicate cancellations between the two anomalous contributions combine to yield the correct answer.

The ability to screen representatives of each minimal model correlator using only $M_-(x)$ in this system corresponds to the ability to choose representatives of the form $V_{1,s}(x)$ for all operators in the model. This is a general feature of the $(3,p)$ models, but will not hold for more general models as we now demonstrate.

\subsubsection*{Tricritical Ising model}
The tricritical Ising model corresponds to $(p,p')=(5,4)$. The independent operators are
\begin{equation}
    M_{1,1}=M_{3,4} \; , \hspace{.1 in} M_{1,2}=M_{3,3} \; , \hspace{.1 in} M_{1,3}=M_{3,2} \;, \hspace{.1 in} M_{1,4}=M_{3,1} \; , \hspace{.1 in} M_{2,2}=M_{2,3} \; , \hspace{.1 in} M_{2,4}=M_{2,1} \; .
\end{equation}
The selection rule on momenta in the linear dilaton system is
\begin{equation}
    \sum n +10n_+ -8n_-= 2 \; . 
\end{equation}
There are in principle 15 separate OPE's to check, but the basic problem can be illustrated using the four-point function $\langle M_{22}M_{22}M_{22}M_{22}\rangle $. It is easy to see that no representative of this correlation function can be screened using only $M_-(x)$. If we deform by both marginal operators $M_{\pm}(x)$, then it is possible to screen all five representatives of this correlator. For instance, the correlation function
\begin{equation} \label{eq:TriCritCorr1}
    \langle V_{2,2}(x_1)V_{2,2}(x_{2})V_{2,2}(x_3)V_{2,3}(x_4)\int d^2y e^{2i\alpha_+\phi(y)}\int d^2w e^{2i\alpha_-\phi(w)}  \rangle 
\end{equation}
satisfies the momentum constraint. Unfortunately, the correlator
\begin{equation} \label{eq:TriCritCorr2}
    \langle V_{2,2}(x_1)V_{2,2}(x_{2})V_{2,2}(x_3)V_{2,3}(x_4)\prod_{i=1}^{1+4k}\int d^2y_i e^{2i\alpha_+\phi(y_i)}\prod_{j=1}^{1+5k}\int d^2w_j e^{2i\alpha_-\phi(w_j)}  \rangle 
\end{equation}
also satisfies the momentum constraint for any positive $k$. This term would arise as a contribution to the four-point function from a higher order in perturbation theory. If it were to contribute, then the total answer in the deformed linear dilaton model would not match that of the minimal model. The same basic puzzle arises in any model with $p'\neq 2,3$ when we are seemingly forced to deform by both marginal operators in order to screen all relevant four-point functions. 

\begin{table}[h]
\centering
\begin{tabular}{ |c|c|c|c|c|c|c|c|c|c|c| } 
 \hline
  $n$&  $s=-2$& $s=-1$&$s=0$&$s=1$&$s=2$&$s=3$&$s=4$&$s=5$&$s=6$&$s=7$ \\ \hline
    $r=0$  &$-7$& $-3$&$1$&$5$&$9$&$13$&$17$&$21$ &$25$& $29$\\ \hline
     $r=1$  &$-12$& $-8$&$-4$&$0\cellcolor{teal}$&$4\cellcolor{cyan}$&$8 \cellcolor{green}$&$12\cellcolor{lime}$&$16$&$20$&$24$ \\ \hline
     $r=2$  &$-17$& $-13$&$-9$&$-5\cellcolor{violet}$&$-1\cellcolor{olive}$&$3\cellcolor{olive} $&$7\cellcolor{violet}$&$11$&$15$&$19$ \\ \hline
     $r=3$  &$-22$& $-18$&$-14$&$-10 \cellcolor{lime}$&$-6\cellcolor{green}$&$-2 \cellcolor{cyan} $&$2\cellcolor{teal}$&$6$&$10$&$14$ \\ \hline
\end{tabular}
\caption{Values of the momentum $n$ of some operators in the $(5,4)$ compact linear dilaton.}
\end{table}

The fusion rule associated to this correlator is
\begin{equation}
    M_{2,2}\times M_{2,2} = M_{1,1} + M_{1,2}+M_{1,3} + M_{1,4} \; . 
\end{equation}
Factorizing on three-point functions, we find the constraints for exchange:
\begin{equation}\label{eq:TriCrit}
    n=4-10n_++8n_-  \; , \hspace{.5 in} n=2+10m_+-8m_- \; . 
\end{equation}
In table \ref{table:TriCrit1} we list the lowest lying values of $n$ given by  \eqref{eq:TriCrit}. The values of $n_\pm,m_{\pm}$ depicted in this table are special since, for fixed $n$, they satisfy $n_++m_+=1$ and $n_-+m_-=1$. Not coincidentally, these values of $n$ are precisely those expected from the minimal model fusion rules. This is related to the fact that the correlator \eqref{eq:TriCritCorr1} has one insertion of $M_+(x)$ and one insertion of $M_-(x)$. In fact, given a four-point function with a fixed number of screening charges, determination of the possible operator exchanges amounts to determining the pairs of three-point functions that can be simultaneously screened using all partitions of the available screening charges. 

\begin{table}[h]
\centering
\begin{tabular}{ |c|c|c|c| } 
\hline
$(n_+,n_-)$&$4-10n_++8n_-$& $(m_+,m_-)$ &$2+10m_+-8m_-$\\\hline
$(0,0)$ &$4$&$(1,1)$&$4$ \\ \hline
$(0,1)$ &$12$&$(1,0)$&$12$ \\ \hline
$(1,0)$ &$-6$&$(0,1)$&$-6$ \\ \hline
$(1,1)$ &$2$&$(0,0)$&$2$ \\ \hline
\end{tabular}
\caption{Low-lying operator exchanges corresponding to the correlator \eqref{eq:TriCritCorr1}}\label{table:TriCrit1}
\end{table}

To see this, note that taking the difference of the two equations in \eqref{eq:TriCrit} yields
\begin{equation}
    2=10(n_++m_+)-8(n_-+m_-) \; . 
\end{equation}
There are an infinite number of solutions to this equation, all of the form
\begin{equation}\label{eq:TriSols1}
    n_++m_+=1+4k \;, \hspace{.5 in} n_-+m_-=1+5k \; . 
\end{equation}
Taking the sum of the equations in \eqref{eq:TriCrit} and using \eqref{eq:TriSols1} gives an infinite set of solutions 
\begin{equation}
    n=4 -10n_+ +8n_-  
\end{equation}
with $n_{\pm}$ unconstrained since $m_{\pm}$ enters with coefficient $1$ in \eqref{eq:TriSols1}. This point of view makes it obvious that a correction like  \eqref{eq:TriCritCorr2} will be problematic since it allows for many more potential operator exchanges than are desired. 

It seems plausible that corrections like \eqref{eq:TriCritCorr2} do vanish due to the BRST structure inherent in the model. Indeed, these terms all involve one or more insertions of the form
\begin{equation}\label{eq:HigherOrd1}
   \left( \int d^2w e^{2i\alpha_-\phi(w)} \right)^p \; . 
\end{equation}
Schematically, when we factorize this expression (inside correlation functions) using the techniques of section \ref{sec:Integrals}, the result  involves a line integral of the form
\begin{equation}\label{eq:HigherOrd2}
    \left(\int dz e^{2i\alpha_-\phi_L(z)}\right)^p \; . 
\end{equation}
This is most obvious in the holomorphic Coulomb gas formalism, where the analog of \eqref{eq:HigherOrd1} is literally \eqref{eq:HigherOrd2}. Now \eqref{eq:HigherOrd2} is of the form $Q_-^p$, and $Q_-$ is a differential satisfying $Q^{p-s}Q^s=0$. This suggests that all higher order contributions might vanish as desired. This argument, even if correct, obviously needs to be made more precise.

\subsection{Summary of fusion rules and truncation}
In this section we provide a brief summary of the derivation of the minimal model fusion rules within the BRST quotiented compact timelike Liouville description.
The first important point is that the minimal model fusion rule \textit{cannot} be read off from the $U(1)$ selection rule for a single three-point function. For two fixed minimal model exponentials, infinitely many three-point functions can be screened. In fact, some representatives of the three-point functions which should vanish outside of the fusion range are finite and nonzero. The resolution of this puzzle relies on the fact that factorization of a four-point function $\langle V_{\alpha_1}(x_1)V_{\alpha_2}(x_2)V_{\alpha_3}(x_3)V_{\alpha_4}(x_4)\rangle$ in a model with a background charge asymmetry involves a particular asymmetric combination of three-point function coefficients
\begin{equation}\label{eq:AsymmetricOPE}
    C(\alpha_1,\alpha_2, -2iQ -\alpha)C(\alpha,\alpha_3,\alpha_4) \; .
\end{equation}
In particular, a four-point function of identical operators is not related to squares of OPE coefficients. This means that it is possible for the Liouville description to reproduce the minimal model fusion rules without getting all representations of the three-point functions correct: it is enough for either term in \eqref{eq:AsymmetricOPE} to vanish individually. Said differently, one must be able to simultaneously screen both three-point functions in \eqref{eq:AsymmetricOPE} using the same combination of screening charges used to screen the four-point function if the operator $V_{\alpha}(x)$ is to contribute.

In order to actually determine the fusion rules, it is enough to consider the four-point function with a single reflection operator insertion: 
\begin{equation}
\langle V_{\alpha_1}(x_1)V_{\alpha_2}(x_2)V_{\alpha_3}(x_3)\tilde{V}_{\alpha_4}(x_4) \rangle \sim \sum_{\mathcal{O}} \;  \langle V_{\alpha_1}(x_1)V_{\alpha_2}(x_2) |\mathcal{O}\rangle \langle \tilde{\mathcal{O}}| V_{\alpha_3}(x_3)\tilde{V}_{\alpha_4}(x_4)\rangle \;  . 
\end{equation}
 This four-point function requires some \textit{minimal} number of screening charges, which must be distributed onto the three-point functions upon factorization: 
\begin{equation}\label{eq:Partitions}
\langle V_{\alpha_1}(x_1)V_{\alpha_2}(x_2)V_{\alpha_3}(x_3)\tilde{V}_{\alpha_4}(x_4) S_+^{N_+}S_-^{N_-}\rangle_{\text{free}}
 \rightarrow \langle V_{\alpha_1}(x_1)V_{\alpha_2}(x_2) S_+^{n_1}S_-^{m_1}|\mathcal{O}\rangle \langle \tilde{\mathcal{O}}| V_{\alpha_3}(x_3)\tilde{V}_{\alpha_4}(x_4) S_+^{n_2}S_-^{m_2} \rangle \; .  
\end{equation}
The different (non-vanishing) partitions of the screening charges satisfying $n_1+n_2=N_+$ and \\ $m_1+m_2=N_-$ pick out the correct operators that appear in the minimal model fusion rules. Roughly speaking, the momentum of the operator $\mathcal{O}$ enters the selection rule with a positive sign in one three-point function and with a negative sign  in the other three-point function (the one involving $\tilde{\mathcal{O}}$). This is responsible for truncation from above and from below. 

At lowest order in perturbation theory (using the minimal number of screening charges), the range of operators arising in \eqref{eq:Partitions} matches the fusion rules of the minimal model. However, this range is enlarged if $N_+$ and $N_-$ are larger than the minimal value needed to screen the four-point function. It therefore appears that truncation of perturbation theory is necessary in order to provide an honest derivation of the minimal model results within the linear dilaton theory.

One can artificially enforce truncation of perturbation theory in the $(2,p)$ models by deforming by the single marginal operator $M_-(x)$. In this case, a representative of each minimal model correlator can be screened using a unique number of screening charges and the fusion rules can be rigorously derived. This is not the case for the generic $(p,p')$ model: in order to screen a representative of all minimal model correlation functions one must make use of  both marginal operators. In order to rigorously connect with the minimal models, it must be demonstrated that perturbation theory truncates in these theories. It seems plausible that the peculiar BRST structure inherent in the model is responsible for the truncation, but this needs to be explored more carefully and is left to future work.

\section{Conclusions, questions and future work }
We have demonstrated that the deformed compact timelike linear dilaton (with a peculiar fractional-winding spectrum and a BRST quotient) reproduces minimal model observables. While the analysis does present a coherent derivation of the Coulomb gas rules from standard operations in quantum field theory, we feel that the discussion could be slightly improved. 

We evaluated the torus partition function of the model using a trace in the BRST-quotiented Hilbert space, but it should also be possible to reproduce the answer through the functional integral. This would seem to require a discrete  term in the action capable of introducing the appropriate minus signs in the sum over instantons. There is no candidate term available in the model as defined in this paper, but it seems possible that we might have neglected a discrete gauge field in focusing primarily on genus zero observables. Indeed, a discrete gauge field coupled to a sum over spin structures \cite{AlvarezGaume:1987vm} seems natural if we view the Coulomb gas scalar as the bosonization of the original Feigin-Fuchs fermionic resolution. Understanding this point would simplify the analysis on higher genus surfaces. It would also be nice to understand the precise relation (including the winding states and correlation functions) between our construction and the $SL(2,\mathbb{R})$ quantum Hamiltonian reduction. 
The question remains whether or not to deform by both marginal operators in the general $(p,p')$ model. Doing so would seem to involve contributions from infinitely many orders in perturbation theory, but there is possibility for truncation due to the BRST structure.   We also feel that a more direct explanation for the reflection identification is needed, and might arise from a more thorough study of the zero mode quantum mechanics.  

The Coulomb gas formalism on surfaces with boundary seems to be relatively unexplored (although see \cite{Schulze:1996qm,Kawai:2002vd,Kawai:2002pz}), and might be used to simplify certain calculations in the minimal string. It would also be useful to understand Distler's derivation of topological gravity using the $\eta\xi$ system, paying close attention to the global issues referred to in section \ref{sec:EtaXi}. Finally, the primary motivation for the paper was to understand and rigorously derive the connection between the JT model of two dimensional gravity and the worldsheet description of the $(2,p)$ minimal string. We hope to address this problem soon.

\subsection*{Acknowledgments}
DK would like to thank Lorenz Eberhardt, Shota Komatsu, Shu-Heng Shao and Douglas Stanford for many useful conversations and for comments on the draft. DK would especially like to thank Nathan Seiberg for suggesting the problem and for vital collaboration at an early stage of the project.
DK gratefully acknowledges support from DOE grant DE-SC0009988, the Chooljian family and the Adler Family Fund.
RM is supported in part by Simons Investigator Award \#620869.

\bibliographystyle{apsrev4-1long}
\bibliography{Bib.bib}
\end{document}